\documentclass[a4paper,11pt]{article}
\pdfoutput=1 

\usepackage{jcappub} 

\usepackage[T1]{fontenc} 

\usepackage{graphicx}
\usepackage{float}
\usepackage{hyperref}
\usepackage{xcolor, xstring}
\usepackage{amssymb, amsmath, mathtools}
\usepackage{listings}
\usepackage{lscape}
\usepackage{lipsum}
\usepackage{multicol}
\usepackage{physics}
\usepackage{natbib}
\usepackage{txfonts}
\usepackage[small]{caption}
\usepackage{xspace}
\usepackage{comment}

\newcommand{\pot}[2]{#1 \times 10^{#2}}

\newcommand{\boostO}{\hat{\mathcal{O}}_x}

\newcommand{\DiffO}{\hat{\mathcal{D}}_x}

\newcommand{\diff}{\mathop{}\!\mathrm{d}}

\newcommand{\Gspec}{{G}}
\newcommand{\Yspec}{{{Y}}}

\newcommand{\Mspec}{{M}}

\newcommand{\Litebird}{{\it LiteBIRD}\xspace}

\newcommand{\PICO}{{\it PICO}\xspace}
\newcommand{\PIXIE}{{\it PIXIE}\xspace}

\newcommand{\COBEF}{{\it COBE/FIRAS}\xspace}

\newcommand{\Mpc}{{\rm Mpc}}

\newcommand{\expf}[1]{{{\rm e}^{#1}}}

\newcommand{\Tz}{{T_{z}}}

\newcommand{\Planck}{\textit{Planck}\xspace}

\newcommand{\vgh}{{\hat{\boldsymbol\gamma}}}

\newcommand{\xc}{x_{\rm c}}

\newcommand{\id}{{\,\rm d}}

\newcommand{\beq}{\begin{equation}}   %

\newcommand{\eeq}{\end{equation}}   %

\newcommand{\beqa}{\begin{eqnarray}}   %

\newcommand{\eeqa}{\end{eqnarray}}   %

\newcommand{\bealf}[1]{\begin{align} #1 \end{align}}

\newcommand{\beal}{\begin{align}}
\newcommand{\enal}{\end{align}}

\newcommand{\bspl}{\begin{split}}

\newcommand{\espl}{\end{split}}

\newcommand{\bsub}{\begin{subequations}}

\newcommand{\esub}{\end{subequations}}

\newcommand{\bmulti}{\begin{multline}}   %

\newcommand{\beqm}{\begin{mathletters}}   %

\newcommand{\eeqm}{\end{mathletters}}   %

\newcommand{\kB}{k_{\rm B}}

\newcommand{\me}{m_{\rm e}}

\newcommand{\vek} [1]{\mbox{\boldmath${#1}$\unboldmath}}

\newcommand{\Thz}{\theta_{z}}

\usepackage{bbm}

\usepackage{cancel}

\title{Spectro-spatial evolution of the CMB III: transfer functions, power spectra and Fisher forecasts}

\author[a]{Thomas Kite}

\author[b,c]{, Andrea Ravenni}

\author[a]{and Jens Chluba}

\affiliation[a]{Jodrell Bank Centre for Astrophysics, School of Physics and Astronomy, The University of Manchester, Oxford Road, Manchester, M13 9PL, U.K.}

\affiliation[b]{Dipartimento di Fisica e Astronomia ``Galileo Galilei'', Università degli Studi di Padova, via Marzolo 8, I-35131, Padova, Italy.}
\affiliation[c]{INFN, Sezione di Padova, via Marzolo 8, I-35131, Padova, Italy.}

\emailAdd{Thomas.Kite@Manchester.ac.uk}
\emailAdd{Andrea.Ravenni@unipd.it}
\emailAdd{Jens.Chluba@Manchester.ac.uk}

\date{Dec 2022}

\begin{document}

\abstract{In this paper, we provide the first computations for the distortion transfer functions of the cosmic microwave background (CMB) in the perturbed Universe, following up on paper I and II in this series. We illustrate the physical effects inherent to the solutions, discussing and demonstrating various limiting cases for the perturbed photon spectrum. We clarify the relationship between distortion transfer functions and the photon spectrum itself, providing the machinery that can then compute constrainable CMB signal power spectra including spectral distortions for single energy injection and decaying particle scenarios. Our results show that the $\mu \times T$ and $y\times T$ power spectra reach levels that can be constrained with current and future CMB experiments without violating existing constraints from \COBEF. The amplitude of the cross-correlation signal directly depends on the average distortion level, therefore establishing a novel fundamental link between the state of the primordial plasma from redshift $10^3 \lesssim z\lesssim \pot{3}{6}$ and the frequency-dependent CMB sky. This provides
a new method to constrain average early energy release using CMB imagers. As an example we derive constraints on single energy release and decaying particle scenarios. This shows that \Litebird may be able to improve the energy release limits of \COBEF by up to a factor of $\simeq 2.5$, while \PICO could tighten the constraints by more than one order of magnitude. The signals considered here could furthermore provide a significant challenge to reaching cosmic variance-limited constraints on primordial non-Gaussianity from distortion anisotropy studies. Our work further highlights the immense potential for a synergistic spectroscopic approach to future CMB measurements and analyses.}
\maketitle

\section{Introduction}
The study of perturbations in the primordial plasma has delivered a wealth of cosmological information in the past two decades. Through a combination of theoretical and numerical tools it has been possible to yield not only strong constraints on the initial conditions that seed these perturbations, but also tight limits on the exact constituents of the cosmic inventory \cite{WMAP_params, Planck2015params}. All this insight into the Universe's primordial origins ensued from observations of the photon anisotropies at the last scattering surface, an avenue of discovery in turn made possible by the tight coupling between photons and the rest of the plasma mediated via the baryonic components of the fluid \citep{Sunyaev1970, Peebles1970, Silk1968, Hu1995CMBanalytic}.

While traditional approaches to studying the early Universe via the Einstein-Boltzmann equations \cite{Ma1995, Seljak1997, Hu1997} capture many aspects of the problem, it is arguable that an entirely novel dimension is still \textit{on the table}. 
In its complete form, the photon phase space distribution carries dependence on time, spatial coordinates and momentum. Through various manipulations (Fourier transforms and spherical harmonic projections) and assumptions (e.g., Gaussian perturbations) these degrees of freedom are captured with wavenumber $k$ and Legendre moment $\ell$. The momentum of the distribution is usually only crudely captured by modelling the frequency spectrum as a blackbody with varying temperature -- a consequence of assuming that all energy is thermalised instantaneously in most primordial scenarios. It is well known, however, that the primordial photon spectrum has a greater diversity of spectral shapes at the background level, known as Spectral Distortions (SDs) \cite{Zeldovich1969, Sunyaev1970mu, Illarionov1975, Illarionov1975b, Danese1982, Burigana1993, Hu1993}.

In the first paper of this series \citep[][henceforth `paper I']{Chluba:spectro_spatial_I}, we generalised and expanded the traditional average Boltzmann hierarchy to also span the dimension offered through spectral dependence. By understanding the photon frequency hierarchy as a discretised sum over new basis functions, $Y_n(x)$, of dimensionless frequency $x$, we can accurately model the evolution of the photon spectrum including the residual-era.
In the second paper \citep[][henceforth `paper II']{Chluba:spectro_spatial_II} we use this discretised formalism to extend the spatial Boltzmann hierarchy, thus completing the triad of variables, leaving no information unexplored in the photon sector of the primordial plasma. This allows us to include the main effects relevant to the evolution of primordial distortion anisotropies, namely, Doppler and potential driving, anisotropic heating, perturbed thermalisation and the full spectral evolution from $y\rightarrow \mu\rightarrow T$ across cosmic history.

We previously showed this method works for the evolution of the background spectrum by replicating the average thermalisation Green's function \citep{Chluba2013Green, Lucca2020}. In this paper (Sect~\ref{sec:anisotropic_photon_spectrum}), we apply this formalism to the evolution of anisotropic photon spectra. By studying numerical solutions for the spectrum we show that in the presence of average distortions there are three dominant sources of anisotropies. Firstly, and perhaps most familiar, is Doppler boosting of the background spectrum, whether this originates from potential decay or baryonic Doppler driving \cite{Chluba:2x2}. The boost operator, $\boostO=-x\partial_x$, is also responsible for the cosmic microwave background (CMB) temperature and distortion dipole induced by our own motion \citep{Danese1977, Balashev2015, deZotti2015}. Secondly we have direct anisotropic heating, where the same mechanism causing a global source of energy will inevitably have some patch-to-patch variations (i.e. via variations in local clocks). Finally there is a source of anisotropies associated with the diffusion of the background spectrum, modulated by local temperature patches [see Eq.~\eqref{eq:evol_1_final_Yi_1st_ord}].

A crucial step for interpreting perturbed spectra in terms of SED (i.e., spectral energy distribution) amplitudes is discussed in Sect.~\ref{sec:change_of_basis}. Essentially, there are many ways of describing a spectrum as a series of coefficients, an ambiguity which is important for relating the modelled spectrum and observations (see Sect.~\ref{sec:forecasts_basis_discussion}). If one took the SED amplitudes in the $Y_n$ basis at face value they would falsely imply that almost no $y$ and $\mu$ distortions are present. In reality because of mutual cancellations and non-trivial overlaps between the modes it is possible to compress the information by \textit{projecting} out the usual SD amplitudes, using only a few residual modes to capture the rest (see paper I). We use this Principle Component Analysis (PCA) technique to show results in a reliable way, which is motivated by the observational procedure.

With these clear definitions of SED amplitude we can calculate transfer functions for different spectral modes (Sect.~\ref{sec:transfer_functions}), which thus allows us to present power spectra for the photon spectrum (see Sect~\ref{sec:power_spectrum}). This direct link to what would be seen across the CMB sky is a big step in SD cosmology, since we can now infer properties of the background spectrum from the SD anisotropies, and therefore place limits on primordial energy release. Furthermore, we argue it is possible to place limits on the time and details of the injection by studying the shapes and relative heights of the {\it SD acoustic peaks}.
We demonstrate this technique by presenting forecasted constraints on single energy injection and particle decay (Sect.~\ref{sec:forecasts}). With current data from \Planck we forecast independent and novel limits which are comparable with \COBEF. With future missions like \Litebird and \PICO it is possible to push the limits to be an order of magnitude better than \COBEF, and with potentially much more discriminatory power as to the cause of injection. This opens the exciting opportunity for full spectro-spatial explorations of early-universe and particle physics, bringing CMB anisotropy and spectral distortion science together.
In future, this synergy will be further explored and demonstrated, firstly by focusing on the detailed evolution of non-Gaussian perturbations and secondly with detailed forecasts based on \Planck data using realistic sky simulation.

\section{Generalized photon Boltzmann hierarchy}

\subsection{Brief recap of the important equations from paper I+II}
\label{sec:recap_equations}
For convenience we briefly summarise the bottom line results from the companion papers, which we refer to for more details and clarification of notation. The treatment of paper I introduces a new set of spectral shapes which in addition to the usual shapes form a sufficiently complete basis to model spectral evolution, as seen by comparing to full binned calculations while reducing the number of equations by at two or three orders of magnitude \citep{Chluba2011therm, Acharya2021large}. In the new formalism, the photon moments are packaged together with SD moments in a vector, $\vek{y}$, with convention $\vek{y}=(\Theta, y, y_1, ..., y_n, \mu )^{\rm T}$. These spectral parameters decompose the distortion SED into temperature shift, $G(x)$, $y$-distortion, $Y(x)$, $n^{\rm th}$ boost of $Y$, $Y_n(x)=(1/4)^n\boostO^n Y(x)$, and $\mu$-distortion, $M(x)$. The boost operator is simply $\boostO=-x\partial_x$ with dimensionless frequency variable $x=h\nu/\kB\Tz$, where the reference temperature variables scales as $\Tz\propto (1+z)$.

The treatment of paper II generalises and extends the standard spatial Boltzmann hierarchy for early-universe perturbations \citep{Ma1995, Seljak1997, Hu1997} to describe the full spectro-spatial evolution of the photon field. As such, many equations remain the same as for the standard Boltzmann hierarchy, unless otherwise stated.\footnote{In comparison to \citep{Hu1997} we use $\Theta^{\rm Hu}_\ell=(2\ell+1)\Theta_\ell$, which also is the definition used in \cite{Ma1995}. For $\Phi$ and $\Psi$, we follow the sign convention of \citep{Hu1997}, which means we have $\Phi^{\rm Ma}=-\Phi$ as defined in \cite{Ma1995}.} Specifically, the gravitational potentials, matter densities and velocities and neutrino perturbations remain unchanged. Spanning the same basis as mentioned above, we define a heating vector $\vek{Q}$ and thermalisation vector $\vek{D}$. The former usually only has one non-zero entry contributing to the $y$-distortion amplitude, while the latter sources $\Theta$ from $\mu$ to capture the effect of photon production processes. The Kompaneets operator, describing the Compton scattering process in the spectral diffusion problem of the local monopole spectrum, is cast into the same vector space and can thus be represented by a scattering matrix, $M_{\rm K}$, which gradually converts $Y(x)$ to $M(x)$ along a sequence of intermediate $Y_n(x)$ spectra. A similar description exists for the Doppler boosting operator; however, by construction this appears in the equation as $\vek{b} = M_{\rm B}\vek{y} + (1,0,0,...)^{\rm T}$, where we have added an inhomogeneous contribution to the $\Theta$ component arising from boosts on the background blackbody. $M_{\rm D}=M_{\rm B}(M_{\rm B}-3 I)$ is similarly associated with the boost operator, being the matrix counterpart of the diffusion operator $\DiffO=x^{-2}\partial_x x^4\partial_x$. Both $M_K$ and $M_{\rm B}$ for various representations have been explained in paper I \& II and can be found at \url{www.chluba.de/CosmoTherm}, together with several illustrating videos.

Superscript $X^{(n)}$ shall indicate the order of perturbation, while subscript $X_\ell \equiv \sum_{m=-\ell}^{m=\ell}X_{\ell m}Y_{\ell m}$ indicates the angular moment of the variable\footnote{In rare cases of denoting the angular moment of one of the $Y_n(x)$ amplitudes we will use $y_{n,\ell}$ as the convention, and similarly will rename the y-distortion amplitude $y_{0,\ell}$. In many cases we simply label the $\ell$ explicitly for clarity.} (e.g. $\vek{y}^{(1)}_2$ is the summed quadrupole of the photon vector at first perturbed order). We shall use $\tilde{\vek{y}}_\ell$ for the corresponding Legendre coefficient. We furthermore use conformal time $\eta$ to describe the evolution.
The photon equations in this extended Boltzmann hierarchy are then given by (see paper II)
\bsub
\label{eq:evol_1_final}
\bealf{
\label{eq:evol_1_final_a}
\frac{\partial \vek{y}^{(0)}_0}{\partial \eta}
&=\tau' \Thz\left[M_{\rm K}\,\vek{y}^{(0)}_0+\vek{D}^{(0)}_0\right]+\frac{{\vek{Q}'}^{(0)}}{4},
\\
\label{eq:evol_1_final_Yi_1st_ord}
\frac{\partial \vek{y}^{(1)}}{\partial \eta}+\vgh\cdot \nabla \vek{y}^{(1)}
&=-\vek{b}^{(0)}_0\left(\frac{\partial \Phi^{(1)}}{\partial \eta}+ \vgh\cdot \nabla\Psi^{(1)} \right)
+\tau'\left[\vek{y}_0^{(1)}+\frac{1}{10}\,\vek{y}_2^{(1)}-
\vek{y}^{(1)}+\beta^{(1)}\chi\,\vek{b}^{(0)}_0\right]
+\frac{{\vek{Q}'}^{(1)}}{4}
\\ \nonumber
&
\!\!\!\!\!\!\!\!\!\!\!\!\!\!\!\!
+\tau'\Thz\left\{M_{\rm K}\,\vek{y}^{(1)}_0+\vek{D}^{(1)}_0
+\left[\delta_{\rm b}^{(1)}+\Psi^{(1)}\right]\left(M_{\rm K}\,\vek{y}^{(0)}_0+\vek{D}^{(0)}_0\right)
+\Theta^{(1)}_0\left(\vek{D}^{(0)}_0
+
M_{\rm D}\,\vek{y}^{(0)}
-\vek{S}^{(0)}\right)\right\},
}
\bealf{
\vek{D}^{(0)}
&=
\left(
     \gamma_T\xc\,\mu^{(0)}, 0, 0, \ldots, 0, -\gamma_N\xc\,\mu^{(0)}
\right)^T,\qquad
\vek{D}^{(1)}
=
\left(
     \gamma_T\xc\,\mu^{(1)}, 0, 0, \ldots, 0, -\gamma_N\xc\,\mu^{(1)}
\right)^T, \nonumber \\
\dot{\vek{Q}}^{(0)}
&=\left(0, \frac{\dot{Q}^{(0)}_{\rm c}}{\rho_z}, 0, \ldots, 0, 0\right)^T, \qquad
\dot{\vek{Q}}^{(1)}
=\left(0,\frac{\dot{Q}^{(1)}_{\rm c}}{\rho_z}
+\Psi^{(1)}\frac{\dot{Q}^{(0)}_{\rm c}}{\rho_z}, 0, \ldots, 0, 0\right)^T,
 \nonumber \\
\vek{S}^{(0)}
&=\left(0, \delta_{\gamma,0}^{(0)}+
4 \Theta_{\rm e}^{(0)},
-4 \Theta_{\rm e}^{(0)}
, \ldots, 0, 0\right)^T,
 \nonumber
}
\esub
where the first equation describes the effect of energy release on the average CMB spectrum, and the second is for the CMB anisotropies. For details on all terms we refer the reader to paper II, however the most important terms for this paper are discussed in the following paragraphs.

For convenience we also give the Fourier and Legendre transformed form of the equations, where $k$ shall denote the wavenumber of the mode. These equations more closely resemble the traditional implementation of Eq.~\eqref{eq:evol_1_final} in Einstein-Boltzmann solvers \citep[e.g.,][]{CAMB, CLASSCODE}
\bsub
\label{eq:evol_1_final_hierarchy}
\bealf{
\label{eq:evol_1_final_hierarchy_bg}
\frac{\partial \vek{y}^{(0)}_0}{\partial \eta}
&=\tau'\Thz\left[M_{\rm K}\,\vek{y}^{(0)}_0+\vek{D}^{(0)}_0\right]+\frac{{\vek{Q}'}^{(0)}}{4},
\\[1mm]
\label{eq:evol_1_final_hierarchy_0}
\frac{\partial \tilde{\vek{y}}^{(1)}_0}{\partial \eta}
&=-k\,\tilde{\vek{y}}^{(1)}_1\!-\!
\frac{\partial \tilde{\Phi}^{(1)}}{\partial \eta}
\vek{b}^{(0)}_0
+\frac{{\vek{Q}'}^{(1)}}{4}
\\
\nonumber
&\qquad
+\tau'\Thz\left\{
M_{\rm K}\,\tilde{\vek{y}}^{(1)}_0+\vek{D}^{(1)}_0
+\left[\tilde{\delta}_{\rm b}^{(1)}+\tilde{\Psi}^{(1)}\right]\left(M_{\rm K}\,\vek{y}^{(0)}_0+\vek{D}^{(0)}_0\right)
+\tilde{\Theta}^{(1)}_0\left(\vek{D}^{(0)}_0
+
M_{\rm D}\,\vek{y}^{(0)}
-\vek{S}^{(0)}\right)
\right\},
\\
\frac{\partial \tilde{\vek{y}}^{(1)}_1}{\partial \eta}
&=k \,
\left(\frac{1}{3}
\tilde{\vek{y}}_{0}-\frac{2}{3}
\tilde{\vek{y}}_{2}\right)
+\frac{k}{3}\tilde{\Psi}^{(1)} \,\vek{b}^{(0)}_0
-\tau'\left[
\tilde{\vek{y}}^{(1)}_1-\frac{\tilde{\beta}^{(1)}}{3}\vek{b}^{(0)}_0\right],
\\
\frac{\partial \tilde{\vek{y}}^{(1)}_2}{\partial \eta}
&=
k \,
\left(\frac{2}{5}\tilde{\vek{y}}^{(1)}_1-\frac{3}{5}
\tilde{\vek{y}}^{(1)}_3
\right)
-\frac{9}{10}\,\tau'\,\tilde{\vek{y}}^{(1)}_2,
\\
\frac{\partial \tilde{\vek{y}}^{(1)}_{\ell\geq 3}}{\partial \eta}
&=
k \,
\left(\frac{\ell}{2\ell+1}
\tilde{\vek{y}}_{\ell-1}-
\frac{\ell+1}{2\ell+1}
\tilde{\vek{y}}_{\ell+1}\right)
-\tau'\tilde{\vek{y}}^{(1)}_\ell.
}
\esub
and can be solved using stiff ordinary differential equation (ODE) routines \citep{Chluba2010}. The equation set takes a form that is extremely similar to the standard photon brightness temperature equation with the differences that i) the average CMB monopole can evolve, ii) Doppler and potential driving terms now affect various spectral parameters and iii) the local monopole sees new effects from thermalisation process and energy injection. Some first discussion of the expected physical effects was already given in paper II. Here, we will now demonstrate all these using numerical solutions of the transfer functions, and illustrate how they eventually affect the CMB signal power spectra. Note that we have not included polarisation effects in our description of the spectro-spatial problem; however, this should not affect the main conclusions significantly. We have included polarisation effects for the standard $\Theta$, which on the one hand allows us to compare power spectrum solutions with {\tt CLASS}, and on the other hand provides important cross correlations between SD and $E$-modes (see Sect.~\ref{sec:forecasts}).

\subsection{Principal sources of anisotropic distortions}
\label{sec:main_sources}
It will be useful for interpreting the following sections results to pause and discuss some features of Eq.~\eqref{eq:evol_1_final}. Firstly we note the presence of distinct timescales: Thompson terms are weighted by $\tau'$ while Kompaneets and thermalisation terms are weighted by $\tau ' \theta_z$, where $\theta_z=\kB\Tz/\me c^2$ is the dimensionless temperature variable. This implies the former is the dominant interaction, however only affecting higher multipoles of the distribution, leaving the latter as the dominant term for the monopole. We note that $\tau ' \theta_z$ decreases with time, lending it greater importance in the $\mu$-era. Furthermore the production of photons carries an implicit timescale in the critical frequency $\xc$, effectively shutting off photon creation for $z\lesssim \pot{2}{5}$ \citep{Chluba2014}.

Following some mechanism of average energy injection $\vek{Q}^{\prime (0)}$ which forms the background distortion, three main sources of anisotropic distortions are present: boosting, anisotropic heating and perturbed thermalisation.
\begin{itemize}
    \item Firstly the boosted background spectrum $\vek{b}^{(0)}_0$ appears twice in Eq.~\eqref{eq:evol_1_final_Yi_1st_ord}, once as gravitational boosting which is strongly associated with horizon crossing, and again as the Doppler boosting from local baryon velocities $\beta^{(1)}$. This simply sources the boosted spectrum, e.g. for early energy injection times there will be a spectrum resembling $\boostO \Mspec(x)$ sourced in local patches.
    One hallmark of the boosting effect is that early time injection yields $\boostO \Mspec(x)$, which gives a same-sign combination of $y^{(1)}$ and $\mu^{(1)}$ from performing the PCA projection. On the contrary late time injection yields $\boostO \Yspec(x)$, which gives an opposite-sign combination of $y^{(1)}$ and $\mu^{(1)}$.
    \item The second source is from direct anisotropic heating, which can be from modulations of the background heating $\simeq \Psi^{(1)} Q_{\rm c}^{\prime (0)}$ or from an explicit model dependent heating term $Q_{\rm c}^{\prime (1)}$ (below we will consider the heating from decaying particles which is thus modulated by $\delta_{\rm dm}$). These terms arise momentarily from energy injection, and then undergo thermalisation through $M_{\rm K}\vek{y}_0^{(1)} + \vek{D}_0^{(1)}$.
    There is one more term following this behaviour other than these two explicit ones: while arising from Kompaneets scattering, the term $\Theta^{(1)}_0 \dot{Q}^{(0)}_{\rm c}/4\tau'\theta_z = \Theta^{(1)}_0\left(\Theta_{\rm e}-\Theta_{\rm eq}\right) \in \Theta^{(1)}_0 \vek{S}$ in practice resembles a modulation to heating. It arises from terms associated with electron heating, and importantly carries the inverse time scale $\tau'\theta_z$ which makes it manifest at late times unlike other scattering terms. Physically this is because at early times the electrons quickly reaches equilibrium with photons ($\Theta_{\rm e}\approx\Theta_{\rm eq}$). At late times however we see this term change the details of energy injection to the local photon patch. To avoid the risk of introducing a misnomer we clarify: this term does not inject energy, but simply changes which spectral shape is excited, with a shift between $\Yspec(x)$ and $Y_1(x)$.
    \item The third and final source is perturbed thermalisation, including perturbed scattering effects $\propto \left[ \delta_{\rm b}^{(1)} + \Psi^{(1)} \right]M_{\rm K}\vek{y}_0^{(0)}$ and perturbed emission $\propto\left[ \delta_{\rm b}^{(1)} + \Psi^{(1)} + \Theta_0^{(1)} \right]\vek{D}_0^{(0)}$. These simply modify the local thermalisation timescale of $M_{\rm K}\vek{y}_0^{(1)} + \vek{D}_0^{(1)}$ according to the average spectrum. Also within perturbed scattering we find $M_{\rm D} \vek{y}^{(0)}-\vek{S}^{(0)}$, once the aforementioned heating term has been extracted. This part of perturbed scattering sources a local spectral shape resembling the diffusion operator applied to the background together with a shift from $\Yspec$ to $Y_1$ according to $\Theta_{\rm eq}$. All of these effects are typically important at earlier times only.
\end{itemize}
These three sources are shown in Fig.~\ref{fig:aniso_spec_buildup}. Furthermore see Sect.~\ref{sec:physics_switches} for more details on these \textit{Physics switches} which we make extensive use of to distil the physical picture throughout the paper.

We provide a disclaimer for the choice of groupings both for sources and for the switches introduced later: there is no unique choice of this decomposition, and many terms fit into multiple categories from a physical point of view. Consider for example the heating term which has been extracted from $\vek{S}^{(0)}$ whose origin is in the Physics of Kompaneets scattering, however its behaviour can be thought of as a form of anisotropic heating. Even the term $\Psi^{(1)}Q'_{\rm c}$ is associated with local thermalisation efficiency, and is not a direct form of energy injection per se. Generally, considering this paper is largely concerned with the presentation of numerical results, we have taken a qualitative view of bottom line behaviour rather than a fundamental view of the underlying Physics when choosing our grouping of terms.

\subsection{Broad picture for the anisotropic photon spectrum}
\begin{figure}
\centering
\includegraphics[width=1.0\columnwidth]{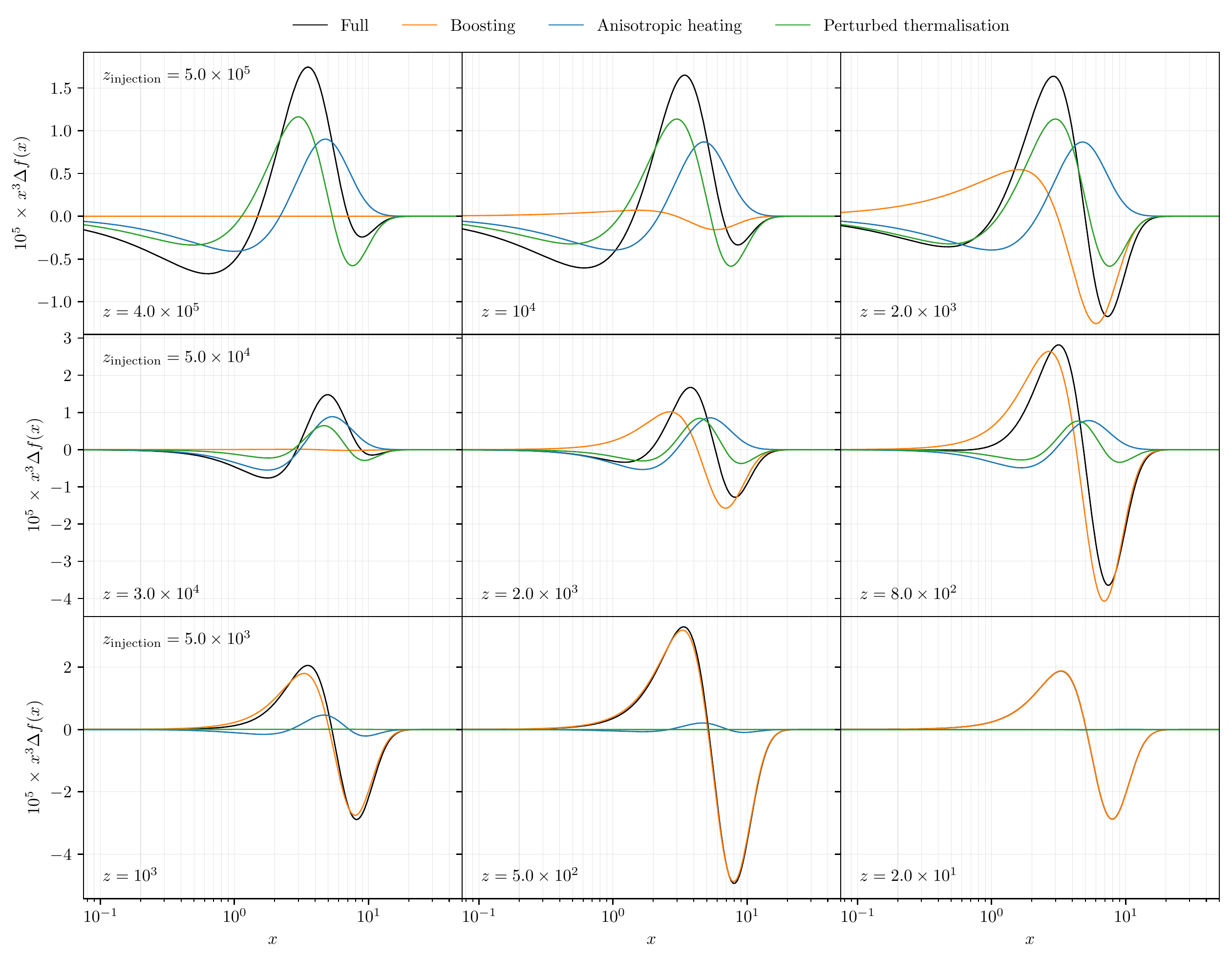}
\caption{Snapshots of the SD spectrum for $\ell=0$ and $k=0.01\,{\rm Mpc}^{-1}$ arising from several injection redshifts. The rows show, from top to bottom, injection at $z=\pot{5}{5}$, $z=\pot{5}{4}$ and $z=\pot{5}{3}$. Left to right show different stages of evolution for each injection scenario. Coloured lines in each panel show the spectrum with only one class of source terms included.}
\label{fig:aniso_spec_buildup}
\end{figure}
\label{sec:anisotropic_photon_spectrum}
\begin{figure}
\centering
\includegraphics[width=1.0\columnwidth]{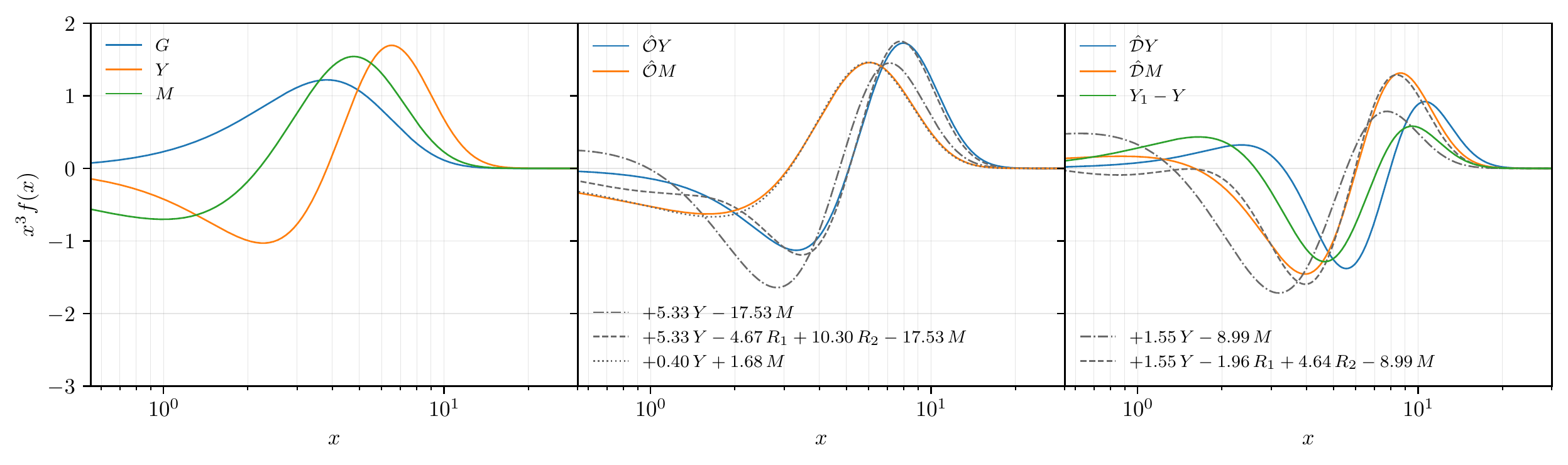}
\caption{A figure showing the main types of spectral shape. Note that they have been re-scaled by arbitrary constants to make them all comparable. The leftmost panel shows the usual shapes familiar from average frequency evolution, and also the limiting case for anisotropic injection before the $y$-era. The middle panel shows the boosted spectra for both $\Yspec$ and $\Mspec$. The rightmost panel shows the diffusion operator acting on $\Yspec$ and $\Mspec$ as well as $Y_1-\Yspec$, shapes that all emerge from early time thermalisation effects and late time anisotropic heating (recognisable by their three peak structure. Some numerical approximations are shown in gray lines.}
\label{fig:spectral_shapes_demonstration}
\end{figure}
In paper I, we showed how the extended $Y_n$ frequency basis can accurately capture the evolution of the background spectrum by reproducing much more expensive binned frequency calculations within {\tt CosmoTherm} \cite{Chluba2013Green, Chluba2015GreensII}. The goal in this paper is to apply this new basis to the generalised Boltzmann hierarchy and explore the evolution of the anisotropic photon spectrum.
In contrast to the background spectrum, this will depend on wavenumber $k$ and angular scale $\ell$, as is familiar from usual early-universe perturbation theory, a fact which renders the usual binned spectral treatments prohibitively expensive.

To study the three main anisotropic distortion sources we numerically solve scenarios with $\Delta\rho/\rho = 10^{-5}$ Dirac-$\delta$ energy injection at fiducial times $z=\pot{5}{5}$ ($\mu$-era), $z=\pot{5}{4}$ (residual-era), $z=\pot{5}{3}$ ($y$-era). In Fig.~\ref{fig:aniso_spec_buildup} we show the corresponding spectrum for $k=0.01\,{\rm Mpc}^{-1}$ at three time slices, showing the time dependence of different sources. Note that we show the purely distorted spectrum with energy dimensions $x^3 \Delta f(x)$, meaning we have subtracted the local temperature shift $\Theta^{(1)}$. This is common throughout the paper to avoid inflationary perturbations dominating the figures (typically $\simeq 10^5$ larger).

Typically speaking the leftmost panels will show the spectrum shortly after the energy injection (corresponding to super-horizon state for the top two rows). The rightmost panel shows the spectrum at late times -- around recombination or later. We see that in all cases the boosting sources grow strongly form left to right, starting at horizon crossing (gravitational boosting) and continuing sub-horizon (baryonic Doppler boosting). The other two sources are only important for early injection times, and dominate over the boosting sources deep in the $\mu$-era. Notice that the earliest injection times yield anisotropic spectra with unfamiliar three-peak structure arising from both $M_{\rm D}\vek{y}^{(0)}$ and $Y_1-\Yspec$ (see perturbed thermalisation term), and cannot be easily recognised as a simple $y$ or $\mu$ spectrum. Anisotropic heating on the other hand initially sources $y$ (with a small $y_1$ correction), which then has the opportunity thermalise via the equivalent terms to the average spectrum [see first terms in second row of Eq.~\eqref{eq:evol_1_final_Yi_1st_ord} in comparison to Eq.~\eqref{eq:evol_1_final_a}], and thus follows the same evolution as the average distortion picture. For example, the spectra from anisotropic heating cross the zero at $x\simeq 1$ and $x\simeq 2$ for $\mu$ and residual-era injection respectively, emulating the usual three era picture. The anisotropic heating spectrum in the third row however does not correspond simply to a $y$ distortion due to the term $\propto \Theta_0^{(1)}\left[Y_1(x)-\Yspec(x)\right]\in \Theta_0^{(1)}\vek{S}^{(0)}$, which has no opportunity to thermalise in the late Universe.

All of these spectral shapes can be recognised in Fig.~\ref{fig:spectral_shapes_demonstration}, where each of the important operators are demonstrated. For example, we can see that spectral shapes with the three peak structure arise from $\DiffO$ (early injection perturbed thermalisation) and $Y_1-\Yspec$ (late time anisotropic heating). In the middle panel we show boosted spectra, where it is important to note $\boostO \Mspec$ projects onto a same-sign mix of $\Yspec$ and $\Mspec$ and is well captured by these two numbers. On the other hand, $\boostO \Yspec$ gives an opposite-sign mix, and additionally needs around two residual modes to converge (see Sect.~\ref{sec:change_of_basis}). This dependence on residual modes will manifest later in late time injection power spectra (see Fig.~\ref{fig:zh_power_spectra_high_resid}). Likewise, $\DiffO \Mspec$ requires at least two residual modes to converge, however, we will see later that early time injection power spectra do not in fact make as heavy use of residual mode information, likely from some cancellation of residual modes with other sources. Comparing individual SED amplitudes here can be somewhat misleading considering the different energies they carry (this leads to $\mu$ often being $\approx 1.401/0.25=5.6$ times larger than $y$), but we can loosely assert that boosting the average $y$ distortion -- the dominant late time behaviour -- yields a $\mu$ amplitude four times the size of $y$, however $\DiffO \Mspec$ carries closer to six times as much $\mu$ as $y$. This will be further exacerbated by anisotropic heating thermalising to a $\mu$ distortion. We will verify later that the early universe injection will yield much stronger $\mu\times\mu$ correlations than $y\times y$ (see Fig.~\ref{fig:zh_distortion_auto_spectra}).

\subsection{Convergence of the photon spectrum}
\begin{figure}
\centering
\includegraphics[width=1.0\columnwidth]{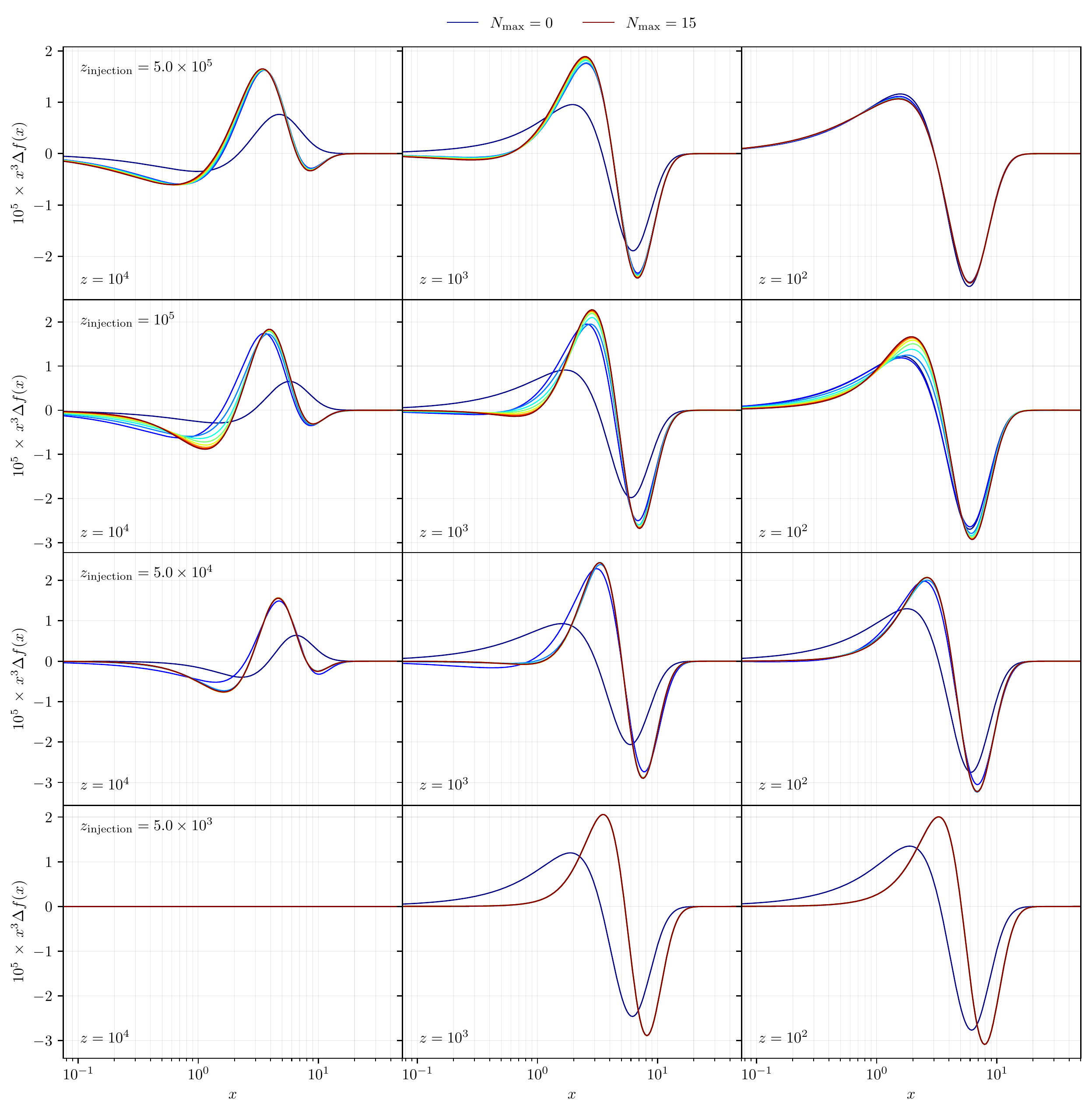}
\caption{Snapshots of the SD spectrum for $\ell=0$ and $k=0.01\,{\rm Mpc}^{-1}$ arising from several injection redshifts. Within each panel, we show the solution when varying the size of the computation basis. For  $z_{\rm injection}\lesssim \pot{5}{3}$ and $z_{\rm injection}\gtrsim \pot{5}{5}$ convergence is extremely rapid, while for intermediate cases, a basis with $Y_{15}$ starts to shows its limitations.}
\label{fig:aniso_spec_convergence}
\end{figure}
When studying the background spectrum there is the luxury of comparing to the full {\tt CosmoTherm} calculation (see Sect.~2.8 in the companion paper I). This has demonstrated that the ODE representation of the thermalisation problem is highly accurate and captures the main physical features of the full treatment.
For anisotropies, however, the parameter space grows in many dimensions, and a direct {\tt CosmoTherm} convergence benchmarks would be quite expensive. Despite this limitation we can expect the anisotropic treatment to perform well.
It can be seen in Eq.~\eqref{eq:evol_1_final} that the sources of anisotropies arise either from direct sources of $y$, or the matrix forms of the boost and diffusion operators ($\boostO\rightarrow M_{\rm B}$, $\DiffO\rightarrow M_{\rm D}$), precisely the operators around which the spectral basis was constructed. The bottom line is that the process of boosting is captured exactly in this formalism, not approximately, as long as the dominant part of a spectrum is relying mostly on $N<N_{\rm max}$. More concretely, the only boosted SEDs not directly contained in the basis are $\boostO Y_{N_{\rm max}}(x)$ and $\boostO M(x)$. For the first case we note that $y_{N_{\rm max}}$ becomes smaller for growing $N_{\rm max}$, with $y_{15}$ only seeing significant contributions in very narrow windows of the residual-era ($z\approx 10^5$ providing a worst case scenario). For the second case we have shown $\boostO M(x)$ to map extremely well back into the basis even with $N_{\rm max}\leq 1$ (see paper I).
This all means that the accuracy of the anisotropic evolution will typically be limited by the accuracy of the average evolution.

In Fig.~\ref{fig:aniso_spec_convergence} we show the photon spectrum for $0\leq N_{\rm max} \leq 15$ at various single energy injection redshifts (from top to bottom row: $z_{\rm injection}=\pot{5}{5},~10^5,~\pot{5}{4},~\pot{5}{3}$). From left to right we show different stages of the evolution, from the stable super-horizon state through to late post-recombination evolution. We can immediately see that earlier injection times see poorer convergence than later times, with $10^5$ performing the worst as expected. This is due to the only late time source being the boost operator, which given the arguments above is well captured in this basis. The greater diversity of sources for the $\mu$-era injection puts more strain on the numerical method, however, we note that between $N_{\rm max}=15$ and $N_{\rm max}=13$ we only see sub-percent changes in the right most panels of Fig.~\ref{fig:aniso_spec_convergence}. This statement depends on the moment you observe the spectrum, which is why we opted to study the spectra in Fig.~\ref{fig:aniso_spec_convergence} at the same moments of time in a given column, unlike Fig.~\ref{fig:aniso_spec_buildup} where we prioritised elucidating physical sources at various moments of evolution. We note that in Fig.~\ref{fig:power_spectrum_convergence_1} and Fig.~\ref{fig:power_spectrum_convergence_2} we perform a similar convergence analysis for the power spectra which gives a less time dependent sense of the performance of the basis.

Notably the spectrum for $y$-era injection (bottom row) is captured almost exactly even with just $Y_1$ since the expected limiting case $\boostO Y = 4Y_1$ is precisely captured in that basis (late energy injection sees very little contribution from other sources, see Fig.~\ref{fig:aniso_spec_buildup}).
The $\mu$-era injection (top row of Fig.~\ref{fig:aniso_spec_convergence}) shows similarly good result. The main source in this era is less clear than for late times, but is some mix of $\boostO \Mspec(x)$, $\DiffO \Mspec(x)$, $\Mspec(x)$ and $Y_1(x)-\Yspec(x)$, all of which are captured well in this basis.
It should be highlighted that the transition from $T$ to $\mu$-era is slower in this formalism compared to usual numerical solutions and thus a study of convergence here is not the whole picture. In essence some small thermalisation from $\mu^{(0)}$ to $\Theta^{(0)}$ occurred where we would not have expected any; however, the correction is small and can likely be eliminated with further improvements of the thermalisation treatment (see paper I for discussion).

\subsection{Change of basis}
\label{sec:change_of_basis}
\begin{figure}
\centering
\includegraphics[width=1.0\columnwidth]{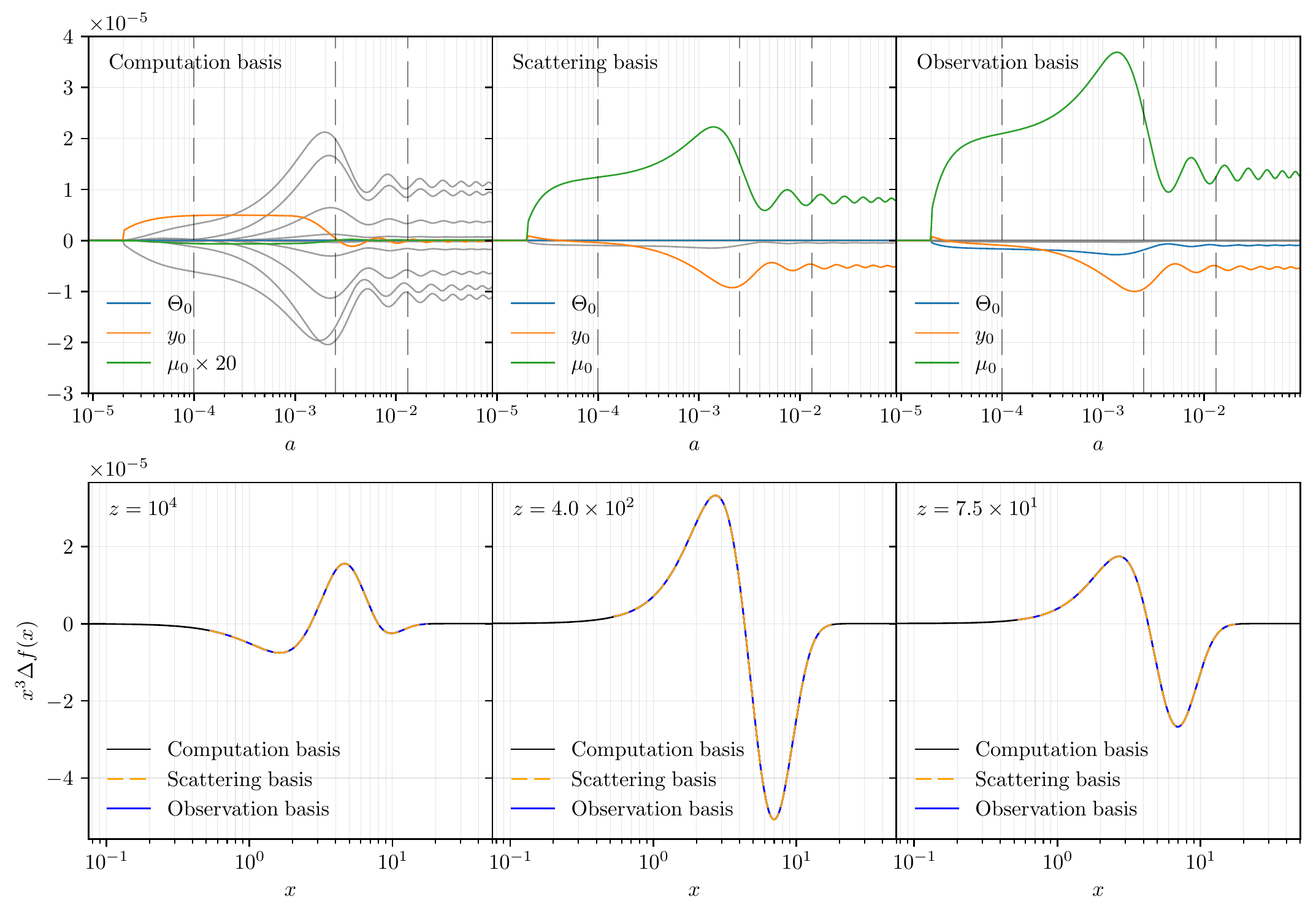}
\caption{Differences between monopole SED transfer functions in various bases (upper panels) and the corresponding SD signals at three snapshots (lower panels). The transfer functions correspond to waves with $k=0.01\,{\rm Mpc}^{-1}$ following an energy injection of $\Delta\rho/\rho=10^{-5}$ at $z=\pot{5}{4}$. The amplitude of $\Theta$ is subtracted prior to changing basis, to avoid hiding the signal under primordial fluctuations five orders of magnitude larger. In this way, we highlight the fictitious $\Theta$ from the unconstrained change of basis (observation basis). The solid/gray lines in each panel represent the residual modes for the given basis. In the computation basis, these higher order contributions exceed the $y$ and $\mu$ transfer functions, while for the scattering and observation basis, they contribute marginally, as anticipated. In spite of the drastic SED parameter differences for the three basis, the basis-independence of the SD signal is demonstrated in the lower panels.}
\label{fig:basis_comp}
\end{figure}
It is explained in paper I that there is a large degree of degeneracy between the $Y_n(x)$ spectral shapes and the more common $\Gspec(x)$, $\Yspec(x)$ and $\Mspec(x)$. Upon solving the average spectral evolution this led to seemingly very different energy branching ratios for SED amplitudes that otherwise converged to the same expected resulting photon spectrum. This problematic disconnect between branching ratios (or soon transfer functions) can be remedied by performing a \textit{change of basis} in which a new set of basis function SEDs are chosen that better suit the physics being studied.

In particular one can \textit{prioritise} the regular SED shapes by first performing a projection onto $\Gspec(x)$, $\Yspec(x)$ and $\Mspec(x)$, and subsequently constructing a set of orthogonal SED through PCA. We will construct two such bases: firstly the {\it observation basis}, which as the name suggests performs a PCA on the binned frequency space\footnote{We take a fiducial binning from $\nu_{\rm min}=30\,{\rm GHz}$ to $\nu_{\rm max}=1000 \,{\rm GHz}$ with steps $\Delta\nu=1 \,{\rm GHz}$} to allow direct comparison to results that are akin to what would be obtain in real measurements. This basis was first considered in \cite{Chluba2013PCA} and compresses the accessible signal information significantly. Secondly, the {\it scattering basis}, which is constructed as before, but with the additional constraint of photon number, so no other SED will project to $\Gspec(x)$ and vice-versa. This final basis highlights a the fact that subtracting the theoretically accurate $\Gspec(x)$ from an observationally acquired spectrum is usually not possible provided the finite binning of an experiment. 
In this light the $Y_n(x)$ basis will be called the {\it computation basis}, since it has been constructed to model the boosting/diffusion/scattering processes dictating the evolution of photons in plasma.
This new vocabulary emphasises the {\it basis-independence} of the CMB spectrum anisotropies in contrast to the {\it basis-dependence} of SED amplitudes -- a fact which should be present in the readers mind when interpreting any results in the subsequent sections in this paper, especially for detection prospects (Sect.~\ref{sec:forecasts}).

While transfer functions will be discussed extensively in Sect.~\ref{sec:transfer_functions}, we show a single example in Fig.~\ref{fig:basis_comp} to illustrate the difference in basis choice. Focusing on the upper row first, the upper left panel shows the computation basis with residual modes $y^{(1)}_{n,0}$ typically dominating over the $y^{(1)}_{\ell=0}$ and $\mu^{(1)}_{\ell=0}$ amplitudes. This is understood since the the background spectrum consists of a wide mix of $y^{(0)}_{n}$ following energy injection in the residual-era, as these get boosted to perturbed spectra with $y^{(1)}_{n+1,1}$. Only for $y^{(0)}_{N_{\rm max}}$ and $\mu^{(0)}$ does this boosting mix directly into $y^{(1)}_{\ell}$ or $\mu^{(1)}_{\ell}$. The upper middle panel in the upper row of Fig.~\ref{fig:basis_comp} shows the results of casting to the scattering basis -- projecting the spectrum back onto the main SED and using residual modes to capture the remaining signal. This can indeed be seen to give the same spectral shape while compressing the information to the usual SD amplitudes and just a small contribution from residual modes (see lower panels). Finally the upper right panel shows the observation basis, representing what could be seen with a binned observation of the sky. Most notably it can be seen that some $\Theta^{(1)}_{\ell}$ is generated by counteracting an increase in $\mu^{(1)}_{\ell}$, a result of inferring spectral shapes from a limited window of visibility.\footnote{This effect is familiar in different settings \citep{Chluba2013Green, Lucca2020}.} However, the representation of the signal is independent of these parametrisation aspects (lower panels in Fig.~\ref{fig:basis_comp}).

\section{Numerical solutions for distortion transfer functions}
\label{sec:transfer_functions}
With some understanding of SED transfer functions and how these map to a corresponding distorted photon spectrum, we are now in the position to gain a more intuitive understanding of the behavior of distortion anisotropies and their evolution. Like for the thermalisation Green's function it is instructive to first consider single redshift injections of average energy from which we will then distill some of the physics for distortion modes at various scales. All results in the following section are shown in the scattering basis (see Fig.~\ref{fig:basis_comp} for illustration), meaning that $\mu$ and $y$ are usually representative of the spectrum, with only minor contributions from the residual modes which will not be highlighted here. The primordial temperature fluctuations have not been subtracted this time, allowing for comparison of relative phases between the SED parameter amplitudes.

\subsection{Numerical setup}
To solve the coupled system of Boltzmann equations we extend the anisotropy module of {\tt CosmoTherm} \citep{Chluba2011therm}. We set adiabatic initial conditions for the standard perturbations while the distortion parameters are initially set to zero, given that no initial inflationary distortion signals are expected. The ODE system is solved using a sixth order Gear's method with adaptive time-stepping. This method is stiffly-stable and does not require any separate treatment in the tight-coupling regime. The corresponding solver was implemented to solve the cosmological recombination problem \citep{Chluba2010, Chluba2010b}. A relative precision of $\simeq 10^{-4}$ is requested and redshift is used as the main time-variable.

We truncated the multipole hierarchy following \citep{Ma1995}. Depending on the scale, we include a varying number of multipoles for CMB temperature and polarisation anisotropies, neutrinos and the distortion parameters. We find that $\ell_{\rm max}=15$ is sufficient to achieve accurate power spectrum results; however, for the transfer functions in this section we expand this greatly (up to $\ell_{\rm max}=100$) to ensure no \textit{reflected} energy in the shown time intervals. We do not include reionisation or perturbed recombination effects in our treatment. Also, polarisation effects are only treated carefully for the temperature perturbations, not for the distortion parameters. However, these approximations are not expected to change the overall picture significantly.

\subsubsection{Switching the physics}
\label{sec:physics_switches}
For the results presented below, it is instructive to switch on/off various physical effects. The goal is to illustrate the effect on the distortion anisotropies, so in all cases, we do not modify the standard perturbation equations for temperature and polarisation terms. We introduce various physical switches in relation to the sources mentioned in Sect.~\ref{sec:main_sources} (also see Fig.~\ref{fig:aniso_spec_buildup}): Doppler/potential boosting, perturbed emission/scattering, and anisotropic heating.
\begin{itemize}
    \item Referring to Eq.~\eqref{eq:evol_1_final_hierarchy}, 
switching off \textbf{Doppler} boosting (here sometimes also referred to as Doppler driving) means we drop the term $\propto \tilde{\beta}^{(1)}\vek{b}^{(0)}_0/3$ in the dipole equations of the distortions.
    \item Similarly, to switch off \textbf{potential} driving we drop the terms $-\partial_\eta \tilde{\Phi}^{(1)}
\vek{b}^{(0)}_0$ and $k\tilde{\Psi}^{(1)} \,\vek{b}^{(0)}_0/3$ in the monopole and dipole equations of the distortions. These two switches are presented together simply as \textbf{boosting}.
    \item Perturbed \textbf{emission} off means not accounting for the group of terms $\propto \left[\tilde{\delta}_{\rm b}^{(1)} + \tilde{\Psi}^{(1)} +\tilde{\Theta}^{(1)}_0 \right]\vek{D}^{(0)}_0$ in the monopole distortion equation.
    \item Similarly perturbed \textbf{scattering} off means not accounting for the group of terms $$\propto \left[\tilde{\delta}_{\rm b}^{(1)} + \tilde{\Psi}^{(1)} \right] M_{\rm K}\,\vek{y}^{(0)}_0 +\tilde{\Theta}^{(1)}_0\left(M_{\rm D}\vek{y}_0^{(0)} - \vek{S}^{(0)} \right)$$ in the monopole distortion equation, \textit{except} the aforementioned terms within $\vek{S}^{(0)}$ which are deemed anisotropic heating (see Sect.~\ref{sec:main_sources}). These previous two switches together make up \textbf{perturbed thermalisation}.
    \item Finally neglecting \textbf{anisotropic heating} means omitting all terms within ${\vek{Q}'}^{(1)}/4$, and also the terms $\propto Q_{\rm c}'/4\tau'\theta_z$ within $\vek{S}^{(0)}$.
\end{itemize}
The purely spatial thermalisation terms $M_{\rm K}\,\tilde{\vek{y}}^{(1)}_0+\vek{D}^{(1)}_0$ are always switched on, meaning there is a similar evolution for spatial spectra as for average spectra. This is not to say that all sources will undergo a simple thermalisation process, since terms like $M_{\rm D}\,\vek{y}^{(0)}$ will continuously source from the average spectrum, and boosting sources typically only occur after the early-time thermalisation window (this is true for $k$ modes which influence the CMB power spectrum).

One small clarification about the nomenclature of the perturbed thermalisation terms is in order. Physically, the thermalisation process requires the combined action of Compton scattering and DC/BR emission and absorption \citep{Sunyaev1970mu, Danese1982, Burigana1991, Hu1993}. When we say `perturbed scattering', we mean `perturbed Compton scattering' as opposed to `perturbed Thomson scattering', which has no effect on the spectral shape but would only slightly modify the Thomson visibility function, leading to a higher order effect \citep[e.g.,][]{Senatore2009}. The term `perturbed emission' is indeed somewhat misleading as it includes the change in the balance between DC/BR emission and Compton up-scattering, which ultimately defines the distortion visibility \citep{Chluba2011therm, Chluba2014}. This latter effect was estimated by \citep{Zegeye2022} in the context of primordial non-Gaussianity, and originates from changes in the thermalisation efficiency around $z_\mu\simeq \pot{2}{6}$ due to the presence of perturbations. To not confuse it with `perturbed thermalisation' (which includes all terms), we shall choose to use `perturbed emission' instead of e.g., `perturbed thermalisation efficiency' or `perturbed visibility'.

\begin{figure}
	\centering
	\includegraphics[width=0.96\columnwidth]{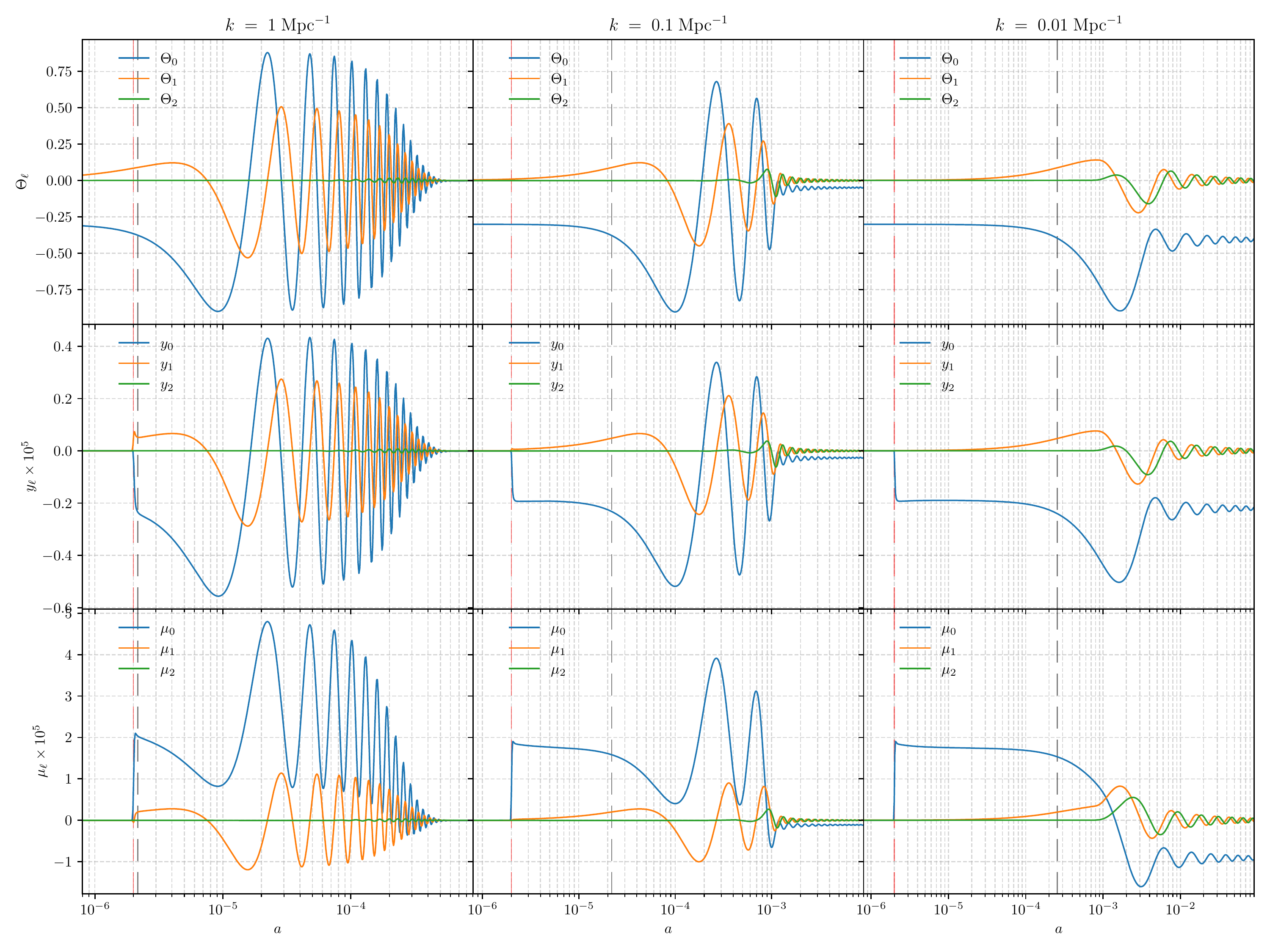}
	\caption{CMB transfer functions for single heating at $z_{\rm injection}=\pot{5}{5}$  with $\Delta \rho_\gamma/\rho_\gamma=10^{-5}$ and wavenumber $k$ as labeled. This leads to an average distortion that freezes with $\Theta^{(0)}\approx\pot{4.9}{-6}$, $y^{(0)} \approx 0.0$ and $\mu^{(0)}\approx \pot{1.1}{-5}$ at $z\lesssim 10^5$. The numerical solutions are computed using $N_{\rm max}=15$ with a rotation to the scattering basis (Sect.~\ref{sec:change_of_basis}). Dashed vertical lines show times of horizon crossing (gray) and energy injection (red).}
\label{fig:zh_5e+5_transfer}
\end{figure}
\begin{figure}
	\centering
	\includegraphics[width=1.0\columnwidth]{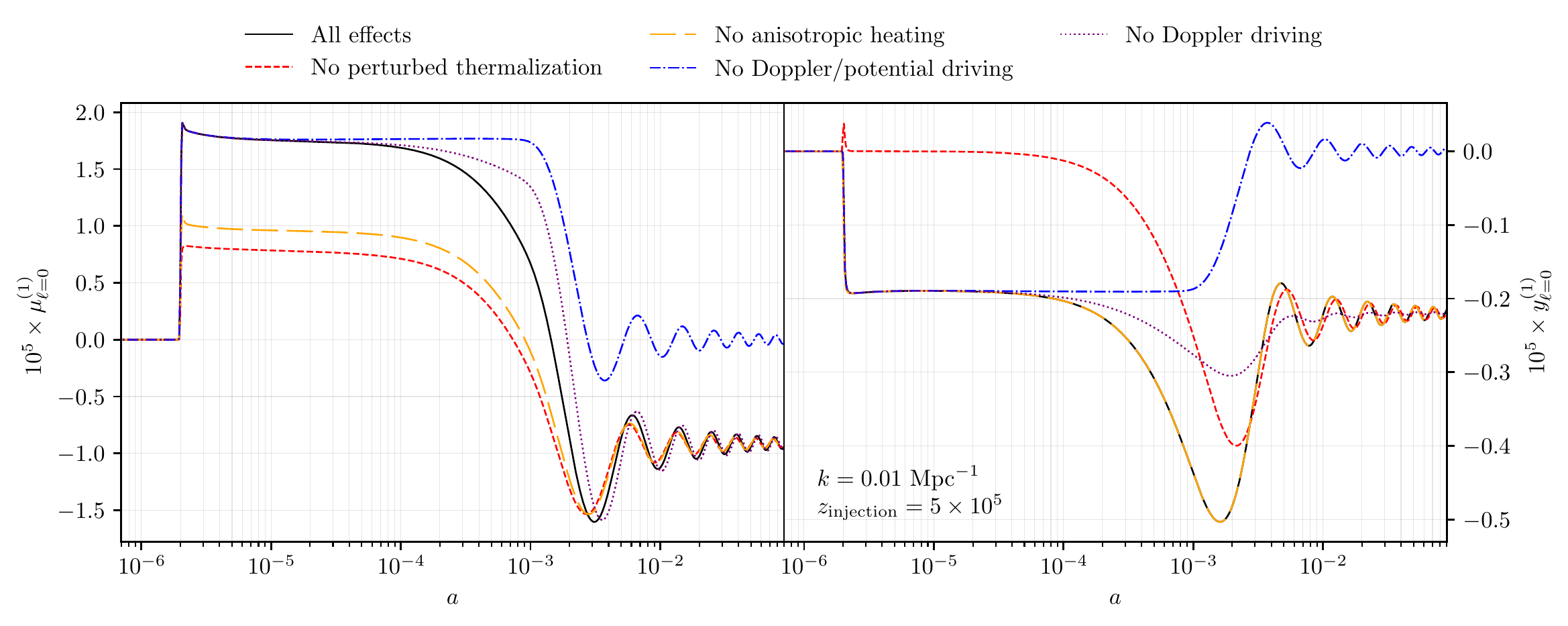}
	\caption{Same as Fig.~\ref{fig:zh_5e+5_transfer}, but focusing on $\mu^{(1)}_0$ (left column) and $y^{(1)}_0$ (right column) for $k=0.01\,{\rm Mpc}^{-1}$. Perturbed thermalisation effects (here dominated by DC and BR emission effects) are switched off for the red/dashed line, showing a reduction of around half the initial local distortion, however illustrating that the overall late evolution is not affected significantly by these corrections. Excluding the effect of Doppler and potential driving for the distortion anisotropies shows that without potential driving the modes simply decay and oscillate around a zero point. Doppler driving adds a small correction to this picture, most important around recombination. Switching off perturbed thermalisation leads the local $y$ distortion to only show an initial transient spike, which rapidly thermalises through $M_{\rm K} \vek{y}_0^{(1)} + \vek{D}_0^{(1)}$ (never switched off). Anisotropic heating (orange) contributes about $1/2$ of the initial $\mu^{(1)}$ distortion, and does not source $y^{(1)}$. Finally, only neglecting Doppler driving shows clearly how the slow part of the evolution is dominated by potential decay (purple/dash-double-dotted line).}
\label{fig:switches_zh_5e+5}
\end{figure}
\begin{figure}
	\centering
	\includegraphics[width=1.0\columnwidth]{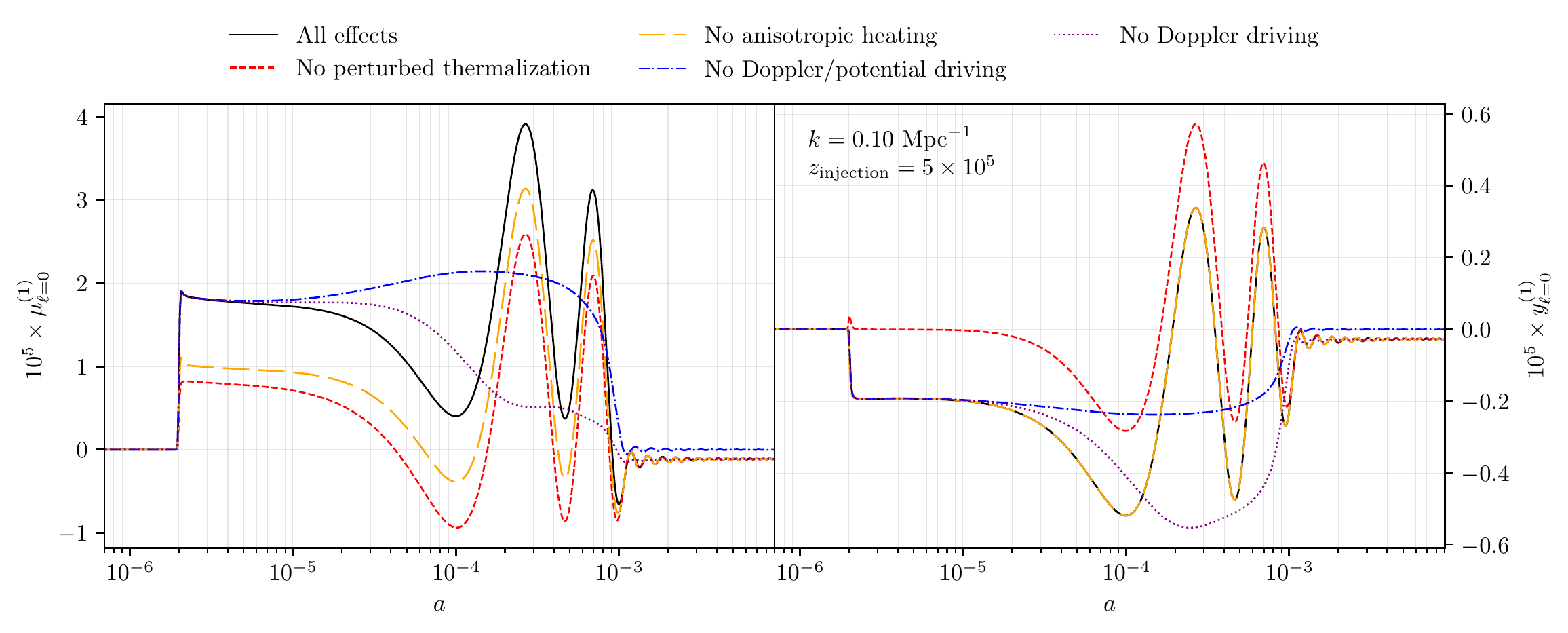}
	\caption{Same as Fig.~\ref{fig:switches_zh_5e+5}, but for $k=0.1~{\rm Mpc}^{-1}$.}
\label{fig:switches_k_0.1_zh_5e+5}
\end{figure}
\begin{figure}
	\centering
    \includegraphics[width=1.0\columnwidth]{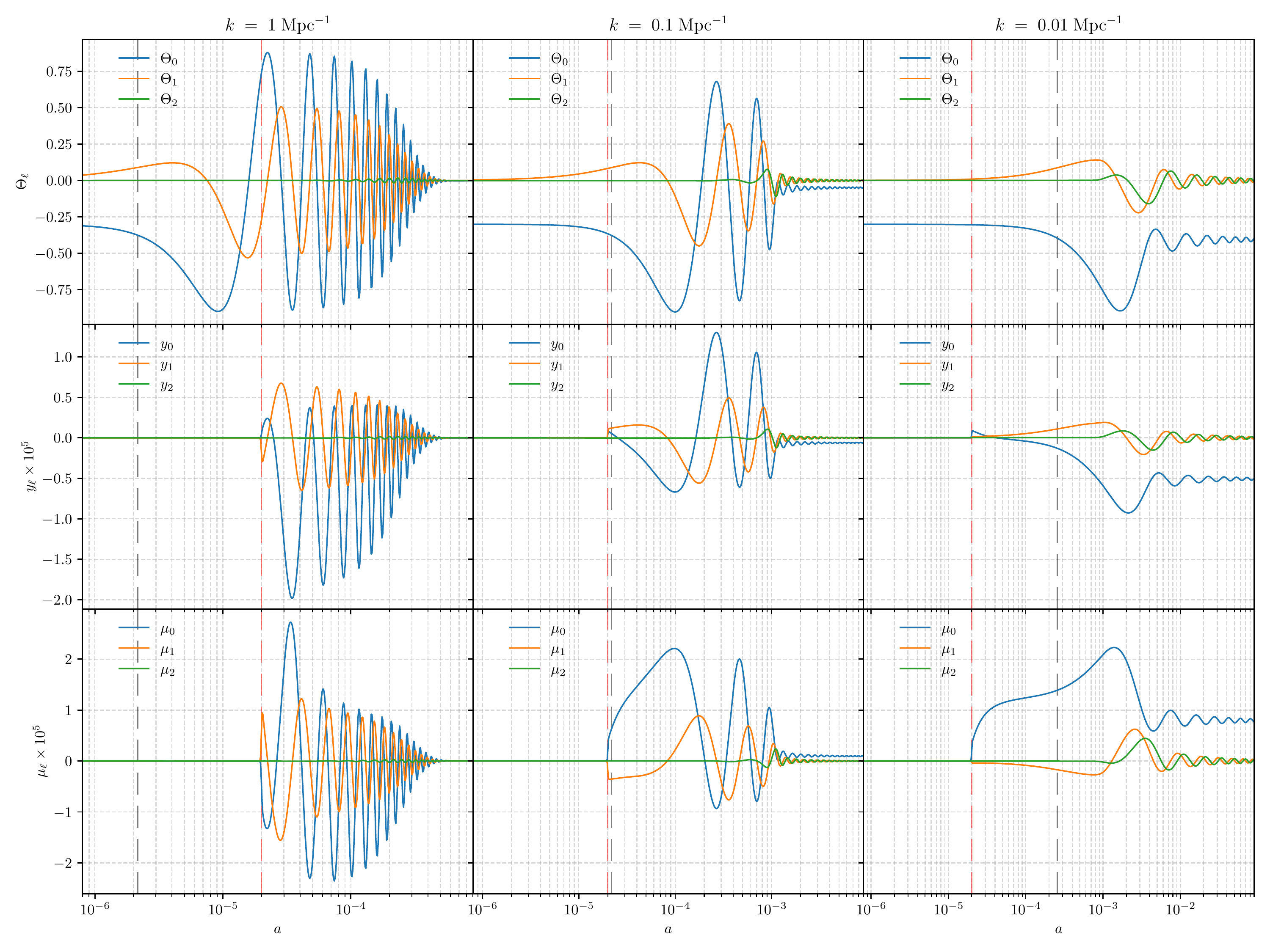}
	\caption{CMB transfer functions for single heating at $z_{\rm injection}=\pot{5}{4}$ with $\Delta \rho_\gamma/\rho_\gamma=10^{-5}$ and wavenumber $k$ as labeled. This leads to an average distortion that freezes with $\Theta^{(0)}=\pot{7.5}{-9}$, $y^{(0)}=\pot{3.9}{-7}$ and $\mu^{(0)}=\pot{1.2}{-5}$ at $z\lesssim 8000$. The numerical solutions are computed using $N_{\rm max}=15$ with a rotation to the scattering basis. Dashed vertical lines show times of horizon crossing (gray) and energy injection (red).}
\label{fig:zh_5e+4_transfer}
\end{figure}
\begin{figure}
	\centering
	\includegraphics[width=1.0\columnwidth]{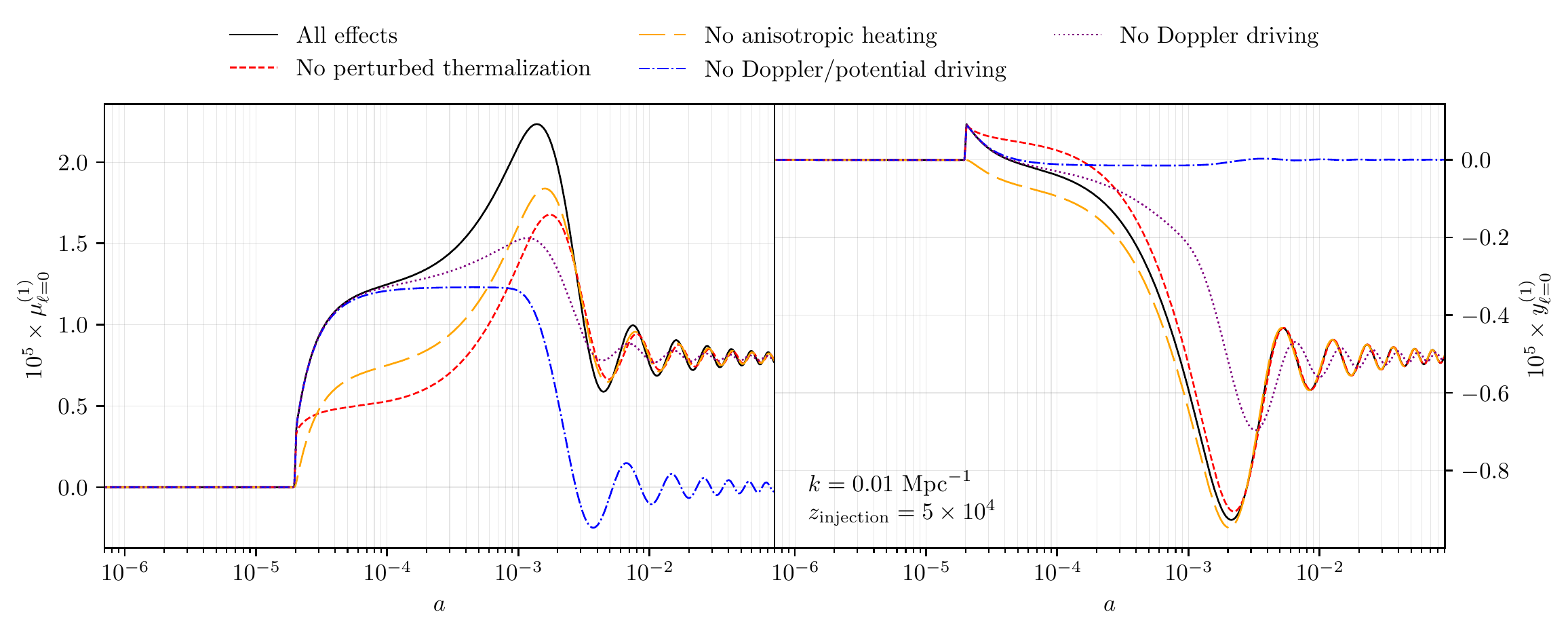}
	\caption{Same as Fig.~\ref{fig:zh_5e+4_transfer}, but focusing on $\mu^{(1)}_0$ (left column) and $y^{(1)}_0$ (right column) for $k=0.01\,{\rm Mpc}^{-1}$. Similar general trends hold to Fig.~\ref{fig:switches_zh_5e+5}. Perturbed thermalisation and anisotropic heating effects again show equal importance to super-horizon evolution, while boosting is now more important and drives opposite signs for $\mu^{(1)}$ and $y^{(1)}$.}
\label{fig:switches_zh_5e+4}
\end{figure}
\begin{figure}
	\centering
	\includegraphics[width=1.0\columnwidth]{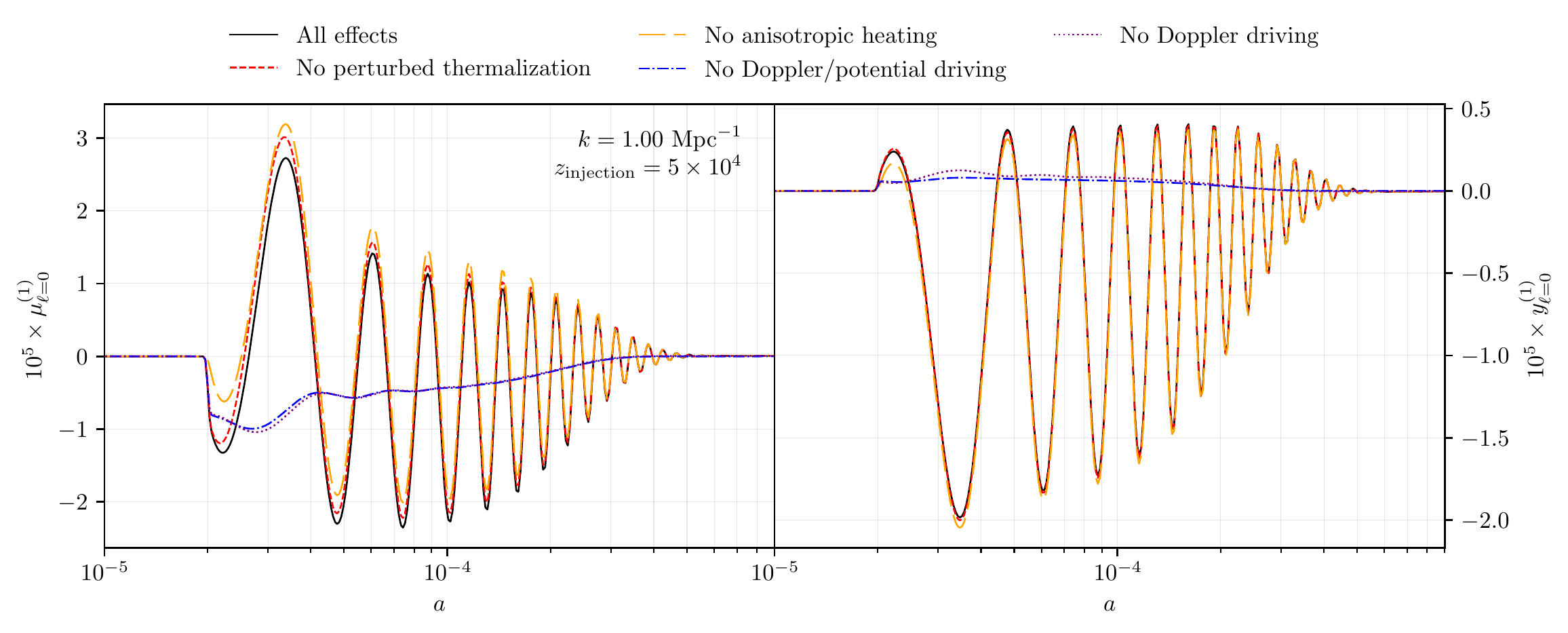}
	\caption{Same as Fig.~\ref{fig:switches_zh_5e+4}, but for $k=1~{\rm Mpc}^{-1}$.}
\label{fig:switches_k_1_zh_5e+4}
\end{figure}
\begin{figure}
	\centering
    \includegraphics[width=1.0\columnwidth]{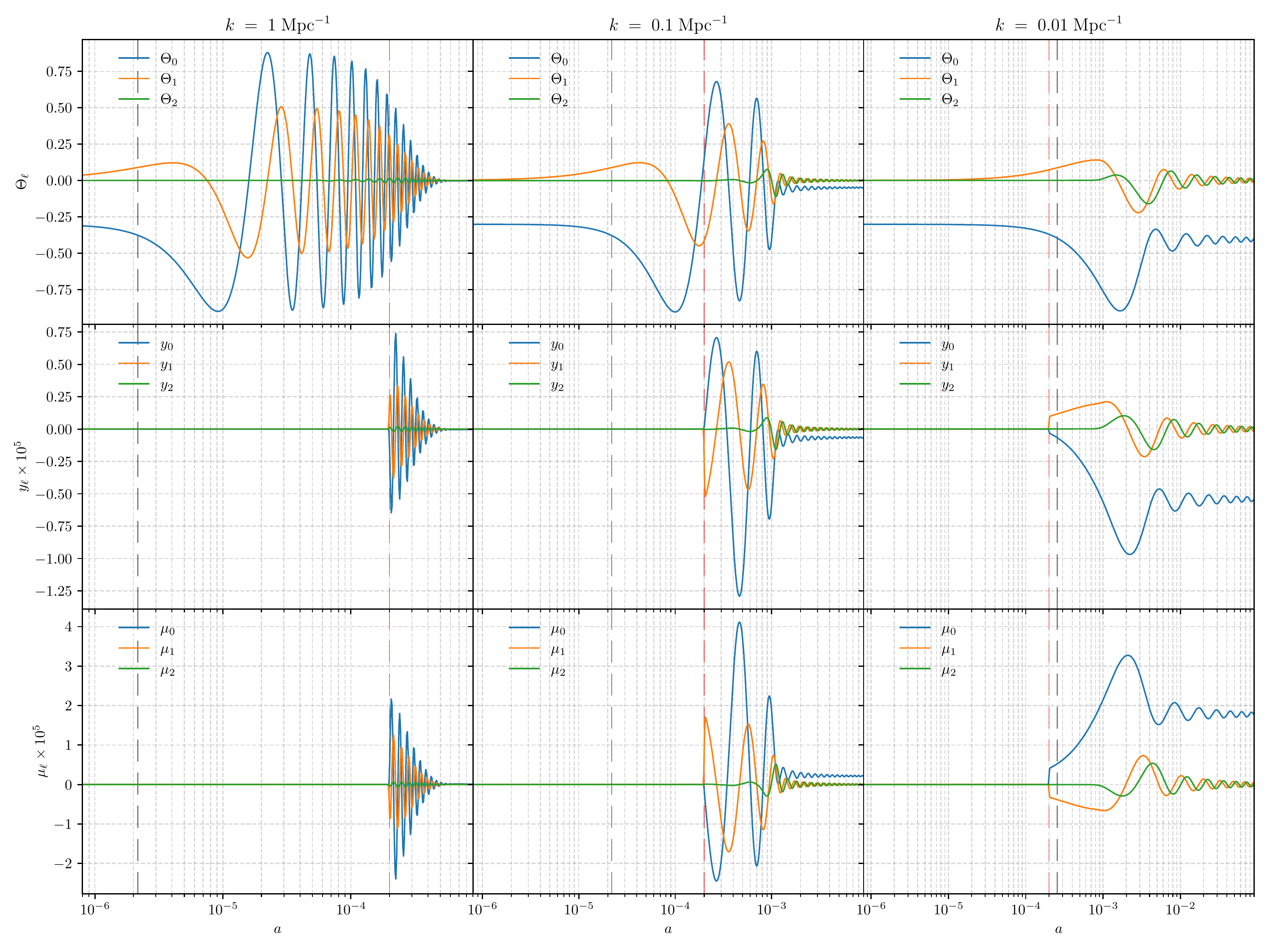}
	\caption{CMB transfer functions for single heating at $z_{\rm injection}=\pot{5}{3}$ with $\Delta \rho_\gamma/\rho_\gamma=10^{-5}$ and wavenumber $k$ as labeled. This leads to an average distortion that freezes with $\Theta^{(0)}\approx\pot{3.5}{-12}$, $y^{(0)}\approx\pot{2.5}{-6}$ and $\mu^{(0)}\approx\pot{2.0}{-7}$. The numerical solutions are computed using $N_{\rm max}=15$ with a rotation to the scattering basis (Sect.~\ref{sec:change_of_basis}). Dashed vertical lines show times of horizon crossing (gray) and energy injection (red).}
\label{fig:zh_5e+3_transfer}
\end{figure}
\begin{figure}
	\centering
	\includegraphics[width=1.0\columnwidth]{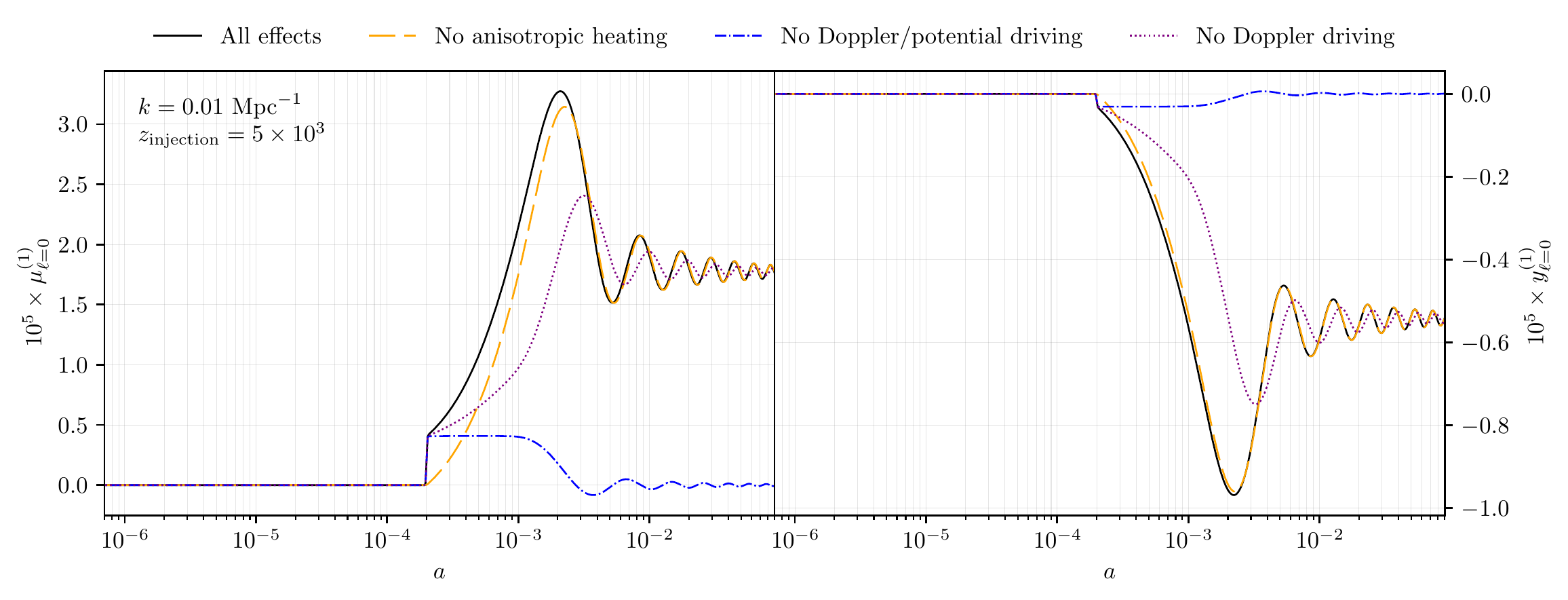}
	\caption{Same as Fig.~\ref{fig:zh_5e+3_transfer}, but focusing on $\mu^{(1)}_0$ (left column) and $y^{(1)}_0$ (right column) for $k=0.01\,{\rm Mpc}^{-1}$. Perturbedd thermalisation is not shown now, since it has no effect (verified independently). Anisotropic heating sources both $\mu^{(1)}$ and $y^{(1)}$ via projections of $\Yspec-Y_1$. Other than this boosting is the dominant source, showing the usual behaviour with the opposite sign mix of $\mu^{(1)}$ and $y^{(1)}$.}
\label{fig:switches_zh_5e+3}
\end{figure}
\vspace{2mm}
\subsection{Anisotropies for energy injection in the $\mu$-era}
\label{sec:transfer_mu}
We begin our analysis by considering average energy release deep into the $\mu$-era ($z>\pot{3}{5}$). We illustrate the transfer functions for $\Theta^{(1)}$, $y^{(1)}$ and $\mu^{(1)}$ in the various columns of Fig.~\ref{fig:zh_5e+5_transfer} for a single heating occurring at redshift $z_{\rm injection}=\pot{5}{5}$, varying the wavenumber of the mode in the columns. In this figure (as well as Fig.~\ref{fig:zh_5e+4_transfer} and Fig.~\ref{fig:zh_5e+3_transfer}) the transfer functions for $\Theta^{(1)}$ behave all as expected and well-known for adiabatic perturbations \citep[e.g.,][]{Hu1995CMBanalytic}. Similarly, as expected, distortion anisotropies only become visible after the average distortion is present. Additionally Fig.~\ref{fig:switches_zh_5e+5} and Fig.~\ref{fig:switches_k_0.1_zh_5e+5} show the $k=0.01~{\rm Mpc}^{-1}$ and $k=0.1~{\rm Mpc}^{-1}$ modes with various physics switches on/off.

Focusing on the early evolution we see that following the creation of an average distortion there is both a local monopole $y$ and $\mu$ sourced. We can see by inspecting the upper-left panel of Fig.~\ref{fig:aniso_spec_buildup} that this is equal parts from anisotropic heating and the perturbed scattering terms, as verified in Fig.~\ref{fig:switches_zh_5e+5}. This evolution quickly reaches an equilibrium state, where the mode then waits till horizon crossing, upon which the boosting effects from gravitational potential decay and doppler boosting begin. These negatively drive both $\mu$ and $y$, where the equal sign is characteristic of $\boostO \Mspec(x)$ [we will see opposite sign mixes later from $\boostO \Yspec(x)$]. At late times the distortion SED transfer functions oscillate around a varying mean, mostly driven by the gravitational potentials, with small corrections from baryonic Doppler boosts, again as seen from Fig.~\ref{fig:switches_zh_5e+5}. Note, however, that the time of recombination receives large contributions from baryonic Doppler driving (red line in Fig.~\ref{fig:switches_zh_5e+5}), which makes it an important source to CMB power spectra. By further inspecting Fig.~\ref{fig:switches_k_0.1_zh_5e+5} we can distinguish that the oscillations in the tight coupling phase are associated with potential driving, while the baryonic Doppler boosts mainly contribute at horizon crossing.

There is a small transient phase of evolution before reaching the superhorizon equilibrium state (also seen well in the right panel of Fig.~\ref{fig:switches_zh_5e+5}). This can be seen as the equivalent thermalisation process to what we see for average distortions $M_{\rm K}\,\tilde{\vek{y}}^{(1)}_0+\vek{D}^{(1)}_0$, except with a slowing effect captured in, e.g., the term $\propto \left[\tilde{\delta}_{\rm b}^{(1)} + \tilde{\Psi}^{(1)}\right] M_{\rm K}\,\vek{y}^{(0)}_0$. This signifies a small delay to the conversion of $y$ to $\mu$ with respect to the average distortion, since $\tilde{\delta}_{\rm b}^{(1)} + \tilde{\Psi}^{(1)}<0$ for adiabatic modes. Because for the considered case, the conversion to $\mu$ is extremely rapid, this manifests in a small peak in the $\mu_0$ and $y_0$ transfer functions before reaching its super-horizon plateau. For later injections, this evolution will be more visible since the conversion from $y$ to $\mu$ is less rapid (see Fig.~\ref{fig:zh_5e+4_transfer}).

Focusing on the late evolution, broadly speaking, we can see that aside from minor phase differences the transfer functions of the respective multipoles of all spectral parameters behave similarly. This is expected since the main driver during the late phase is Doppler driving and decaying potentials, which source the distortion anisotropies in very much the same way to the temperature anisotropies. This also means that the distortion-temperature correlations should be significant, as we further demonstrate below.

Regardless of what occurs super-horizon, horizon-crossing will drive a source of both $\mu$ and $y$ anisotropies (noticeable shortly after the gray vertical lines). This boosting typically occurs long after the ceasing of thermalisation (for $k$-modes relevant to CMB power spectra), and will become the dominant sources for late injection (see Fig.~\ref{fig:zh_5e+3_transfer}).

We also mention that one source of $y$-distortion anisotropies is from the shift in the average CMB temperature by thermalisation. This comes from the Doppler boost of $\Theta^{(0)}$ ($\boostO\Gspec=\Yspec+3\Gspec$) and for the early injection considered here is found to cause $y_0^{(1)}\simeq - 10^{-7}\zeta$. At this level, several other terms will become important so that we leave a more detailed investigation to the future. We note, however, that this $y$-distortion mode could in principle allow us to test changes to the temperature-redshift relation caused at late phases of the cosmic history. To leading order, the expected signal can be thought of as a mismatch of the average CMB spectrum and the spectrum of the CMB anisotropies due to the independent evolution of the average spectrum \citep{Chluba2014TRR}. In addition, entropy production right after the Big Bang Nucleosythesis era could be tested, which given current CMB anisotropy constraints on the helium abundance could still accommodate $\Delta \rho_\gamma/\rho_\gamma \simeq 0.01$ \citep{Steigman2007, Steigman2009}.

\vspace{2mm}
\subsection{Anisotropies for energy injection in the residual distortion era}
\label{sec:transfer_functions_r_era}
We next consider injection at $z_{\rm injection}=\pot{5}{4}$, an approximate midpoint of the residual distortion era ($10^4\lesssim z \lesssim \pot{3}{5}$). The average distortion now has a non-vanishing $y$-distortion contribution -- amounting to $\simeq 50\%$ of total energy for this redshift.
Needless to say, the transfer functions for $\Theta^{(1)}$ remain unchanged, but are shown again in Fig.~\ref{fig:zh_5e+4_transfer} for convenience.

The distortion transfer functions all show similar overall behavior as for $z_{\rm injection}=\pot{5}{5}$ except for some subtle yet notable changes. At $z\lesssim \pot{2}{5}$, DC emission and absorption terms become negligible, like for the average evolution \citep{Chluba2014}. However, perturbed scattering effects are still relevant, and in comparable to the anisotropic heating (see Fig.~\ref{fig:aniso_spec_buildup}). The super-horizon evolution now shows both $y^{(1)}$ and $\mu^{(1)}$ contributions from anisotropic heating, however the $M_{\rm D}$ operator together with $(Y_1-\Yspec)$ causes anisotropic $\mu^{(1)}$ to dominate the picture (see Fig.~\ref{fig:switches_zh_5e+4}).

Both the $\mu$ and $y$ transfer functions become highly correlated around horizon crossing, with boosting now carrying more importance compared to earlier injection times. The mix of both $y^{(0)}$ and $\mu^{(0)}$ at background now produces boosted opposite sign mixes of local $y^{(1)}$ and $\mu^{(1)}$, an effect characteristic of late time injection.

One more small detail we can see in this later injection picture is the effect of injecting a distortion while in sub-horizon evolution, like the case of $k=1.0~{\rm Mpc}^{-1}$. We can see by comparing the leftmost column of Fig.~\ref{fig:zh_5e+4_transfer} (compare to Fig.~\ref{fig:zh_5e+5_transfer}) that the oscillations begin immediately following the formation of an average distortion, since many driving sources (e.g. Doppler boosting) are still in effect, and in particular drive with the same frequency in either case. The lack of a super-horizon equilibrium however reduces the noticeable effect of the offset varying mean. Injecting energy close to or after horizon-crossing for smaller $k$ will leave noticeable impacts on the CMB power spectrum, which would be most prominent in the $y$-era. We see this effect in Sect.~\ref{fig:zh_power_spectra}. To see this clearer we include Fig.~\ref{fig:switches_k_1_zh_5e+4}, where we can explicitly see a lack of contribution from Doppler driving, considering the lack of an average distortion at the time of horizon-crossing. The perturbed thermalisation and anisotropic heating terms are still able to cause a slight offset of the oscillation, but it is much less dramatic than for modes with a full super-horizon phase.

\vspace{2mm}
\subsection{Anisotropies for energy injection in the $y$-era}
As a last illustration, we consider distortion anisotropies for injection at $z_{\rm injection}=\pot{5}{3}$, as shown in Fig.~\ref{fig:zh_5e+3_transfer}. The average distortion is mainly a $y$-type signal with a $\simeq 2\%$ energy contribution from $\mu$. At this late stage, none of the perturbed thermalisation effects (i.e., scattering and emission corrections) contribute significantly, and the evolution is dominated by the Doppler and potential driving terms upon horizon-crossing. We see by inspecting Fig.~\ref{fig:switches_zh_5e+3} that perturbed thermalisation gives an initial boost predominantly $\mu$, but potential driving is the dominant source.

We can see that in all cases shown in Fig.~\ref{fig:zh_5e+3_transfer} the distortion transfer functions very quickly become highly correlated at a fixed ratio, i.e., $y^{(1)}_\ell\propto \mu^{(1)}_\ell$. This is expected since there is no spectral evolution and the anisotropies simply follow $\boostO Y$.

The $k=0.01~{\rm Mpc}^{-1}$ mode crosses horizon very soon after the injection time, and as such still receives the potential boosting contribution. Note however that a smaller $k$ could have undergone gravitational decay before an average distortion existed in the Universe. We will see later that some peaks in the CMB power spectrum are hindered by very late injection time, since they receive Doppler driving but not potential decay (see Fig.~\ref{sec:power_spectrum_switches}).

\vspace{2mm}
\subsection{Anisotropic heating from decaying particles}
\label{sec:particle_decay}
All of the discussion presented above only considered an average heating processes at a single redshift. Another interesting case we consider is due to heating by decaying particles, for which two additional aspects become important. Firstly, decaying particle scenarios lead to a more complicated time-dependent evolution of the average distortion \citep[e.g.,][]{Chluba2011therm, Chluba2013fore, Chluba2013PCA}. This will affect the main distortion transfer functions in interesting ways. Secondly, assuming that the decaying particle densities are modulated by perturbations in the cosmic fluid, anisotropic energy release will occur, which directly creates distortion anisotropies [the ${Q'_{\rm c}}^{(1)}\in\vek{Q'}^{(1)}$ term in Eq.~\eqref{eq:evol_1_final_Yi_1st_ord}]. While the average energy release has been used to constrain decaying particle scenarios based on \COBEF data \citep{Sarkar1984, Hu1993b, Chluba2013PCA, Acharya2021large}, the latter effect was never before discussed.
\begin{figure}
	\centering
    \includegraphics[width=1.0\columnwidth]{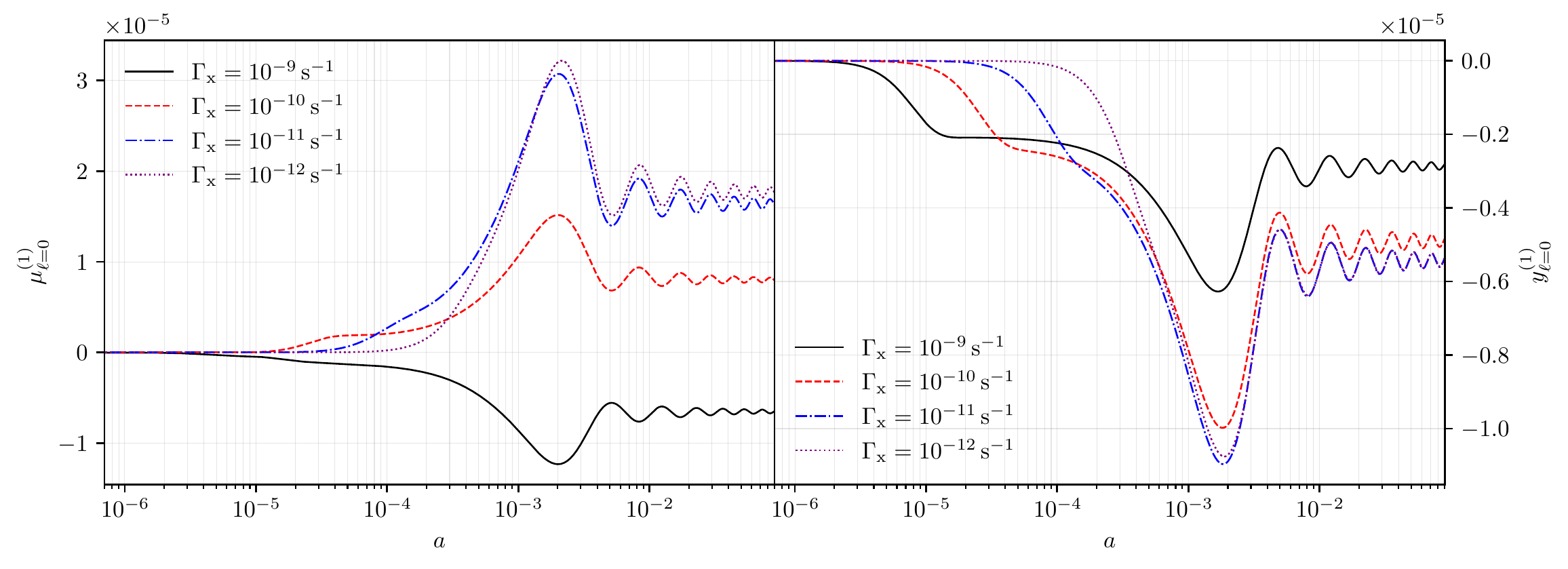}
	\caption{Decaying particle distortion transfer functions for $\mu^{(1)}_{\ell=0}$ (left column) and $y^{(1)}_{\ell=0}$ (right column) at $k=0.01\,{\rm Mpc}^{-1}$ and a total energy release of $\Delta \rho_\gamma/\rho_\gamma=10^{-5}$. In each figure, we varied the lifetime of the particle. Anisotropic heating is {\it not} included here. For the short-lifetimes, perturbed thermalisation effects are visible before horizon-crossing.}
\label{fig:decay_cases}
\end{figure}
\subsubsection{Time-dependent heating effect on the distortion transfer functions}
Following \citep{Chluba2011therm, Bolliet2020PI}, we implemented a simple heating module for decaying particles, assuming a constant lifetime, $t_X=1/\Gamma_X$, and mass of the particle, $m_X$. The average relative heating rate can then be expressed as \citep[see Eq.~(6) of][]{Bolliet2020PI}
\begin{align}
\label{eq:decay_heating}
\frac{\id \mathcal{Q}^{(0)}}{\id t}
&\approx \frac{m_X c^2\,\Gamma_X\,N_X}{\rho_\gamma}=\frac{\rho_{X,0}\,\Gamma_X\,\expf{-\Gamma_X t}}{\rho_{\gamma,0} (1+z)}
\approx \pot{4.85}{3}\,
f_{\rm dm}\,
\left[\frac{\Omega_{\rm cdm}h^2}{0.12}\right]\,\frac{\Gamma_X\,\expf{-\Gamma_X t}}{1+z},
\end{align}
where in the last step we introduced $f_{\rm dm}=\rho_{X,0}/\rho_{\rm cdm,0}$ to allow varying the fraction of dark matter that the particle can make up. Note that calligraphic $\mathcal{Q}$ (as compared to $Q_{\rm c}$) is normalised by $1/\rho_z$, making these expressions match the terms appearing in Eq.~\eqref{eq:evol_1_final}.

In Fig.~\ref{fig:decay_cases}, we illustrate the distortion monopole solutions for various particle lifetimes. We fixed the total energy release to $\Delta \rho_\gamma/\rho_\gamma=10^{-5}$ by adjusting $f_{\rm dm}$. Comparing to the single injection transfer functions above it is clearly visible how different decay rates smoothly vary across different distortion eras, with shorter (longer) lifetimes having the characteristic final same-sign (opposite-sign) combination of $y^{(1)}_{\ell=0}$ and $\mu^{(1)}_{\ell=0}$ from the boosting effects. This is related to the switch of the early (late) average distortion being $M$ ($Y$), as discussed in Sect.~\ref{sec:main_sources}.
For our illustration we focused on $k=0.01\,{\rm Mpc}^{-1}$, however, the overall picture does not change much when varying $k$. We also restricted ourselves to decays in the pre-recombination era, such that we could neglect the direct effects of decay on the ionisation history \citep{Chen2004, Padmanabhan2005}. The latter scenarios can be directly constrained using CMB anisotropies.
\begin{figure}
	\centering
	\includegraphics[width=0.99\columnwidth]{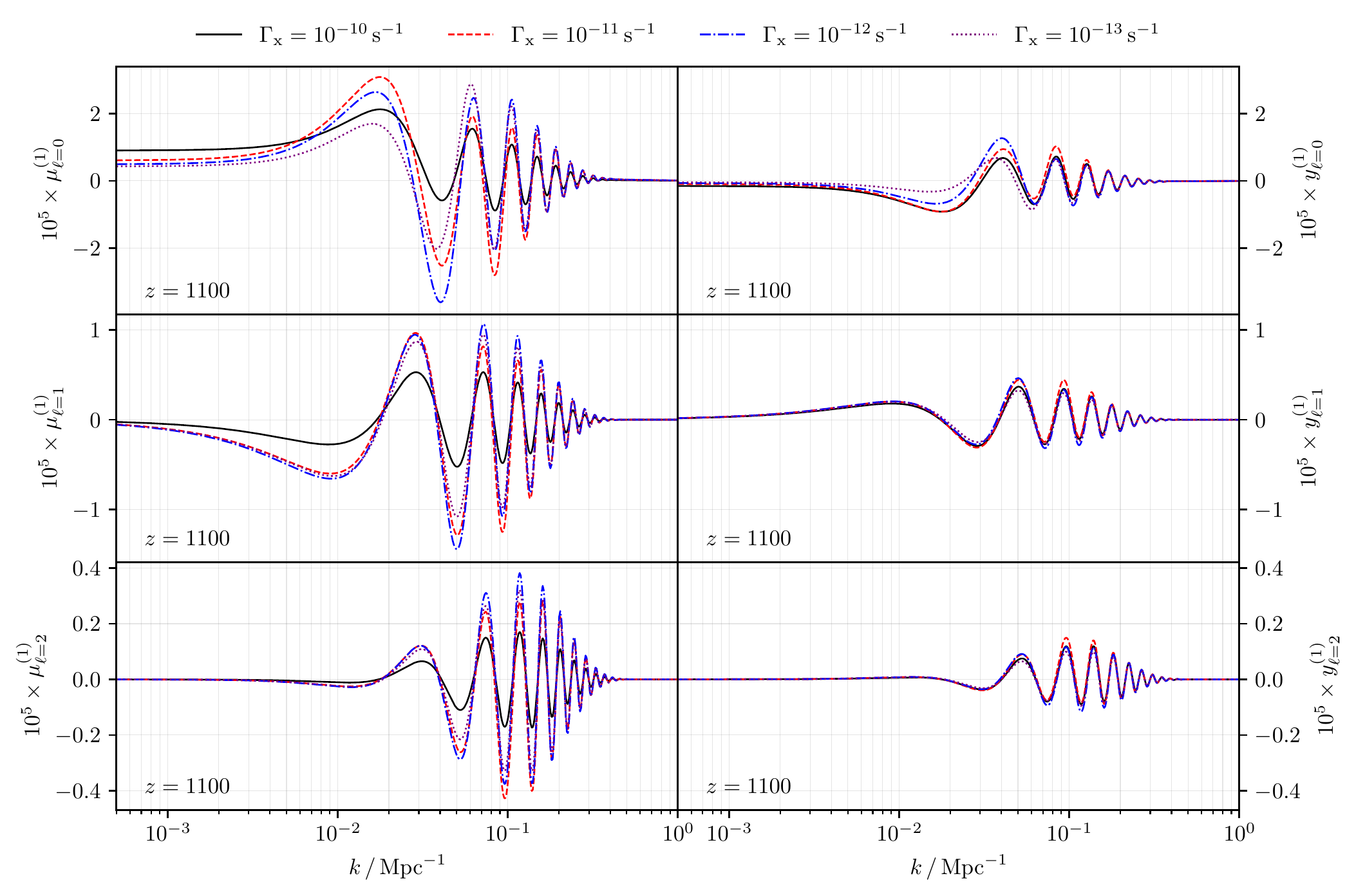}
	\\
	\caption{Snapshot of the first three decaying particle distortion transfer functions for $\mu^{(1)}$ (left column) and $y^{(1)}$ (right column) at $z=1100$ for a total energy release of $\Delta \rho_\gamma/\rho_\gamma=10^{-5}$. Each row shows a different multipole of the SED amplitude. In each figure, we varied the lifetime of the particle.}
	\vspace{-3mm}
\label{fig:perturbed_decay_cases_pert0}
\end{figure}
\begin{figure}
	\centering
	\includegraphics[width=0.99\columnwidth]{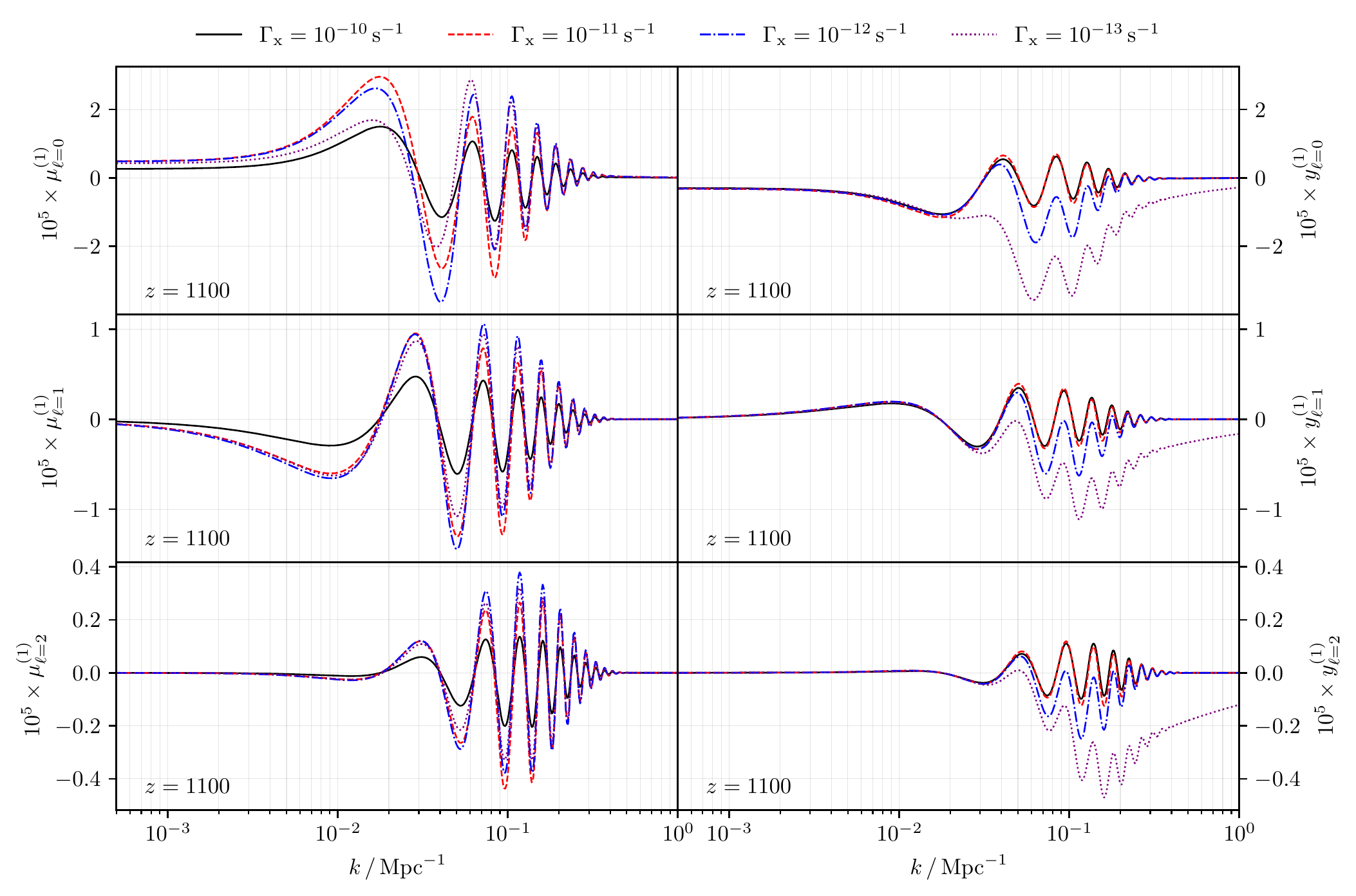}
	\\
	\caption{As for Fig.~\ref{fig:perturbed_decay_cases_pert0} but now perturbed decay was included here, with the effect becoming visible in particular for $y_\ell$ and late times.}
	\vspace{-3mm}
\label{fig:perturbed_decay_cases_pert1}
\end{figure}
\subsubsection{Perturbed decay effect on the distortion transfer functions}
In the previous section, we only consider the isotropic part of the heating process. However, if the decaying particle density is assumed to follow the dark matter distribution, we will also have an anisotropic heating term (see Appendix~\ref{app:pert_decay} for a brief derivation).
\begin{align}
\label{eq:perturbed_decay_heating}
\frac{\id \mathcal{Q}_{\rm c}^{(1)}}{\id t}
&\approx
\delta^{(1)}_{\rm cdm}\,
\frac{\id \mathcal{Q}_{\rm c}^{(0)}}{\id t},
\end{align}
which approximately accounts for the effect of number density modulation that acts alongside the usual modulation of the local time in each Hubble patch $\propto \Psi^{(1)}$ [present for all heating mechanisms as per Eq.~\eqref{eq:evol_1_final}]. We also assume that heating always only affects the local monopole, sourcing $y^{(1)}_{\ell=0}$.

In Fig.~\ref{fig:perturbed_decay_cases_pert0}, we show snapshots of the monopole, dipole and quadrupole $\mu^{(1)}_{\ell}$ and $y^{(1)}_{\ell}$ distortion transfer functions at $z=1100$, thus highlighting relative contributions to the SD power spectra (see Sect.~\ref{sec:power_spectrum}). The lifetime of the decaying particle is varied in each panel. Broadly speaking, the longer the lifetime the larger the contribution of $y^{(1)}$.

In Fig.~\ref{fig:perturbed_decay_cases_pert1} we show the same figure, with perturbed decay included. For the longest lifetimes we can see a significant enhancement directly from the perturbed decay term (e.g., blue/dashed-dotted lines). This effect is not visible in the $\mu^{(1)}$ transfer function, since at these late times there is no chance for the distortion to thermalise (this could be different for much earlier times than $z=1100$, however here we are concerned with CMB power spectra).

\begin{figure}
	\centering
	\includegraphics[width=1.0\columnwidth]{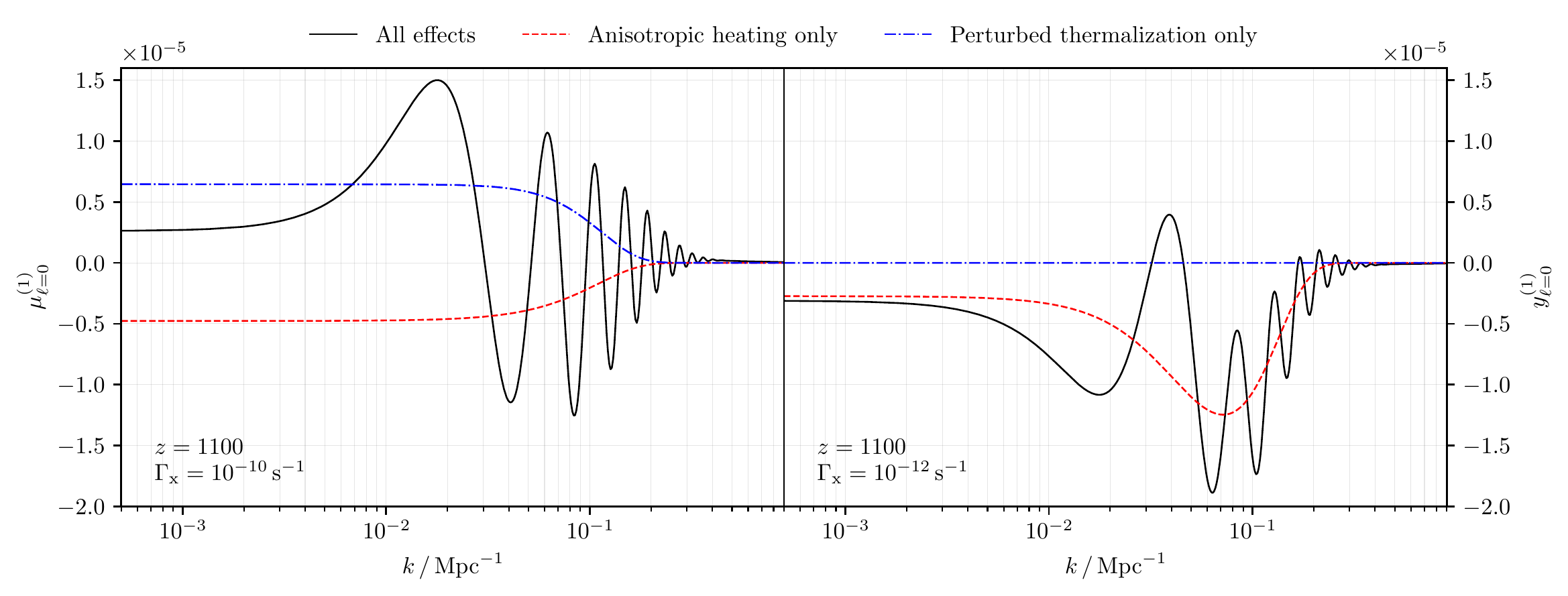}
	\caption{Snapshot of the decaying particle distortion transfer functions for $\mu^{(1)}_0$ (left column) and $y^{(1)}_0$ (right column) at $z=1100$ and a total energy release of $\Delta \rho_\gamma/\rho_\gamma=10^{-5}$. The particle lifetime was fixed as annotated but physical contributions were varied. The numerical solutions are computed using $N_{\rm max}=15$ and the transfer functions are given in the scattering basis.}
\label{fig:decay_cases_phys}
\end{figure}
In contrast to the previous discussions, we find that for the short lifetimes perturbed thermalisation effects contribute noticeably to $\mu^{(1)}_{\ell=0}$ at $k\lesssim 0.01\,{\rm Mpc}^{-1}$, and in fact almost cancel the anisotropic heating effects with the perturbed decay included. To illustrate these last two statements more clearly, in Fig.~\ref{fig:decay_cases_phys} we fixed the lifetimes as annotated but explicitly vary the physics. Perturbed thermalisation decreases rapidly for longer lifetimes (right panel), leaving anisotropic heating as the dominant driving force which enhances the fluctuation amplitude at intermediate and large scales. As the red line in Fig.~\ref{fig:decay_cases_phys} indicates, this contribution, is quite smooth without acoustic oscillations. In the left panel we see a combined effect of perturbed thermalisation and anisotropic heating almost cancelling, contrary to the intuition built in Fig.~\ref{fig:aniso_spec_buildup}. This is due to the aforementioned combination of $\delta_{\rm cdm}^{(1)}+\Psi^{(1)}$ which in adiabatic initial conditions evaluate to reverses the sign of the anisotropic heating had we neglected perturbed decay.

We can anticipate that for lifetimes $t_X\gtrsim 10^{12}\,{\rm s}$, the effects may become even more dramatic; however, in this regime also changes to the ionisation history ought to be included. In this case, the Doppler and potential driving effects will reduce, and pure anisotropic heating terms, leading to $y\times y$-type distortions only, will dominate. We will consider this regime in future work.

\section{CMB power spectra with primordial distortions}
\label{sec:power_spectrum}
Studying the transfer functions in Sect.~\ref{sec:transfer_functions} has revealed many important physical aspects in the evolution of anisotropic photon spectra: three main types of source connect the average distorted spectrum to local distortions patches. At early times the picture is dominated by anisotropic heating and perturbed thermalisation, with late times seeing main contributions from boosting sources. These local distortion patches undergo their own evolution including Thompson scattering and thermalisation terms yielding complex SED transfer functions.

This all tells us that the simple three-era picture of average spectral distortions does not exist in the anisotropic case, or at least not as directly. Studying the various limiting cases of energy injection into the $\mu$- and $y$-eras reveals that a mix of both $\mu^{(1)}$ and $y^{(1)}$ will almost always be present in the anisotropic spectrum. It is feasible that by carefully studying the composition of the anisotropic spectrum one could deduce what composition of $X$, $\boostO X$, $\DiffO X$ and $Y_1(x)-\Yspec(x)$ is present, where $X=\Mspec(x)~{\rm or}~\Yspec(x)$ would give a sense of the origin of the anistropic signal. The bottom line is that there still is a three-era picture, encoded by complex mixes of the SEDs making up the simple picture at background level.

The simplicity lost in the SD description allows us to yield an exciting gain in observational power. Using the formalism described in this paper we can calculate power spectra from the primordial SED perturbations, and thus open the door to apply the conventional tools used in CMB analysis, but now resolving nuanced spectral shapes in place of a simple blackbody. By measuring the cross correlations of temperature with $\mu^{(1)}$ or $y^{(1)}$ we can not only place novel constraints on the total energy release in the primordial plasma, but we can also potentially infer the time of this injection. Furthermore, the precise shape of the spectrum could additionally reveal details of the energy injection itself, with multiple injection or continuous energy release scenarios producing distinct power spectra. We illustrate these points by again studying the range of power spectra arising from single injection events, and contrasting with the particle decay scenarios.

Given the context of observation, the results will now be shown only in the observation basis (see Sect.~\ref{sec:change_of_basis}). We remind the reader that this projects the $Y_n(x)$ SEDs back to $\Gspec(x)$, $\Yspec(x)$ and $\Mspec(x)$, making only small use of the residual modes. The observation basis in particular does not preserve photon number in this projection, since the process of finitely binning and observing the frequency space would not allow for number estimates in real observation (see paper I). Because of this, the basis will typically slightly exaggerate the \textit{physical} $\mu$ amplitude and compensate with a negative temperature shift, as seen in Fig.~\ref{fig:basis_comp}. However, given the \COBEF constraints on average energy release, the latter is too minor to change the temperature fluctuation significantly, unless more minor (second order) effects would be considered. We are thus left with a marginal boost of $\mu$ due to this interplay at the start of the residual distortion era (see \citep{Chluba2013PCA, Lucca2020} and paper I).

Evolving the various SED amplitudes until today can be performed with the usual line-of-sight (LOS) integration by including the modified system in Eq.~\eqref{eq:evol_1_final}. Again we summarise the bottom line from the companion paper II:
\begin{align}
\label{eq:formal_sol_Leg_fin}
    \tilde{\vek{y}}^{(1)}_\ell(\eta_f, k) &= \int_0^{\eta_f} \id \eta \,g(\eta)\, \tilde{\mathcal{\vek{S}}}_\ell(\eta, \eta_f, k),
    \\
\tilde{\mathcal{\vek{S}}}_\ell(\eta, \eta_f, k)&=
\left[
 \tilde{\vek{y}}_0^{(1)}+\tilde{\Psi}^{(1)}\vek{b}^{(0)}_0+
 \left(\frac{\partial \tilde{\Psi}^{(1)}}{\partial \eta}
 -\frac{\partial \tilde{\Phi}^{(1)}}{\partial \eta}
 \right)\frac{\vek{b}^{(0)}_0}{\tau'}
 \right]\,j_\ell(k\Delta\eta)
+\tilde{\beta}^{(1)}\vek{b}^{(0)}_0\,j^{(1,0)}_\ell(k\Delta\eta)
+
\frac{\tilde{\vek{y}}_2^{(1)}}{2}\,j^{(2,0)}_\ell(k\Delta\eta)
\nonumber
\\
\nonumber
&\!\!\!\!\!\!\!\!\!\!\!\!\!\!\!\!\!\!\!\!\!\!\!\!\!\!\!\!
+
\!\left\{\Thz\!\left[M_{\rm K}\,\tilde{\vek{y}}^{(1)}_0+\tilde{\vek{D}}^{(1)}_0
+\left[\tilde{\delta}_{\rm b}^{(1)}+\tilde{\Psi}^{(1)}\right]\left(M_{\rm K}\,\vek{y}^{(0)}_0+\vek{D}^{(0)}_0\right)
+\tilde{\Theta}^{(1)}_0 \left(\vek{D}^{(0)}_0
+
M_{\rm D}\,\vek{y}^{(0)}
-\vek{S}^{(0)}\right)
\right]
+\frac{{\vek{Q}'}^{(1)}}{4\tau'}\right\} j_\ell(k\Delta\eta).
\end{align}
We note again that the first entry in spectral parameter vector $\vek{y}^{(1)}_\ell$ is the standard temperature perturbation (see Sect.~\ref{sec:class_comp}). The other SED amplitudes are all smaller in proportion to the total energy injection. Throughout this section we inject total energy $\Delta\rho/\rho = 10^{-5}$, yielding typical dimensionless power spectra of magnitude $\mathcal{D}_{\ell}^{{\Theta}\mu} \simeq \mathcal{D}_{\ell}^{{\Theta}y} \simeq 10^{-5} \mathcal{D}_{\ell}^{{\Theta\Theta}}$. Given $\mathcal{D}_{\ell}^{{\Theta\Theta}}\simeq 10^{-9}$ at the largest scales in standard $\Lambda$CDM, this implies a typical cross-power spectrum amplitude of $\mathcal{D}_{\ell}^{{\Theta}\mu} \simeq \mathcal{D}_{\ell}^{{\Theta}y} \simeq 10^{-14}$ in dimensionless units. As we discuss in Sect.~\ref{sec:forecasts}, this level is in fact just below the sensitivity of \Planck but already exceeds the sensitivity of \Litebird and \PICO.

To compute the signal power spectra one can apply the standard formula
\begin{align}
C_\ell^{XY}(\eta)
&=
    \frac{2}{\pi}
    \int k^2 \diff k \, P(k)\, \hat{X}_\ell(\eta, k) \, \hat{Y}_\ell(\eta, k),
\end{align}
where the transfer functions for the variables $X$ and $Y$ are used together with the standard curvature power spectrum, $P(k)$. We shall assume the standard cosmological parameters \citep{Planck2015params} in all our computations below. We will present results with the usual normalisation $\mathcal{D}_{\ell}^{XY}=\frac{\ell(\ell+1)}{2\pi} \,C_\ell^{XY}$.

\subsection{Numerical setup}
\label{sec:LOS_integral_details}
The calculation of power spectra using Eq.~\eqref{eq:formal_sol_Leg_fin} can be numerically challenging. Here we provide details on the new implementation of this calculation within {\tt CosmoTherm}, which relied heavily on the advice provided in section V of \cite{callin:how_to_CMB}.

Transfer functions for sufficiently large $k$ undergo Silk damping \cite{Silk1968} long before recombination, and do not impact the CMB power spectrum. On the contrary, modes with low $k$ have not yet crossed horizon even at modern times, and thus also have no influence on the CMB spectrum. We therefore limit our calculations to $\pot{2}{-5}\leq k/{\Mpc}^{-1} \leq 0.5$, with an understanding that the larger (lower) $k$ in this range impact the high (low) $\ell$ power spectrum. In particular, we highlight that $k=0.01\,{\Mpc}^{-1}$ corresponds roughly to the scale of the first peak in the temperature power spectrum, since it reaches its maximum amplitude at recombination (see e.g. Fig.~\ref{fig:basis_comp}). Most figures in Sect.~\ref{sec:transfer_functions} showed this mode, which can be helpful in observing some of the physical effects discussed below.

One of the largest complicating aspects of the power spectrum calculation is the combination of a slow varying source with a rapidly oscillating bessel function under the same integrand. To illustrate this, we schematically\footnote{Note that to fully express Eq.~\eqref{eq:formal_sol_Leg_fin} in this form we would take a summation over $j_\ell(k\Delta\eta)$, $j^{(1,0)}_\ell(k\Delta\eta)$ and $j^{(2,0)}_\ell(k\Delta\eta)$ with their corresponding sources, however here our aim is to clarify the computation and will use the simpler expression with a single source.} write
\begin{equation}
    I = \int_0^{\eta_f} {\rm d}\eta \; g(\eta) \, \mathcal{S}(\eta,k) \, j_\ell(k\Delta\eta),
\end{equation}
where we have split the source $S$ from Eq.~\eqref{eq:formal_sol_Leg_fin} into a source explicitly dependent on perturbed quantities and the leading spherical Bessel function.

Given this decomposition the approach will be to pretabulate a relatively sparse grid of $\mathcal{S}(\eta, k)$ in a relevant region. This greatly simplifies the calculation since the source function varies slowly in log space while also being expensive to calculate -- requiring evolving the primordial perturbations forward from much earlier times. The penalty is increased in this new framework where establishing an accurate background spectrum requires solving even background equations from the time of energy injection, long before relevant scales have crossed horizon. The relevant region for this pretabulation is dictated by the visibility function, which in practice can be seen as restricting the integral limits to concentrate around recombination $z\approx 1100$. This is slightly different for the Integrated Sachs-Wolfe effect, whose terms contain an explicit $1/\tau'$ which can be interpreted of as changing $g(\eta) \rightarrow \exp(\tau(\eta_f)-\tau(\eta))$ and thus stretching the region of importance all the way to modern times. In the calculations shown below we create a pretabulated region with 500 points $k/{\rm Mpc}^{-1}\in\big[ \pot{2}{-5}, 0.5 \big]$ and 1000 points $\eta/{\rm Mpc}\in\big[200, \eta_0\big]$, crucially both being log-spaced.

These 2D grids are then interpolated and used for integration with the Bessel functions, which is best done in linear space since the Bessel function zeros are -- for our purposes -- spaced evenly. To efficiently integrate a highly oscillatory function, in {\tt CosmoTherm} we use Chebyshev integration techniques, and find the integral across $\eta$ converges with $\simeq 2^{10}$ samples. Another large efficiency boost in the code is to cache the values of the necessary spherical Bessel function\footnote{We use the {\tt Boost}-library \url{www.boost.org} to accelerate the computation and achieve high precision.}, knowing that $k\Delta \eta$ falls between 0 and some maximal value $k_{\rm max}(\eta_0-\eta_{\rm min})\approx 0.5\times 15000=7500$.\footnote{This statement is somewhat cosmology-dependent; however, $7500$ is already quite conservative.}
Finally, it is noteworthy that the power spectrum is a smooth function, and not all $j_\ell$ need to be integrated. In practice the sampling can become quite sparse towards high $\ell$, with a cubic spline making up the missing evaluations.

Following the integral across conformal time we are effectively left with $X(\eta_0, k)$, and the integral across $k$ can be performed as required. Here, the benefit of the pretabulated sources has become apparent, since many more points are required to effectively capture the oscillations in $X(\eta_0, k)$ than in $\mathcal{S}(\eta, k)$, again because of the spherical Bessel function in the time integrand. Specifically, we calculate $\gtrsim 4000$ points in $X(\eta_0, k)$ from our original grid of only $500$ $\mathcal{S}(\eta, k)$ points, and have thus reduced the number of Boltzmann hierarchy calculations by an order of magnitude. This is especially noteworthy in this new treatment of the frequency space, where we have an additional $>(\ell_{\rm max}+1)\times (N_{\rm max}+2)$ equations compared to the standard Boltzmann solvers.\footnote{In this, the $+2$ comes from $\mu$ and $y$, while the $+1$ comes from the fact that the SD sector must be solved at the background level too.} For our chosen parameters of $N_{\rm max}=15$ and $\ell_{\rm max}=15$ this amounts to $272$ new equations on top of the $5+2\ell_{\rm max}=35$ needed for the standard calculation ($\Phi$, $\delta_{\rm cdm}$, $u_{\rm cdm}$, $\delta_{\rm b}$, $u_{\rm b}$, $\Theta_{\ell}$, $\nu_{\ell}$). Assuming some form of matrix inversion scaling like $\mathcal{O}(N^3)$ we get a solution taking over $1{\rm h}$ where it would have previously taken $10{\rm s}$ (even an optimistic scaling of $\mathcal{O}(N^2)$ yields a factor of $>10$, giving $13{\rm m}$ in place of $10{\rm s}$). 
In this first implementation of the problem, we have had a focus on accuracy and convergence over efficiency, and therefore shall be content with these performance numbers. We find that increasing any parameters here (e.g. $\ell_{\rm max}$ and $k$ or $\eta$ samples) yields no appreciable change to the final results [see however Appendix~\ref{sec:power_spectrum_convergence} for discussion of convergence across $N_{\rm max}$]. The efficiency can likely be increased however following more optimization similar to what has gone into state-of-the-art Boltzmann solvers like {\tt CAMB} \citep{CAMB} and {\tt CLASS} \citep{CLASSCODE}.

\begin{figure}
\centering
\includegraphics[width=1.0\columnwidth]{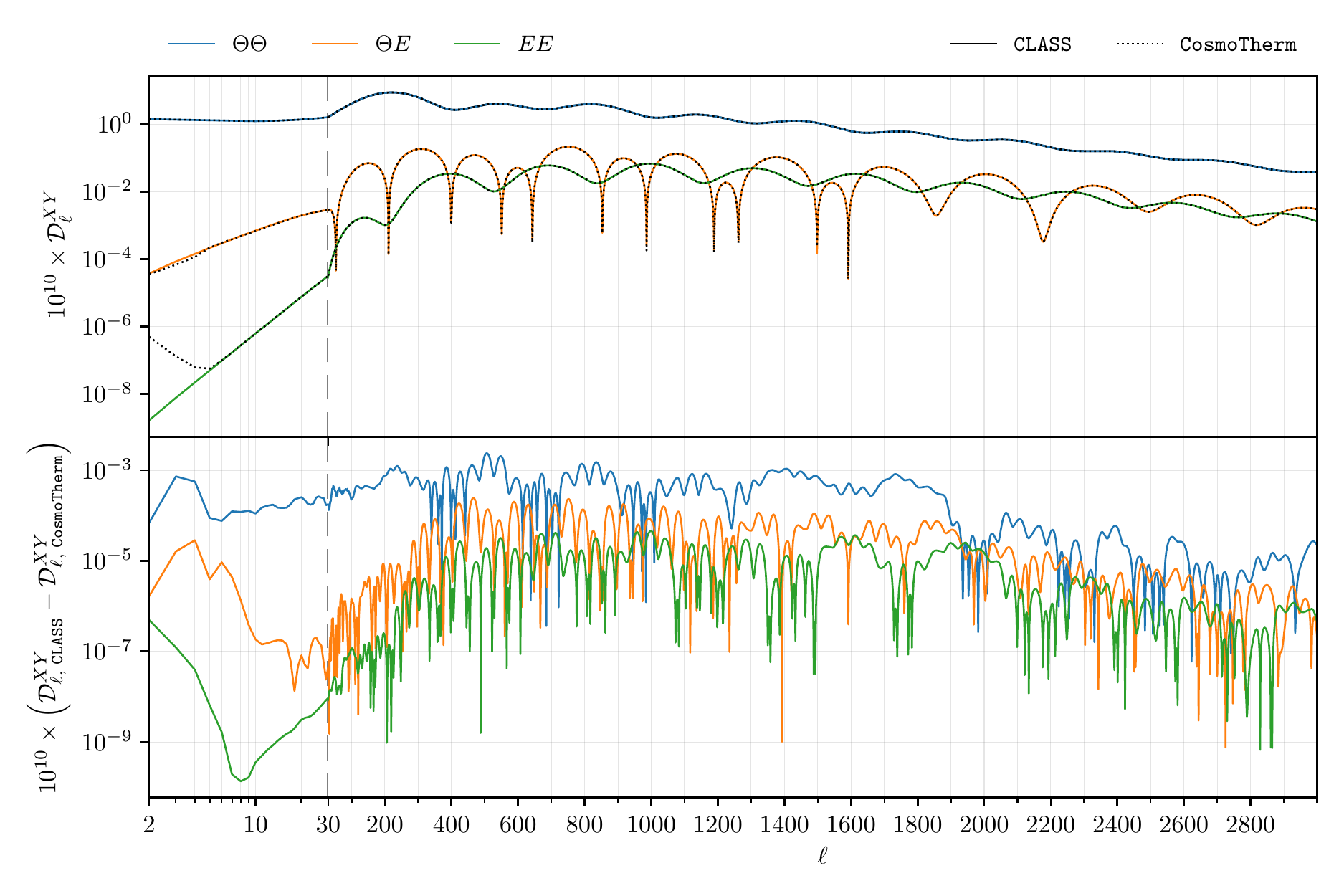}
\\
\caption{A figure showing the standard, albeit dimensionless, CMB power spectra ($\Theta\Theta$, $\Theta E$, $EE$) both in ${\tt CLASS}$ \citep{CLASSCODE} and ${\tt CosmoTherm}$. The top panel shows that only qualitative differences exist for very low $\ell$ in the $E$ mode spectra. The bottom panel reveals through the residuals that differences are below the percent level across the entire $\ell$ range.}
\label{fig:class_comp}
\end{figure}
\subsection{CMB temperature power spectrum benchmark}
\label{sec:class_comp}
The first entry in the photon vector $\vek{y}^{(1)}$ in this implementation reproduces the {\tt CLASS} $C^{\Theta\Theta}_\ell$ power spectrum to high precision as shown in Fig.~\ref{fig:class_comp}. The absolute value of relative differences between ${\tt CosmoTherm}$ and ${\tt CLASS}$ amounts to $0.03\%$ for $\mathcal{D}_\ell^{\Theta\Theta}$ and $0.29\%$ for $\mathcal{D}_\ell^{EE}$ once averaging over $2\leq\ell\leq 3000$ (or $0.02\%$ and $0.03\%$ for averaging residuals without absolute value).\footnote{We included polarisation effects on the temperature equations but removed reionisation effects from {\tt CLASS} for this comparison.} These results are achieved in $\simeq 30$s (wall time) running in parallel over 64 cores, showing some lack of optimisation compared to {\tt CLASS}, however comparable performance can most likely be achieved with further work.
We note that the $\Theta E$ and $EE$ quadrupole appear much larger in ${\tt CosmoTherm}$ than in ${\tt CLASS}$, however this will not impact the forecasts considering the cosmic variance at those scales.

\begin{figure}
\centering
\includegraphics[width=0.98\columnwidth]{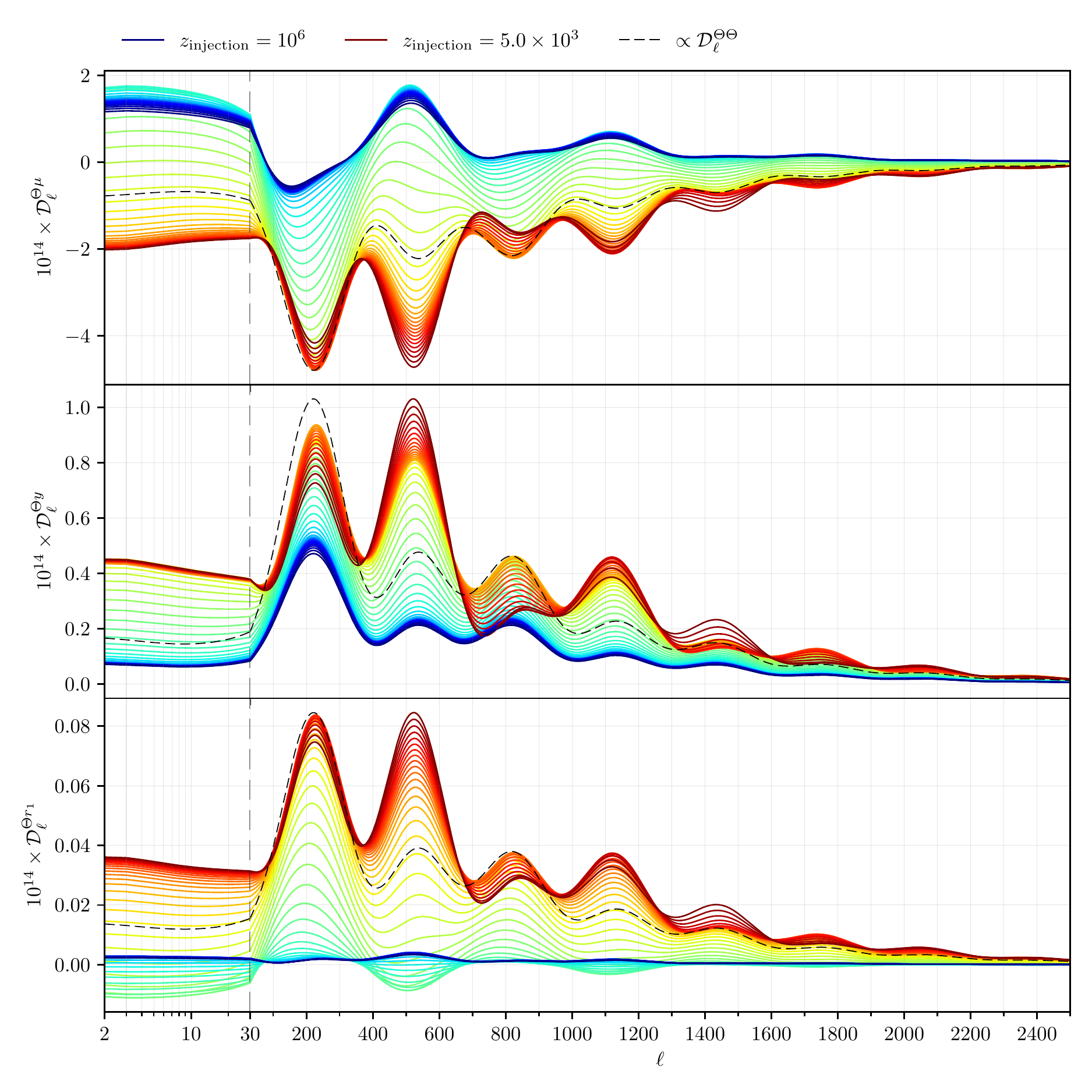}
\\
\caption{The power spectra for $\Theta\times \mu$, $\Theta \times y$ and $\Theta \times r_1$ over a range of $50$ single-injection redshifts. Blue lines show early injection into the $\mu$-era and red lines show late injection in the $y$-era. The vertical dashed line shows a division between log-spaced $\ell$ values (left) and linear-spaced values (right). For reference, we show the familiar $\Theta\times\Theta$ power spectrum (rescaled within each panel). Comparing the acoustic peak structure, we recognize that $\Theta\times\Theta$ and the respective $\Theta\times \mu/y/r_1$ power spectra are in phase, a sign of their common origins (e.g., Doppler boosting).}
\label{fig:zh_power_spectra}
\end{figure}
\begin{figure}
\centering
\includegraphics[width=0.98\columnwidth]{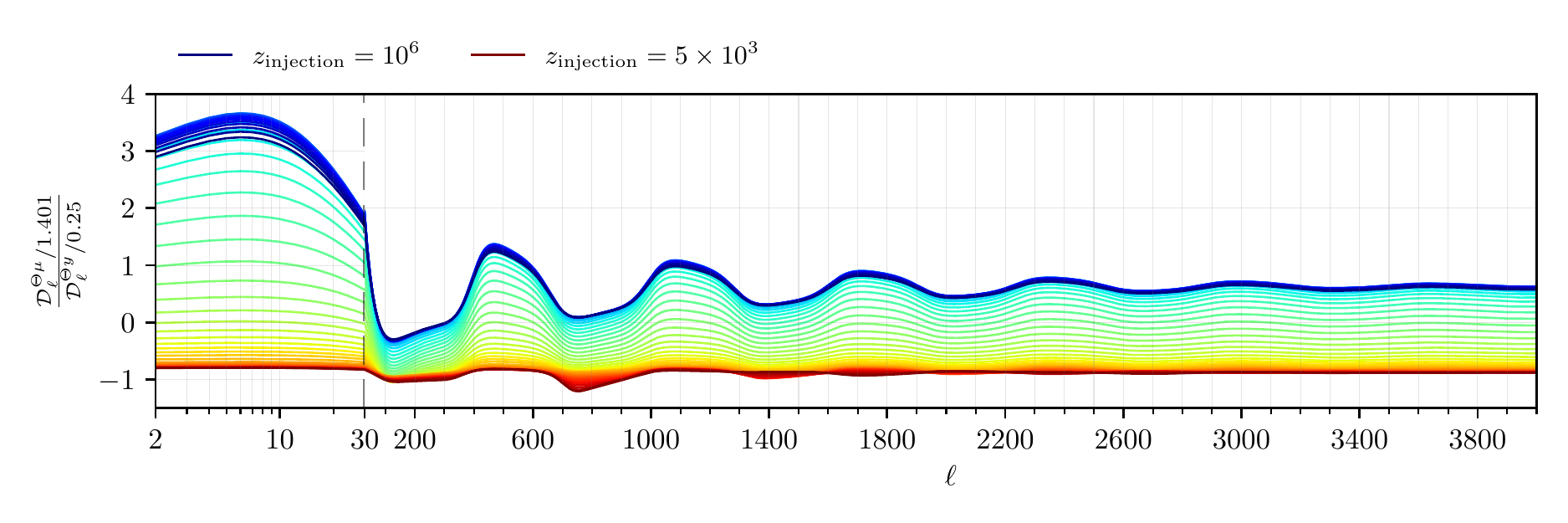}
\\
\caption{A figure showing ratios of power spectra, illustrating the relative composition of the spectrum in terms of $y$ and $\mu$. The additional factors normalise the amplitudes to their energy content. Both the average change in the $\mu/y$ ratio as well as the $\ell$-dependent change provide the means to distinguish energy injection scenarios.}
\label{fig:power_spectra_dist_ratio}
\end{figure}
\subsection{Single injection CMB power spectra}
\label{sec:power_spectrum_single_injection}
We now have all the ingredients to compute the first CMB parameter power spectra. In Fig.~\ref{fig:zh_power_spectra}, we show the $\Theta\times \mu$, $\Theta \times y$ and $\Theta \times r_1$ power spectra for various injection redshifts and $\Delta \rho_\gamma/\rho_\gamma=10^{-5}$. 
A rich acoustic peak structure is revealed, with a clear dependence on the injection epoch.

\begin{figure}
\centering
\includegraphics[width=0.875\columnwidth]{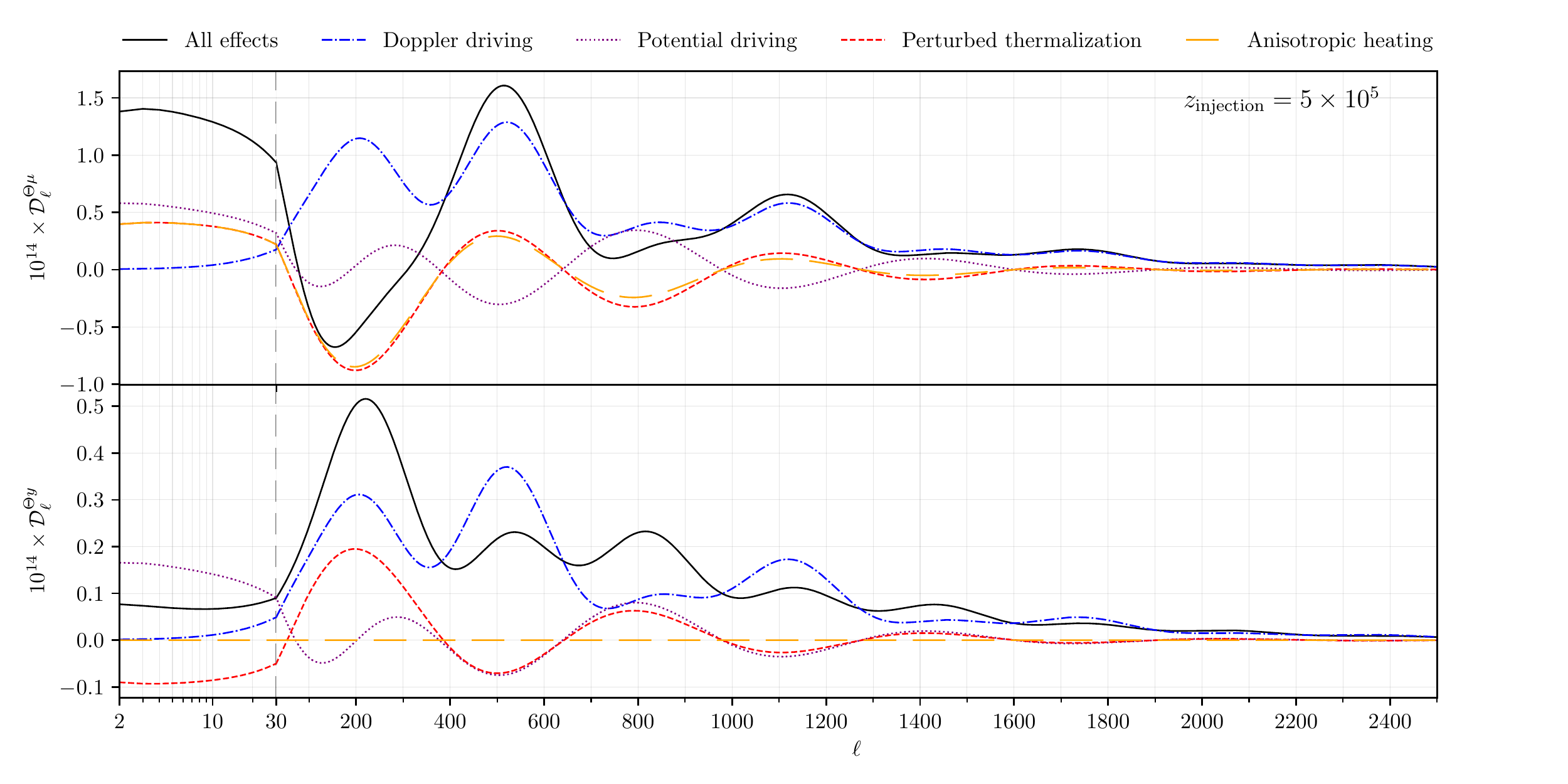}
\\[-2mm]
\includegraphics[width=0.875\columnwidth]{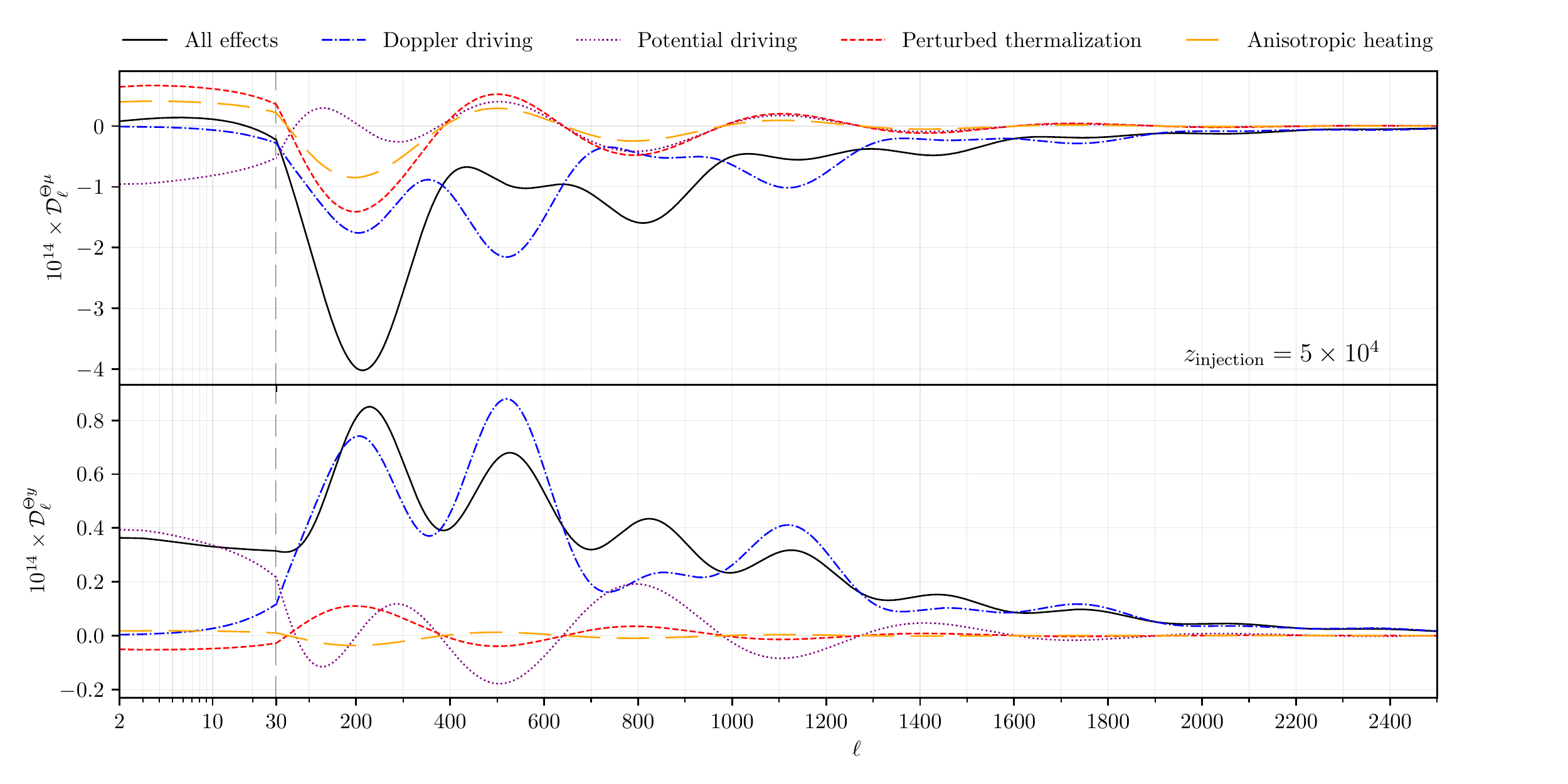}
\\[-2mm]
\includegraphics[width=0.875\columnwidth]{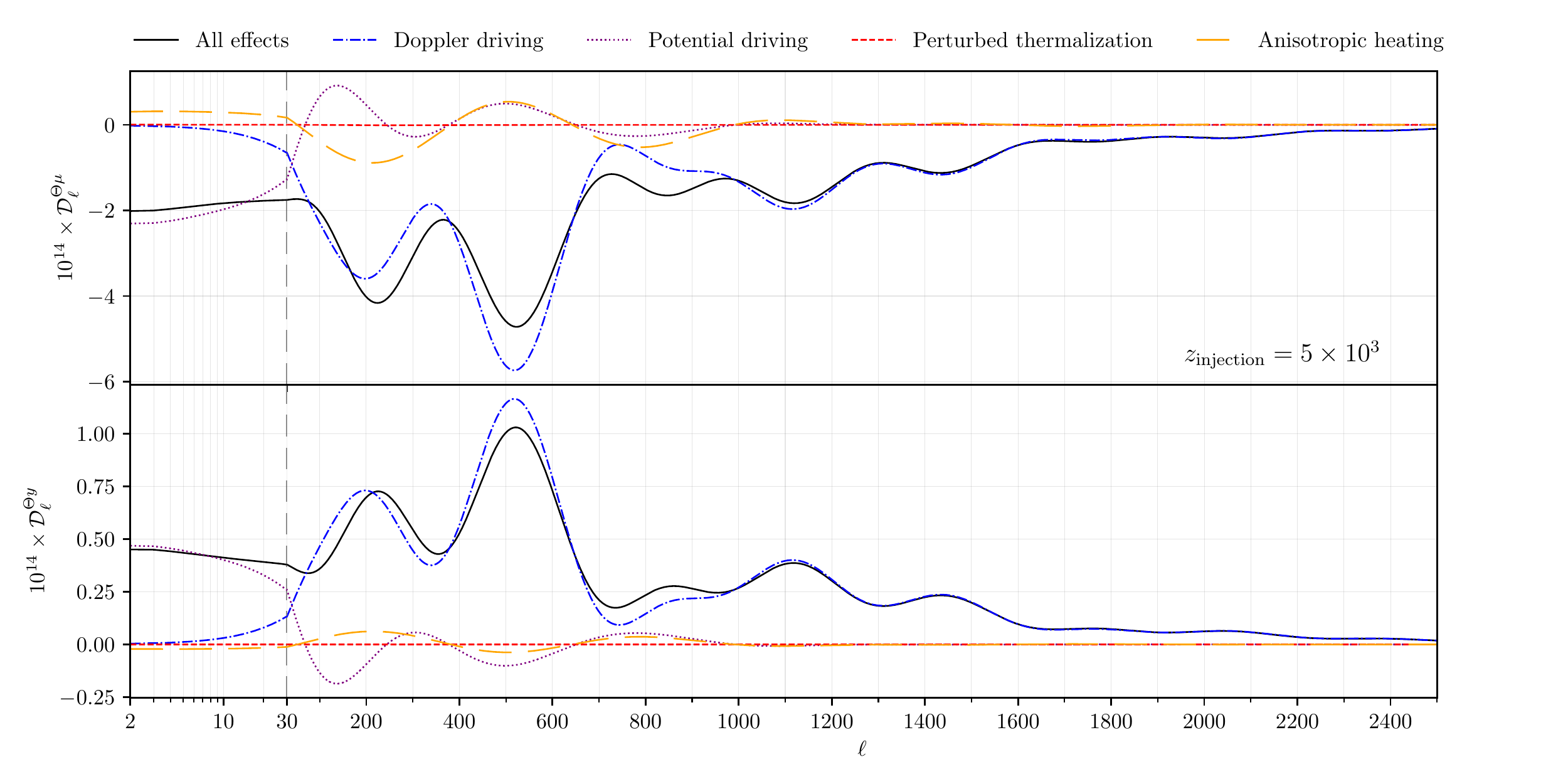}
\\[-2mm]
\caption{Three figures showing the power spectra for $\Theta\times \mu$ and $\Theta\times y$ for the three usual single-injection redshifts corresponding to each main SD era. Different lines indicate the inclusion or exclusion of a physical term (made more explicit in main text). A vertical dashed line shows a division between log-spaced $\ell$ values (left) and linearly-spaced values (right).}
\label{fig:zh_power_spectra_switches}
\end{figure}

Starting with late time injection we can see that the peaks in $\Theta\mu$ and $\Theta y$ are the same shape, with only some negative coefficient relating the two. This is due to boosting as the only source at sufficiently late times, as seen and discussed throughout Sect.~\ref{sec:transfer_functions}. This intuition is reinforced by the ratio of observed $\mu$ and $y$ energies in Fig.~\ref{fig:power_spectra_dist_ratio}, which approaches a consistent value for late time injection. Furthermore the fact that the peaks in the power spectra have similar appearance the usual $\Theta\Theta$ spectrum hints towards the common source of Doppler boosting, which can be verified by inspecting Fig.~\ref{fig:zh_power_spectra_switches}. Finally we note that the low $\ell$ part of the spectrum is similarly due to the late time ISW effect, again familiar from the standard Cosmological picture.

The earlier times are more complicated, with anisotropic heating and perturbed thermalisation taking on more importance and frequently counteracting the minimal contributions from boosting (see Fig.~\ref{fig:switches_zh_5e+5} and compare to top row of Fig.~\ref{fig:aniso_spec_buildup}). This can be immediately seen by how odd peaks are strongly suppressed in the $\Theta \times \mu$ spectrum, indicating a source which is not governed by Doppler peaks. In fact, the prevalence of even peaks hints towards the effect of baryon loading, variables which partly modulate the local thermalisation efficiency. In the $\Theta \times y$ spectrum the Doppler peaks are still appreciable, a consequence of perturbed scattering favouring the creation of $\mu^{(1)}$ through perturbed thermalisation (through both $M_{\rm D} \vek{y}^{(0)}_0$ and $Y_1-\Yspec$) and anisotropic heating thermalising to a $\mu^{(1)}$ spectrum, thus leaving the small boosts as sole contributors to local $y^{(1)}$ distortions.

The amplitude of the $\Theta\times r_1$ power spectrum is roughly one order of magnitude below the $\Theta\times y$, indicating that only about $10\%$ of the SD-energy is contained in this observable. Higher residual distortion power spectra (see Sect.~\ref{sec:higher_single}) drop further in amplitude, indicating fast convergence of the signal model and information.

The range of timings varies quite smoothly in the residual-era, but the spectra start to overlap more at the extremes. This implies a strong level of time sensitivity in observation for residual-era injection, while differentiating the moment injection in, say, the $y$-era will require strong measurements on individual peaks (see discussion in Sect.~\ref{sec:MCMC_prospects}). In the case of the $\mu$-era the discriminating power is quite reduced, with peaks mostly overlapping till injection at $z\lesssim \pot{2}{5}$, the moment thermalisation becomes inefficient and the residual-era begins. Nevertheless, a tomographic picture is revealed at $10^3\lesssim z_{\rm injection}\lesssim  10^{5}$.

The correlations $y\times E$ and $\mu \times E$ are also important for the forecasts (Sect.~\ref{sec:forecasts}) and are shown in Appendix~\ref{app:E_corr}. They are generally more complex and thus less illustrative than the correlations with temperature, hence their omission from main text.
\subsubsection{Isolating various physical effects}
\label{sec:power_spectrum_switches}

The power spectra are complex and composite statistics, where each $\ell$ involves contributions from many $k$ modes which thus encode different times of horizon crossing and thus different relative contributions of aniostropic distortions sources. In order to distil some physics from these data we will again rely on the switches\footnote{We chose to leave the temperature equations unchanged and {\it only} switch distortion drivers} explained in Sect.~\ref{sec:physics_switches}. We furthermore decompose boosting sources into Doppler boosting from baryon velocities and gravitational potential decay (the latter contributing mostly to late time ISW effects).

These switches allow us to dissect the rich features in the acoustic peaks themselves. To isolate a physical effect we calculate the power spectrum with and without the relevant terms in the evolution equations, and plot the difference between the two. For example, the Doppler contribution is found by subtracting the solution without Doppler driving from the full solution.\footnote{There is no way of showing the true isolated effects since the power spectrum is a squared statistic, and thus no simple superposition principle can be used. This technique however is highly illustrative.}

The main point Fig.~\ref{fig:zh_power_spectra_switches} illustrates is that Doppler driving is the dominant effect on the SD power spectra, with only early injection times seeing another comparable term. At these early times we have already seen that anisotropic heating and perturbed thermalisation become large contributors to the SD signal. Potential driving terms are most important at large scales ($\ell \leq 30$--$40$), introducing an integrated Sachs-Wolfe plateau to distortion signals. Although less important at small scales (high $\ell$), the potential driving terms provide important time-dependent information.

One notable feature in the single injection scenario is that for the latest of injection times the first peak starts to wane while the second peak continues its growth. The turn over point in the first peak happens around $z_{\rm injection}\approx \pot{2}{4}$. Similarly we see the third and fourth peak affected by the late injection. We can see that these changes are primarily caused by changes in the potential driving late into the $y$-era. Starting with $z_{\rm injection}=\pot{5}{3}$ we see that potentials don't drive $\ell>1000$, which received contributions from $k$-modes which were deep into the horizon at the time of injection, and thus saw almost no potential driving. Even for $\ell<1000$ we see smaller potential effects with decreasing injection redshift, which are likely caused by some combination of the aforementioned effect \textit{spreading} over $\ell$ and the fact that potential decay is greatly reduced close-to and beyond the matter-radiation transition. 
These potential decay effects are also visible, although less clearly, in the transient effects on the monopole and dipole transfer functions in the central column of Fig.~\ref{fig:zh_5e+4_transfer} (injection near horizon crossing) and Fig.~\ref{fig:zh_5e+3_transfer} (sub-horizon injection). This is in contrast to Fig.~\ref{fig:zh_5e+5_transfer} where the same mode received large boosting from potential decay at the time of horizon crossing, since the average SED amplitudes had been sourced prior.

The anisotropic heating contributions enhance $\mu$ at early injection times, and a mix of both $\mu$ and $y$ for all other times. While this follows the conventional picture in the residual-era -- energy thermalises to some intermediate spectral shape -- it is initially surprising for the late injection times. This is due to the additional anisotropic heating term we identify within $\vek{S}^{(0)}$ (see Sect.~\ref{sec:main_sources}), which sources a spectral shape corresponding to $Y_1(x)-\Yspec(x)$, thus having a nonzero projection onto $\Mspec(x)$.

By individually switching perturbed emission and perturbed scattering (not shown) we can confirm that emission is only ever a small subdominant contribution for the injection times considered here, and furthermore the dominant part of perturbed scattering is the $M_{\rm D}\vek{y}^{(0)}$ and $(Y_1-\Yspec)$ terms, with the other terms simply providing a delaying effect on the natural thermalisation local anisotropies undergo.

\begin{figure}
\centering
\includegraphics[width=0.98\columnwidth]{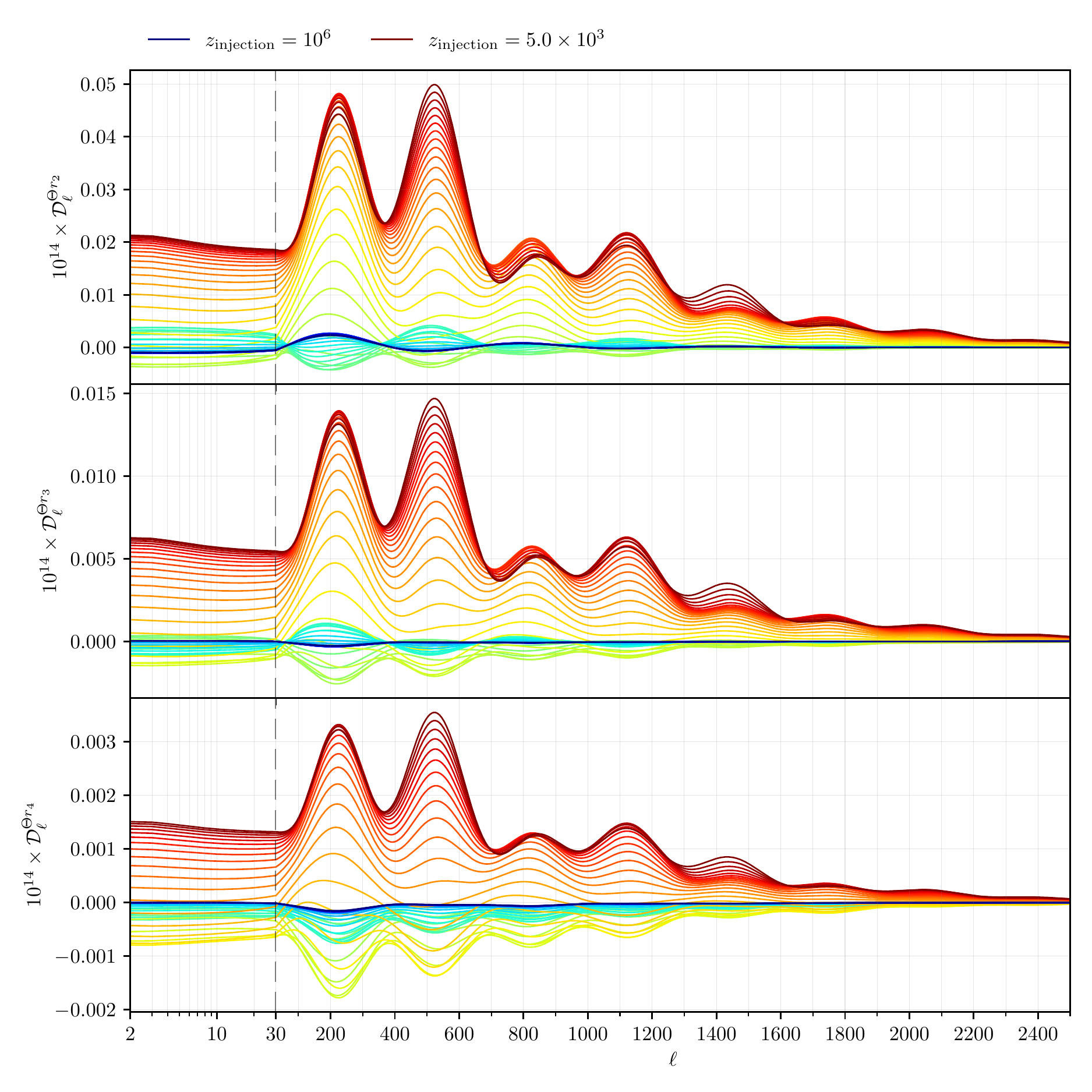}
\\
\caption{As for Fig.~\ref{fig:zh_power_spectra} but this time showing correlations between $\Theta$ the higher residual distortion modes, $r_2$, $r_3$ and $r_4$.}
\label{fig:zh_power_spectra_high_resid}
\end{figure}
\subsubsection{Higher-residual power spectra}
\label{sec:higher_single}
In Fig.~\ref{fig:zh_power_spectra_high_resid} we present the cross correlations for $\Theta\times r_2$/$r_3$/$r_4$, which show the amount of information not captured by the simple decomposition shown above. Importantly the residual modes are rank ordered by their relative importance as can be seen by the decreasing amplitude (they are all normalised similarly to the $y$ distortion, with a relative energy density $E_{r_n}=r_n/4$).

Interestingly, we see how $\Theta\times r_2$ and $\Theta\times r_3$ follow a similar growing shape to the $\Theta\times y$ spectrum for late times. This can be understood by considering that the dominant signal source of power spectra is often the Doppler driving term (see Fig.~\ref{fig:zh_power_spectra_switches}), and upon studying boosts of $\Yspec(x)$ around two residual modes are required for a good fit (see Fig.~\ref{fig:spectral_shapes_demonstration}). In a similar way the $\mu$-era injection makes less use of residual modes, a fact which relates to the decreased importance of the boosting sources. We see the residual modes amplitude drop by around a factor of $5$ for increasing $r_n$, showing the decreasing contributions. The remaining energy content in $r_4$ is quite small, a fact which relates also to convergence within the basis -- we notice small, albeit non-negligible changes in the amplitude of $r_4$ when increasing from, e.g., $N_{\rm max}=13$ to $N_{\rm max}=15$, which currently is close to the limits of our computation. This all hints towards the statement that using roughly 6 numbers ($\Theta$, $\mu$, $y$, $r_{1/2/3}$) is enough information to fully parameterise the photon spectra in the basis chosen here, however $\simeq 18$ are needed in the computation basis to capture the evolution in the most difficult regimes. This statement is, of course, basis dependent (see Sect.~\ref{sec:change_of_basis}), and specifically it does not exclude the possibility of finding an optimised smaller basis for given energy release scenarios and eras. We discuss this possibility and implications in Sect.~\ref{sec:forecasts_basis_discussion}.

\begin{figure}
\centering
\includegraphics[width=1.0\columnwidth]{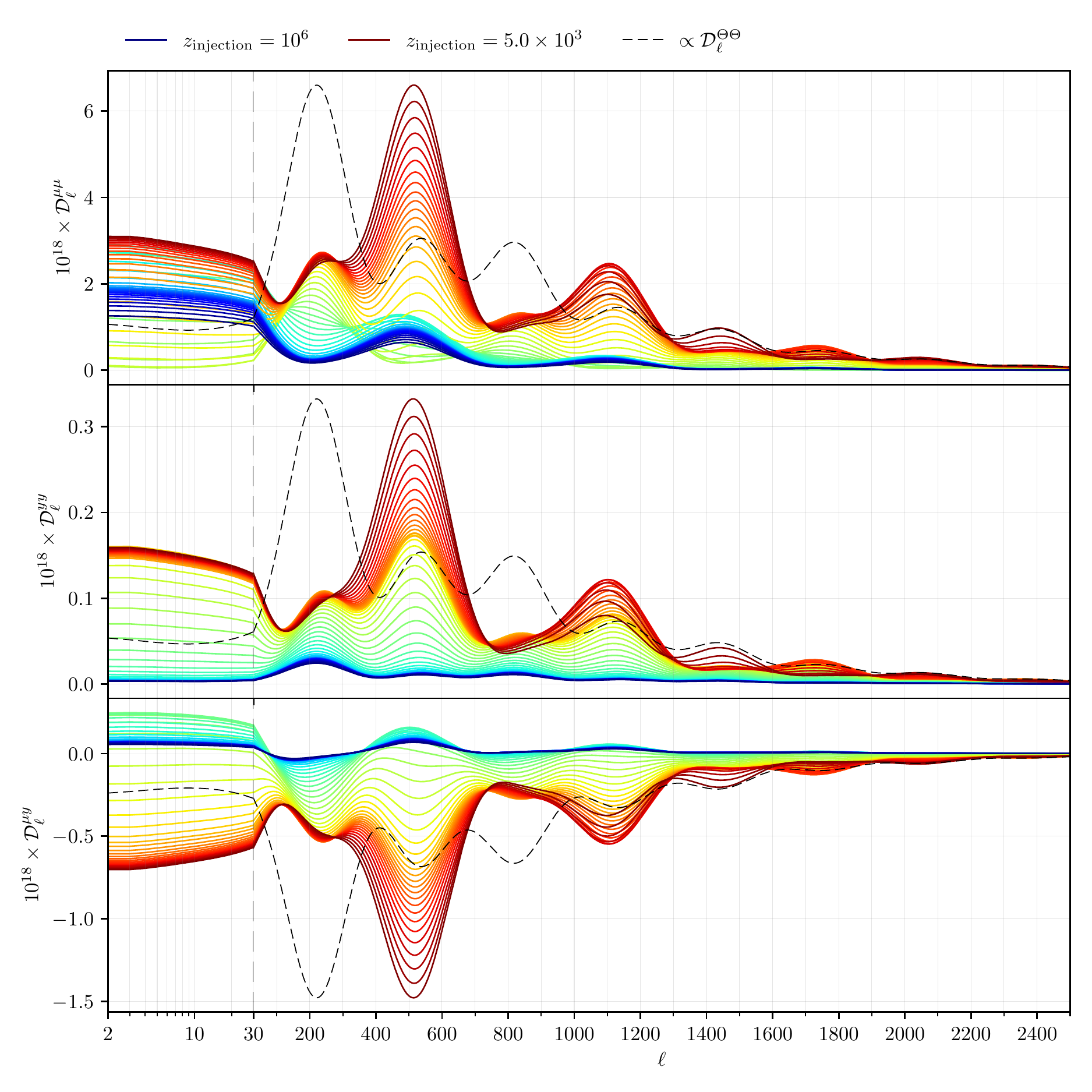}
\caption{A similar figure to Fig.~\ref{fig:zh_power_spectra} but now showing the $\mu\times\mu$ (top), $y\times y$ (middle) and $\mu \times y$ (bottom) spectra. Shown for comparison is the (re-scaled) dimensionless temperature power spectrum.}
\label{fig:zh_distortion_auto_spectra}
\end{figure}
\subsubsection{Distortion auto-power spectra}
Although they are far below the detection prospects of even future imagers (see Sect.~\ref{sec:forecasts}) it is illustrative to study the purely SD power spectra. In Fig.~\ref{fig:zh_distortion_auto_spectra} we show the $\mu\times\mu$, $y \times y$ and $\mu\times y$ spectra, together with a rescaled $\Theta\times\Theta$ spectrum for comparison.

The auto-power spectra show almost exactly the same structure for late injection times (upto an overall scale) since dominant source is boosting, yielding a fixed ratio of of $y^{(1)}$ and $\mu^{(1)}$ amplitudes regardless of the $\ell$. An extension of this is that that the cross spectrum $\mu\times y$ shows a similar shape but with a negative sign, since the boost of $\Yspec(x)$ matches opposite sign mixes of the $y^{(1)}$ and $\mu^{(1)}$.

The first and third peak in $\Theta\times\Theta$ spectrum appears to have no corresponding peak in the SD case. The effect is actually slightly exaggerated -- if the early injection times were amplified for $\Theta\times y$ (blue line in the middle row) then the peaks would in fact be present with the expected ratios, but not for late time injection. In Figs.~\ref{fig:SD_outer_power_spectra_switches_zh5e5}, \ref{fig:SD_outer_power_spectra_switches_zh5e4} and \ref{fig:SD_outer_power_spectra_switches_zh5e3} in Appendix~\ref{app:auto_spectra_physics} we show the effects of physical switches on the distortion spectra in the three characteristic eras. Those figures suggest that this loss of peaks for late time injection occurs due to the missing potential driving terms from energy injection close to horizon-crossing.

Interestingly while there is a strong correlation of $\Theta\times y$ for early injection times, there is a very low correlation of $y\times y$ distortions and a complex pattern of $\mu\times\mu$. In particular at the lowest $\ell$ we see greatly enhanced $\mu\times\mu$ since the super-horizon sources favour production of $\mu^{(1)}$ distortions. The first feature in the $y\times y$ is associated with boosting of those modes at horizon crossing, before which there are no strong sources of anisotropic $y^{(1)}$ [see overall scales in Fig.~\ref{fig:switches_zh_5e+5}].

\begin{figure}
\centering
\includegraphics[width=1.0\columnwidth]{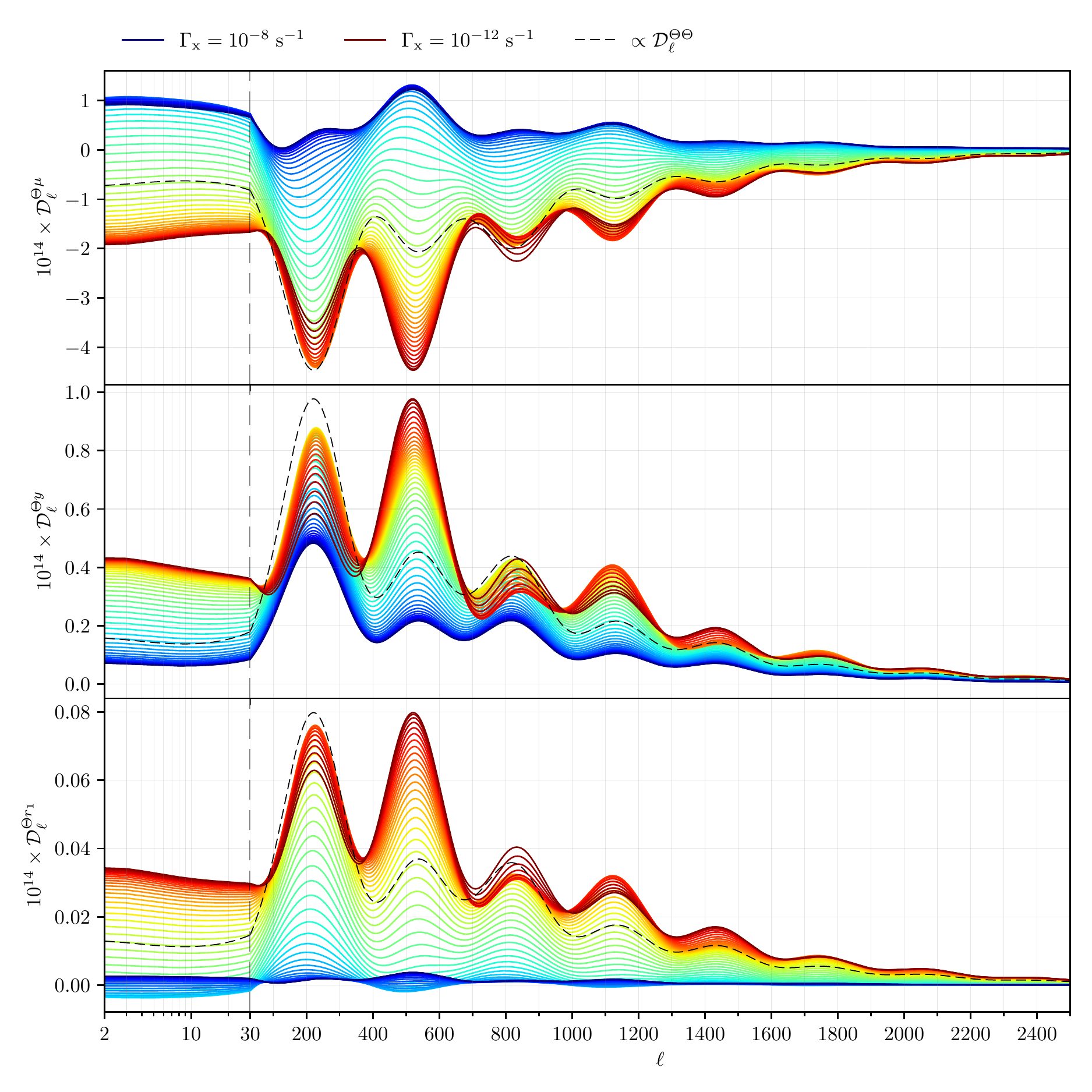}
\caption{A figure showing the power spectra for $\Theta\times \mu$ (top panel) and $\Theta\times y$ (bottom panel) over a range of decaying particle lifetimes. Blue lines show short lifetimes, thus decaying predominantly in the $\mu$-era, while red lines show long lifetimes, therefore decaying predominantly in the $y$-era. A vertical dashed line shows a division between log spaced $\ell$ values (left) and linearly spaced values (right).}
\label{fig:decay_spectra}
\end{figure}
\begin{figure}
\centering
\includegraphics[width=1.0\columnwidth]{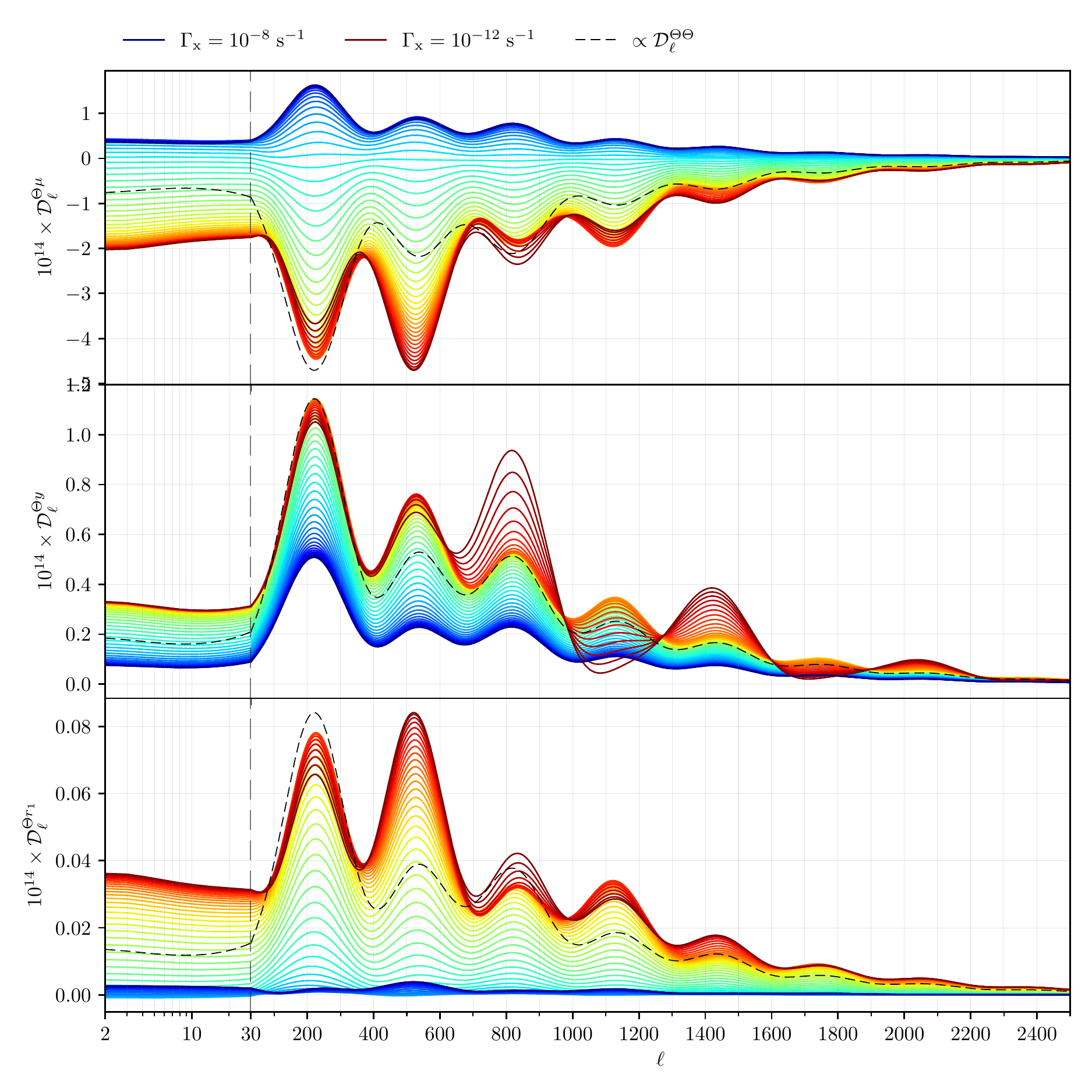}
\caption{Same as Fig.~\ref{fig:decay_spectra}, but now including an anisotropic heating term from the perturbed decay of particles modulated by local matter densities.}
\label{fig:decay_spectra_pert}
\end{figure}
\begin{figure}
\centering
\includegraphics[width=0.98\columnwidth]{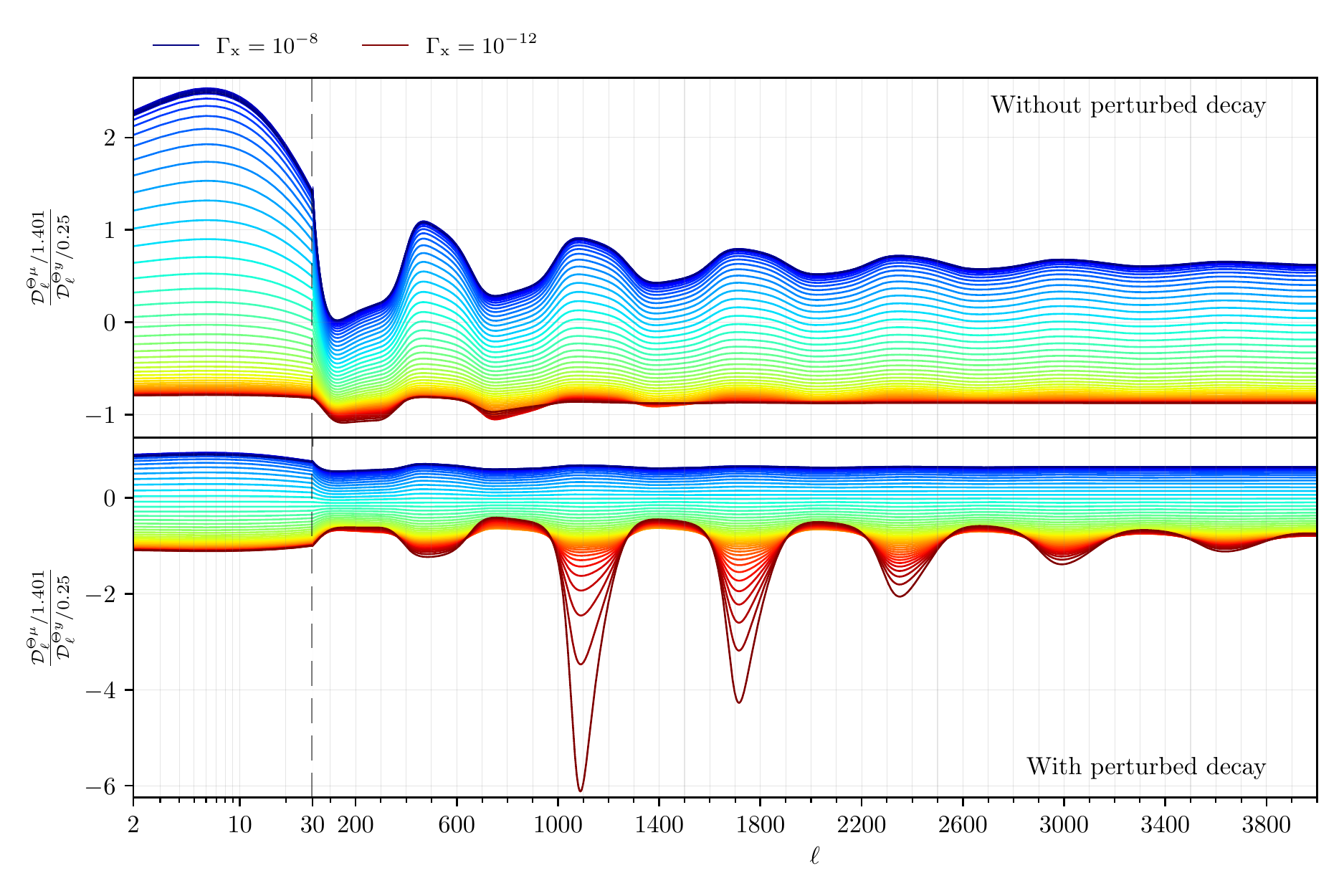}
\\
\caption{As for Fig.~\ref{fig:power_spectra_dist_ratio}, but now for the case of particle decay. Importantly the bottom panel shows the effects of perturbed decay, which affects the peaks $\ell>1000$.}
\label{fig:power_spectra_dist_ratio_decay}
\end{figure}
\subsection{Decaying particle CMB power spectra}
\label{sec:power_spectrum_decay}
In Fig.~\ref{fig:decay_spectra} and Fig.~\ref{fig:decay_spectra_pert}, we show the CMB power spectrum for various particle decay lifetimes, where in the latter figure we include effects of perturbed decay (see Sect.~\ref{sec:particle_decay}). Without perturbed decay the curves resemble those seen in Sect.~\ref{sec:power_spectrum_single_injection}, showing that the single injection scenario, while unphysical, serves as a good illustration of realistic continuous energy injection scenarios if the window of energy creation is sufficiently narrow. The perturbed decay on the other hands changes both the early and late injection scenarios. Due to the adiabatic initial conditions, injection in the $\mu$-era sees a partial cancellation between the new $\propto \delta_{\rm cdm}$ term and the $\propto \Psi$ term within the usual anisotropic heating. This allows boosting to take a more central role in the formation of the anisotropic spectrum, and thus a more dominant first peak in the power spectrum (see Fig.~\ref{fig:zh_power_spectra_switches}). However, the biggest notable feature is the enhancement (reduction) of the odd (even) peaks in the late injection time $\Theta\times y$ spectrum.

The effect of perturbed decay on the spectra is well illustrated by again taking a ratio of the relative $y$ and $\mu$ energy densities as seen through their cross correlation with temperature fluctuations. This is shown in Fig.~\ref{fig:power_spectra_dist_ratio_decay}, where the bottom panel indicates a large enhancement towards the $y$ energy density beyond $\ell=1000$. This is understood since the perturbed decay injects energy directly into $y$, which has no time to boost into a mixed spectrum for the later injection scenarios. This model serves as a motivating example and an enticing hint that a powerful future probe of concrete energy injection mechanisms could be to detect specific enhanced peaks in CMB power spectra.

\vspace{-2mm}
\section{Fisher forecasts}
\label{sec:forecasts}
\vspace{-2mm}
To assess the detectability of the signal and have a mean to compare the prospective constraints on energy injection to the \COBEF \cite{Fixsen1996, Fixsen2009} limits we use a Fisher matrix forecast, which allows us to quickly set a lower bound on parameter errors for a given instrumental configuration.
Here we consider a simplified scenario where the only free parameter is the fractional injected energy $\Delta \rho/\rho$, while all other cosmological parameters and remaining energy-release-model parameters (e.g. redshift of injection or decaying particle lifetime) are fixed.

As observables we consider using all the cross correlations between spectral distortions $\mu$ and $y$ and CMB primary anisotropies $T$ and $E$, neglecting the residual distortion contributions. In this case, the estimate of the $\Delta \rho/\rho$ error reads
\begin{equation}
\label{eq:fisher_error}
    \sigma_{\Delta \rho/\rho}
    =
    \left[
        \sum_\ell \left(\partial_{\Delta \rho/\rho} \hat C_\ell \right)^T \, \Sigma_\ell ^{-1}\,   \partial_{\Delta \rho/\rho} \hat C_\ell
    \right]^{-1/2}.
\end{equation}
Here $\hat C_\ell = \left(\hat C_\ell^{\mu T}, \hat C_\ell^{\mu E}, \hat C_\ell^{y T}, \hat C_\ell^{y E}\right)^T$ is a vector of the observable spectra. To build our intuition we will also show partial results that involve only a subset of spectra; those cases are produced by simply removing the irrelevant entries from $\hat C_\ell$ and from their covariance matrix $\Sigma_\ell$.\footnote{We point out that here we implicitly disregarded $\ell$ couplings even thought they would be non-negligible in an actual survey due to masking and foregrounds. This will be discussed in detail with the analysis of the \Planck maps in future work.}

In principle, additional information on the fractional injected energy could be extracted from the spectral distortion auto and cross-correlations. However, in a real world scenario they are too faint compared to noise and foregrounds to be measured successfully.

To compute the errors, we use the power spectra from the previous sections. Those were all computed using $\Delta \rho/\rho=10^{-5}$, but since the cross-power spectra  considered here simply scale linearly with $\Delta \rho/\rho$, the derivatives in Eq.~\eqref{eq:fisher_error} are trivially obtained. We specify that all limits shown here are calculated assuming a non-detection of the spectra in question.
The elements of the covariance matrix have the formre
\begin{equation}
\label{eq:spectra_covariance}
    \Sigma \left( \hat C_\ell^{\alpha a}, \hat C_\ell^{\beta b} \right)
    =
    \frac{1}{f_\text{sky} (2\ell+1)}
    \left( \hat C_\ell^{\alpha \beta} \hat C_\ell^{a b} + \hat C_\ell^{\alpha b} \hat C_\ell^{\beta a} \right).
\end{equation}
We model each component as $\hat C_\ell^{\alpha \beta} = C_\ell^{\alpha \beta} + N_\ell^{\alpha \beta} $, where the first terms are the theoretical spectrum previously calculated and the $N_\ell$ are the Constrained Internal Linear Combination (CILC) \cite{Remazeilles:2010hq} noise that we will now discuss.

To simulate the impact of foregrounds and instrumental noise on the cross correlations recovered from actual maps, we employ the method outlined in \citep{Cooray:2000xh, Hill:2013baa} according to the implementation of \cite{Ravenni:2020rzd}, to which we refer for the details. Working at power spectrum level, we write, for any $\ell$, the inter-frequency-channel covariance as sum over instrumental noise, foregrounds and cosmological signals $\mathcal{C}_{\ell, \nu \nu'} = N_\ell^\nu \delta_{\nu \nu'} + \sum_{i \in \text{foregrounds}}C_{\ell, \nu \nu'}^i + \sum_{i \in \text{signals}}C_{\ell, \nu \nu'}^i$. The Kronecker-$\delta$ encodes the fact that we take the instrumental noise to be uncorrelated across different channels; the foregrounds encompass dust, synchrotron, free-free, radio and infrared sources \citep{Abitbol:2017vwa, Hill:2013baa, Tegmark:1999ke, Dunkley:2013vu}; the signal are again the ones described previously and we model them as perfectly correlated at all frequencies.
To relate the noise and foreground SEDs to the adimensional theoretical spectra we convert them in thermodynamic units with the standard relation $\Theta = c^2\mathcal{G}(x)^{-1}\nu^{-3}\!/(2h) \, \Delta I_\nu$, using $T_\text{CMB}=2.7255$~K in the conversion. For the SD contributions, this transformation does not remove the frequency-dependence, which is accounted for in the component separation process \citep[e.g.,][]{Remazeilles2022}. As it is now well known \cite{Remazeilles:2018kqd}, deprojecting different spectral shapes is essential to obtain unbiased spectral measurements. Following the rationale of deprojecting stronger signals from the fainter maps, we consider the noise contribution to the temperature power spectrum as obtained with the standard ILC
\begin{equation}
    N_\ell^{TT}
    =
    \left[
        \mathcal{G}(\nu) \,
        \mathcal{C}^{-1}_{\ell, \nu \nu'}
        \mathcal{G}(\nu') 
    \right]^{-1},
\end{equation}
the $y$ (tSZ) spectrum as obtained with CILC deprojecting $T$, and $\mu$ deprojecting both $T$ and $y$, i.e.
\begin{equation}
    N_\ell^{yy}
    =
    \left[
        \left(\mathcal{Y}_0, \mathcal{G}\right)\! (\nu) \,
        \mathcal{C}^{-1}_{\ell, \nu \nu'}
        \left(\mathcal{Y}_0, \mathcal{G}\right)^T\! (\nu')
    \right]^{-1}_{0,0},
\quad
    N_\ell^{\mu\mu}
    =
    \left[
        \left(\mathcal{M}, \mathcal{Y}_0, \mathcal{G}\right)\!(\nu) \,
        \mathcal{C}^{-1}_{\ell, \nu \nu'}
        \left(\mathcal{M}, \mathcal{Y}_0, \mathcal{G}\right)^T\! (\nu')
    \right]^{-1}_{0,0}.
\end{equation}
While the cross correlations like $N_\ell^{\mu y}$ might be important if we were considering the related spectrum as an observable to be analyzed, for which they could constitute a bias \citep{Rotti2022}, they are subdominant in the covariance and thus neglected.\footnote{In fact in \cite{Rotti2022} it was found that using de-projected maps $N_\ell^{\mu T}$ is negligible, even as a bias.} Likewise, we consider the temperature and polarisation power spectra to be de facto cosmic variance limited.
We apply the methodology just described to \Planck \cite{Planck:2018nkj}, which represent the current state of the art, \Litebird \cite{LiteBIRD:2020khw} as a near-future advancement, and \PICO \cite{NASAPICO:2019thw} as more futuristic scenario. In all cases we conservatively assume $f_\text{sky}=0.65$, and set the maximum $\ell$ in the sum in Eq.~\eqref{eq:fisher_error} high enough to saturate the constraints.

\begin{figure}
\centering
\includegraphics[width=0.95\columnwidth]{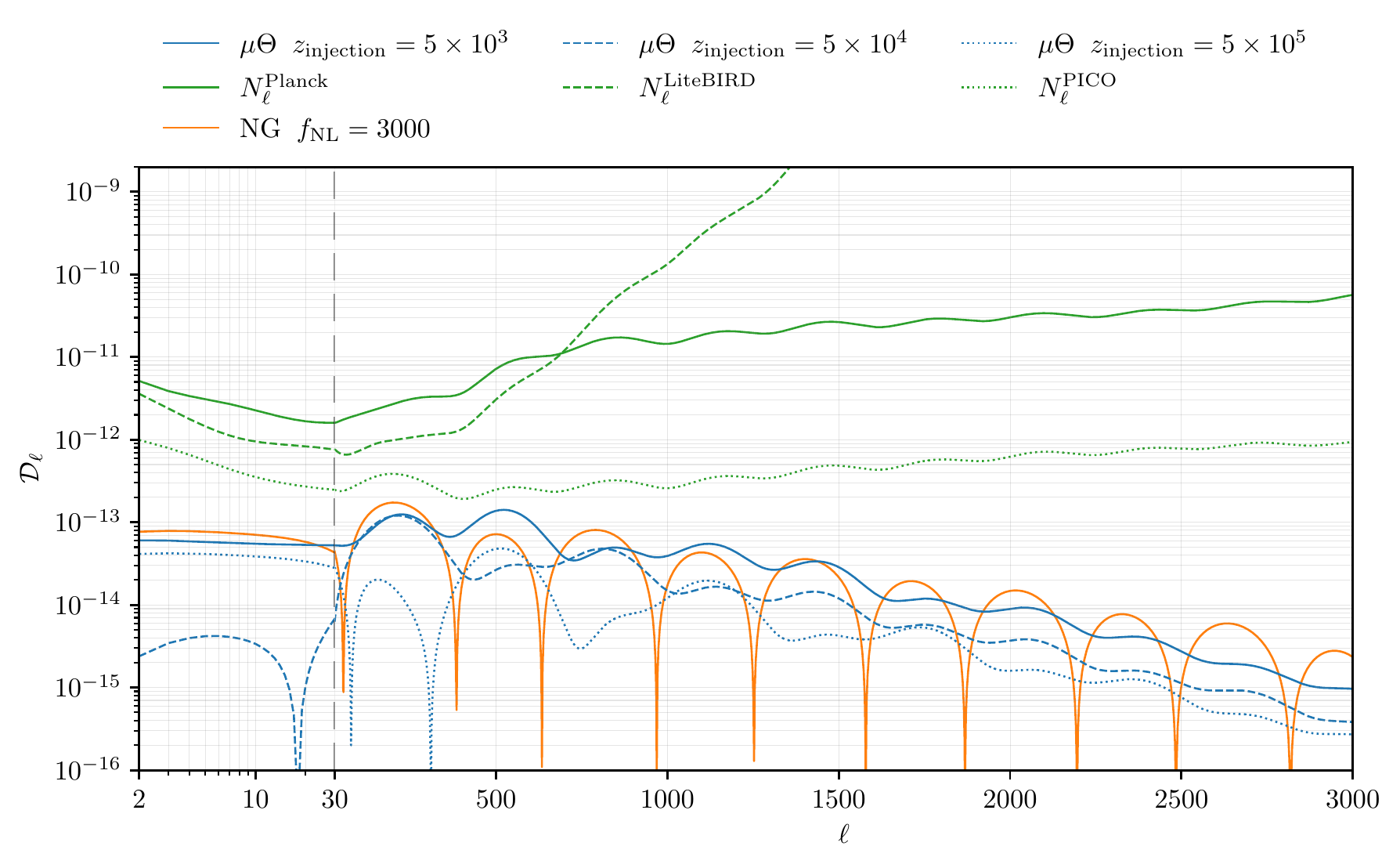}
\caption{A figure showing the expected $\Theta\times\mu$ power spectra (blue) for various injection times with energy release $\Delta \rho/\rho = 3\times10^{-5}$. Also shown for comparison are the noise curves (green) for various instruments and a predicted signal from primordial non-gaussianity (orange).}
\label{fig:forecasts_muT_noise}
\end{figure}

In Fig.~\ref{fig:forecasts_muT_noise} we show (blue) the $\Theta \times \mu$ cross-correlation for three different injection times and a total energy release of $\Delta \rho/\rho = 3\times10^{-5}$, compatible with the $1\sigma$ \COBEF limit. That has to be compared with (green) the square root of the covariance element as defined in Eq.~\eqref{eq:spectra_covariance}. For reference we compare the signal to (orange) the ``standard'' calculation for $\Theta \times \mu$ from primordial non-Gaussianity \cite{Ganc2012, Ravenni2017} with $f_\text{NL}^\text{loc}=3000$, close to the $1\sigma$ \Planck limit \citep{Rotti2022}. We can appreciate that with these specific values of $f_\text{NL}^\text{loc}$ and $\Delta \rho/\rho$ they are comparable in amplitude.

\begin{figure}
\centering
\includegraphics[width=0.95\columnwidth]{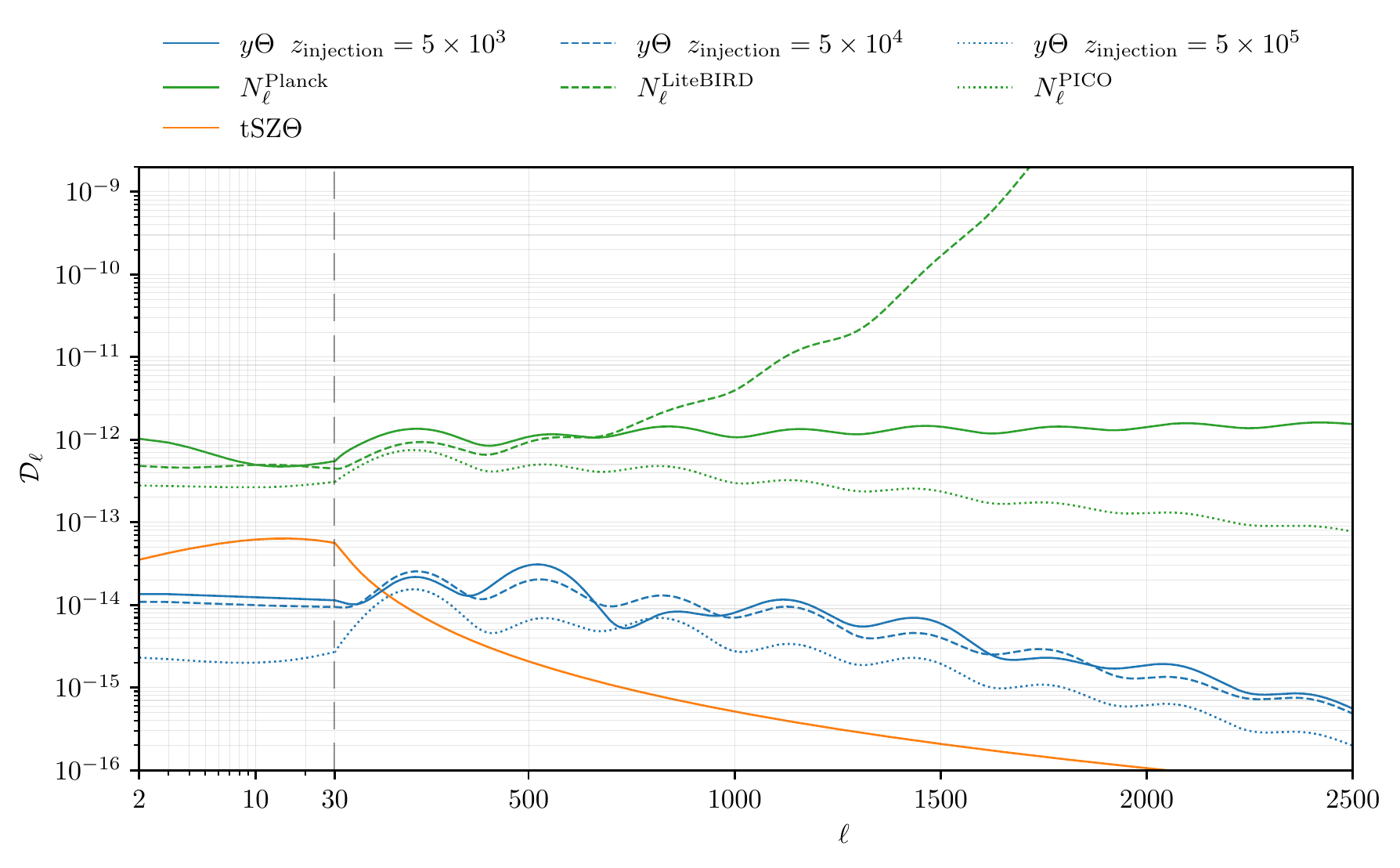}
\caption{A figure showing the expected $\Theta\times y$ power spectra (blue) for various injection times with energy release $\Delta \rho/\rho = 3\times10^{-5}$. Also shown for comparison are the noise curves (green) for various instruments and a predicted signal from the thermal Sunyaev-Zeldovich effect (orange).}
\label{fig:forecasts_yT_noise}
\end{figure}

The same exercise is repeated in Fig.~\ref{fig:forecasts_yT_noise} for the $\Theta \times y$ cross correlation. The only difference is that here we show for reference (orange) the Sunyaev-Zeldovich (SZ) cross correlation with ISW, tSZ$\times \Theta$.
This signal would in principle constitute a bias to the $\Theta \times y$ cross correlation from energy injection. Here we disregard this problem; however, we point out that the vastly different $\ell$ dependence would possibly allow for a successful signal disentanglement.
Conversely, existing primordial distortion anisotropies would provide a noise contribution to SZ searches for the ISW effect \citep{Taburet2011,Creque2016}.
We also specify that this contribution is included in the covariance calculation, but as one can expect, it has a negligible effect on the results.

\begin{figure}
\centering
\includegraphics[width=1.0\columnwidth]{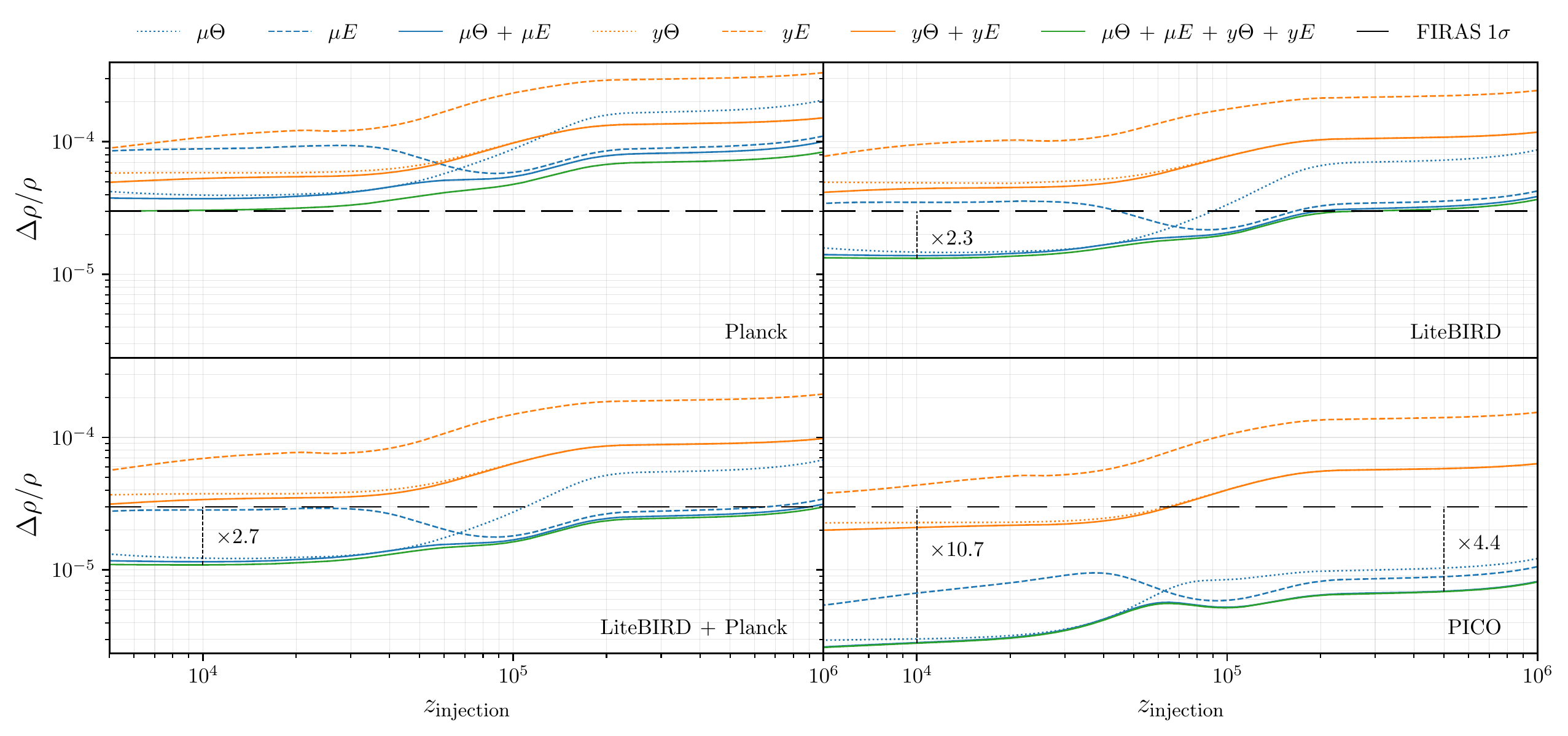}
\caption{Four figures showing the forecasts for time dependent constraints on single energy injection using different cross correlations. Also shown is the \COBEF limit on energy release at a time-independent $\Delta \rho/\rho=\pot{3}{-5}$. How much a given instrument could improve on the \COBEF measurement is shown in multiplicative factors in the plot.}
\label{fig:forecasts_zh}
\end{figure}

In Fig.~\ref{fig:forecasts_zh} we show the constraints on single injection scenario as a function of the injection redshift.
In all the panels we can appreciate a subtle distinctions in three regimes which coincide with the standard $y$, residual and $\mu$-eras.
Generally speaking $\mu$-era injection is less constrainable than the residual and $y$-era injections. In Fig.~\ref{fig:switches_zh_5e+5} we see that the perturbed thermalisation and anisotropic heating sources actually oppose the boosting source for early injection times. The boosting source however flips the sign of its $\mu^{(1)}$ source as the background spectrum contains more contributions of $y^{(0)}$. In Fig.~\ref{fig:switches_zh_5e+4} this leads to an additive effect of boosting for late times. This likely explains both the lack of constraining power at early times as well as the small step around $z\approx \pot{7}{4}$ within each panel of Fig.~\ref{fig:forecasts_zh}.
The other small step occurs around $z\approx \pot{2}{5}$, hinting towards the thermalisation terms becoming inefficient. Similarly a small decrease of constraining power is seen at $z\approx 10^6$ since part of the distortion thermalised to a simple temperature shift.

Combining constraints from $\mu$ and $y$ distortions would allow us to set tight limits on the energy injection throughout the whole post-$T$-era universe history. In particular next generation and futuristic satellites, thanks to their ability to remove foregrounds due to ample frequency coverage, will set constraints exceeding \COBEF'.
Further to $\mu$ and $y$ we could feasibly use the residual $r$ distortions to improve the results further. These however are at least an order of magnitude smaller as seen in Sect.~\ref{sec:power_spectrum}, but could carry details of time dependence. We will carry on a more detailed discussion in the next section.

\begin{figure}
\centering
\includegraphics[width=1.0\columnwidth]{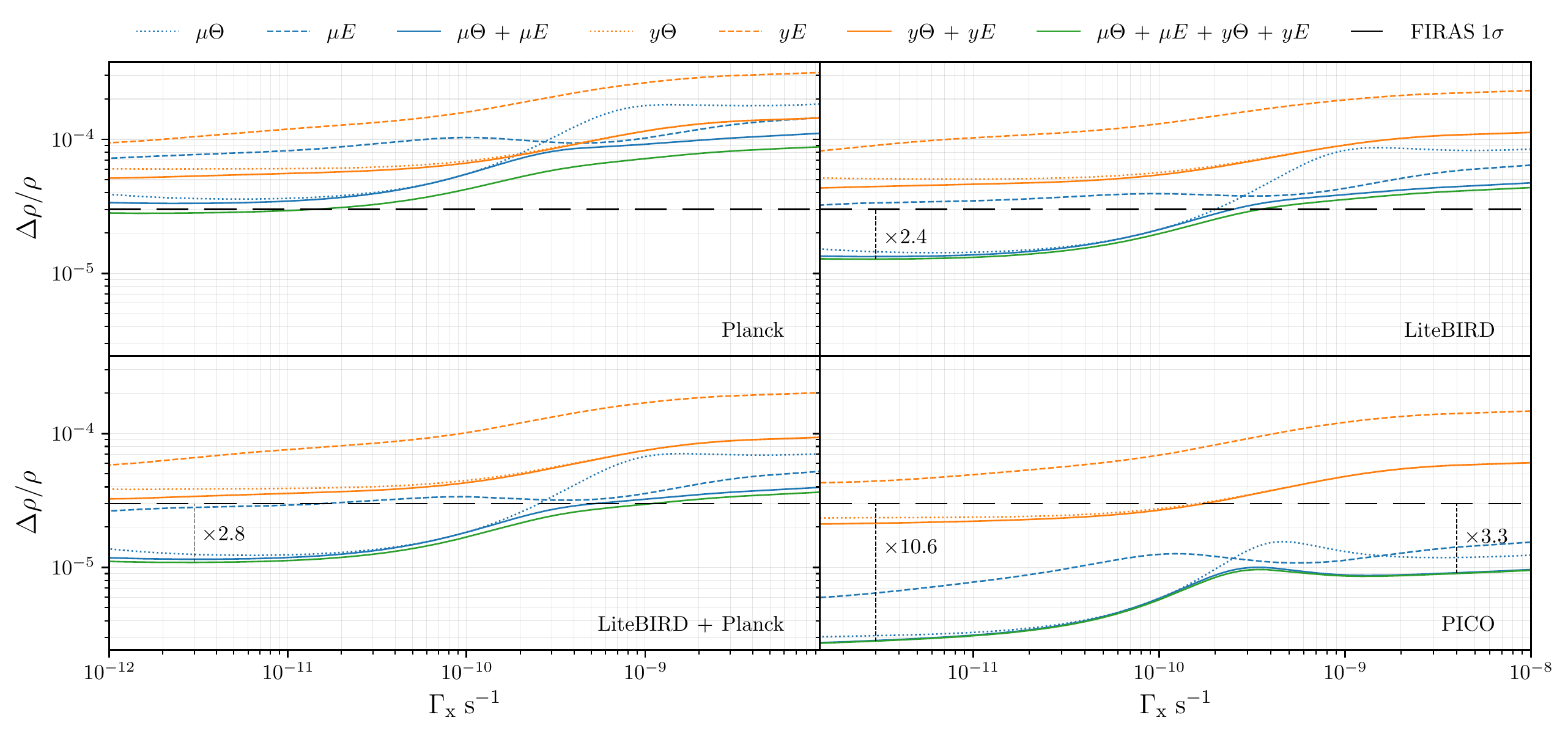}
\caption{Four figures showing the forecasts for decaying particle constraints using different cross correlations. Crucially these constraints ignore the effects of perturbed decay. Also shown is the FIRAS limit on energy release at a time-independent $\Delta \rho/\rho=\pot{3}{-5}$. How much a given instrument could improve on the \COBEF measurement is shown in multiplicative factors in the plot.}
\label{fig:forecasts_gamma_pert0}
\end{figure}
\begin{figure}
\centering
\includegraphics[width=1.0\columnwidth]{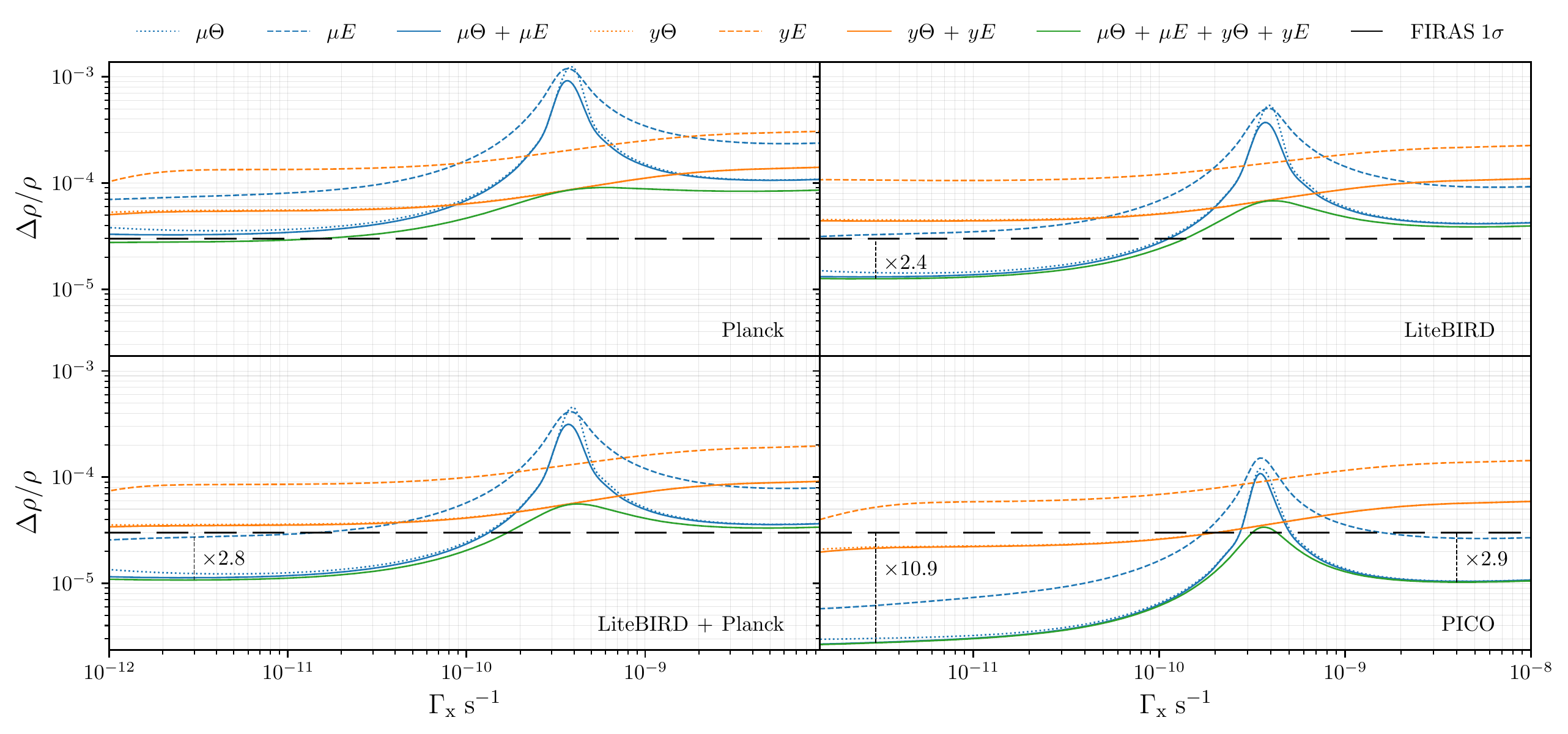}
\caption{As for Fig.~\ref{fig:forecasts_gamma_pert0}, but now including the effects of perturbed decay. The constraints are reduced around the residual distortion era.}
\label{fig:forecasts_gamma_pert1}
\end{figure}

In Fig.~\ref{fig:forecasts_gamma_pert0} and Fig.~\ref{fig:forecasts_gamma_pert1} we show the equivalent constraints for decaying particle scenarios of different lifetimes with and without perturbed decay. As seen above from directly inspecting the power spectra, these models hold many similarities with the single-injection scenarios.
The big differences emerge when including the effects of perturbed decay. Interestingly these serve to decrease constraining power, which may initially be counter intuitive. One problem for constraining this model is that the enhancements from dark matter modulations typically occur for $\ell>1000$ (see Fig.~\ref{fig:power_spectra_dist_ratio_decay}) while the maximum constraining power usually comes from  $100\leq\ell\leq 1000$. Furthermore we previously commented that the combination of $\Psi^{(1)}+\delta_{\rm cdm}^{(1)}$ in adiabatic initial conditions imply some mutual cancellation, thus reducing the overall effect of anisotropic heating.
With anisotropic heating effectively halved (and flipped sign) the early time constraints decrease, and a noticeable peak emerges. This peak is due to the dependence on boosting sources, which between early and late times cross a specific mix of $y^{(0)}$ and $\mu^{(0)}$ which boosts to give no $\mu^{(1)}$ contribution.

This example is illustrative of the high degree of model dependence in constraints when it comes to energy injection modulated directed by local perturbed quantities. On the other hand, the case without perturbed decay illustrates that single injection models do a good job of representing continuous injection mechanisms assuming they have narrow windows.
Overall, our forecasts demonstrate the immense potential for SD anisotropy studies with CMB imagers.

\subsection{Accessing information from the residual distortions }
\label{sec:forecasts_residual_discussion}
As we have seen in Sect.~\ref{sec:power_spectrum}, additional information could be gleaned from the residual distortion signals. In the estimates given above, we neglected this component for several reasons.
Firstly, one does not expect this to improve the detection limits by more than $\simeq 10-20\%$, given that the overall energy contained in the residual spectra is at that level relative to the $y$-distortion.
However, in the residual distortion era, the effect could be slightly more noticeable.
Secondly, the residual distortion modes used in our presentation where constructed for a \PIXIE-like configuration \citep{Kogut2016SPIE, Kogut2019BAAS}. As such, one cannot directly translate these to the experimental configurations assumed in the forecasts, and an experiment-specific analysis would be required. 
Thirdly, for $\mu$ and $y$-type distortion signals, we already have some level of understanding about how foregrounds might affect the constraints \citep{abitbol_pixie, Remazeilles2022, Rotti2022}, but for the residual distortion modes, a more comprehensive analysis is outstanding.
Solving these issues is beyond the scope of this paper, such that we do not attempt further improvements of our forecasts. We note, however, that in particular in the residual distortion era, one expects additional gains. In addition, should one detect a SD signal with future CMB imagers, we expect the residual distortion information to provide more leverage towards distinguishing various distortion signals, as also understood from the average distortion science \citep{Chluba2013fore, Chluba2013PCA}. 

\subsection{Further optimizing and reducing the observation basis}
\label{sec:forecasts_basis_discussion}
We close by remarking on further optimizing the observation basis. The construction so far was motivated by conserving the meaning of the standard CMB signals, $G$, $Y$ and $M$. However, we now understand that for the SD anisotropies the $\boostO \Yspec$, $\boostO \Mspec$, $\DiffO \Mspec$ and $\Yspec-Y_1$ SEDs also play central roles.

A pure $y$-anisotropy can be created by late {\rm perturbed} energy release ($z\lesssim 10^4$), as in the example of decaying particles. Another example would be perturbed dark matter annihilation, SZ clusters or the dissipation of acoustic modes from primordial perturbations, which all intrinsically source $Y$. Pure $\mu$-distortion anisotropies can be sourced by {\rm perturbed} energy release at $z\gtrsim 10^5$ from annihilation/decay or acoustic mode dissipation.

Without imposing a theory prior, one should therefore consider adding $\boostO \Yspec=4Y_1$, $\boostO \Mspec$ and $\DiffO \Mspec$  to the standard distortion shapes relevant to CMB anisotropy analysis. This means, in total six CMB SED amplitudes would have to be determined. Given that $\boostO M$ is extremely well represented by a simple superposition of $Y$ and $M$, large degeneracies would be found. Similarly, $\boostO Y$ does have significant projections onto $G$, $Y$ and $M$, such that without further study it is unclear how much this extended standard SED basis would help. 

However, if one can with certainty assume that the energy release occurred at $z\lesssim 10^4$, our computations have demonstrated (see Fig.~\ref{fig:aniso_spec_convergence}) that one could indeed simply use $G$, $Y$ and $\boostO Y=4Y_1$ for the full analysis. This would ensure that no information is lost, even if the SEDs are not fully independent. In reality, in this regime, one could even get away with simply using a $Y_1$-distortion hierarchy without the need of further SED rotation.
We will investigate the utility of these alternative analysis methods in the future, highlighting that the current observation basis provides a more agnostic approach to the challenge of extracting all the information from the CMB sky.

\subsection{Extracting the time of injection}
\label{sec:MCMC_prospects}
\begin{figure}
\centering
\includegraphics[width=1.0\columnwidth]{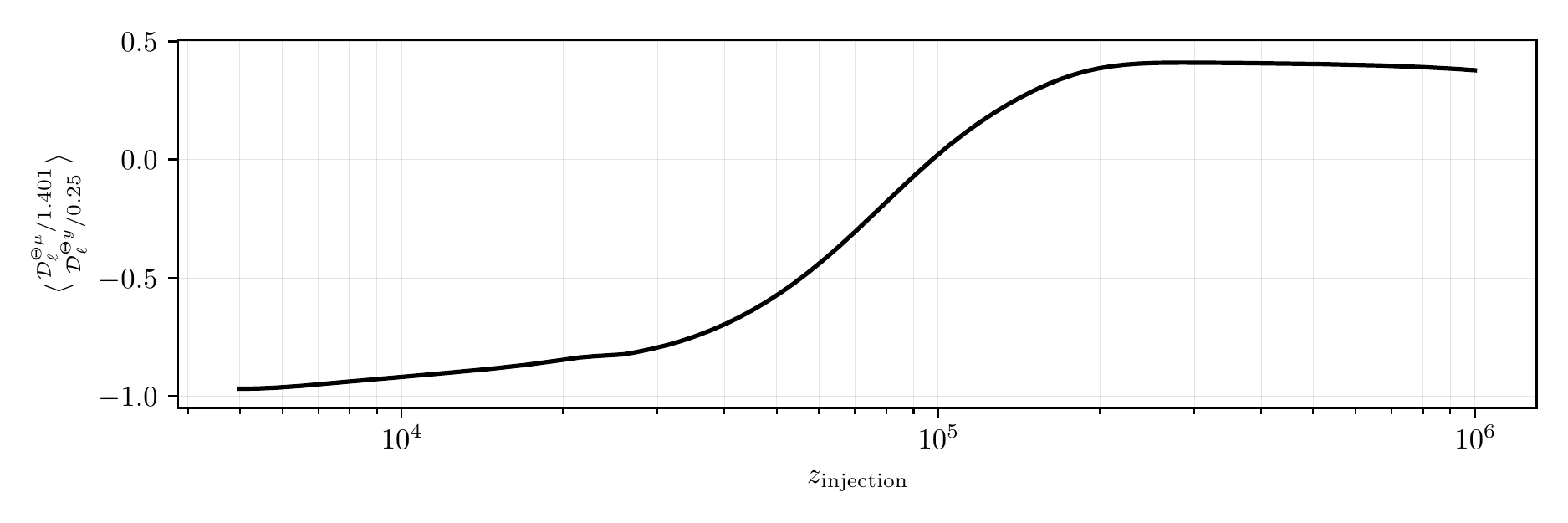}
\caption{A figure showing the average ratio between $\mathcal{D}_\ell^{\Theta\mu}$ and $\mathcal{D}_\ell^{\Theta y}$ for different single injection redshifts with appropriately normalised energy units. The average is taken in the range $100\leq\ell \leq 1000$ where most of the constraining power is.}
\label{fig:av_ratio_zh}
\end{figure}
In much of Sect.~\ref{sec:power_spectrum} we alluded to strong time dependence encoded in the power spectrum. Consider for example Fig.~\ref{fig:power_spectra_dist_ratio} where by taking ratios of the measured CMB power spectra there is a relatively smooth gradient between early and late injection times. To understand what time sensitivity future imagers might have on energy injection times we take an average of this ratio in the range $100\leq\ell \leq 1000$, where we have seen most of the constraining power to reside. This process is shown in Fig.~\ref{fig:av_ratio_zh}, where we can see the previously discussed plateau for early and late injection time, as well as a strong gradient for residual-era injection. Changing the energy release model subtly modifies the shape of the curve, but the general trends within each era hold regardless of exact energy injection mechanism.

The figure simply shows that constrainable metrics exist for time dependence, however, a much more robust approach would be to perform full MCMC searches for a given models parameters. This is motivated by the fact that individual peaks of the spectra could break the plateaus deep into a given era (e.g. the first and second peak in Fig.~\ref{fig:zh_power_spectra} would reveal how late into the $y$-era energy injection occurred). To perform these kind of searches it would be necessary to optimise the numerical treatment. With future work this should be possible, considering that the largest burden in the current code is pushing to $N_{\rm max}=15$, which as seen in Fig.~\ref{fig:power_spectrum_convergence_1} and Fig.~\ref{fig:power_spectrum_convergence_2} yields small percent changes which could be neglected for broad parameter searches.

\section{Discussion and Conclusions}
This paper has taken a step-by-step approach to presenting the spectro-spatial evolution of the photon spectrum. Starting with the photon spectrum, we discussed dominant contributions to the anisotropic spectrum and the different limits they see their greatest importance. We followed with a careful treatment of basis choices for representing these spectra as transfer functions of SED amplitudes. This opened a long discussion on the relevant physics for different injection eras, which culminated in the presentation of distortion power spectra, which show a complex superposition of many physical effects previously analysed. Finally, we used the predicted power spectra to forecast constraints on primordial energy injection which are comparable with or exceed modern limits from \COBEF, thus constituting independent tests of primordial physics.

This detailed step-by-step approach should not distract from the novelty of this formalism and the wealth of opportunities in it: it is now possible to infer valuable information about the background photon spectrum in a potentially more time sensitive way by viewing the spectrum in anisotropic patches of the CMB sky. This allows us to push our understanding of primordial physics beyond the opaque \textit{CMB curtain} and gain insight to $z\simeq \pot{2}{6}$ using well-known CMB imaging techniques.

In essence the time sensitivity arises from the fact that we can distinguish anisotropic heating [creating $\Mspec(x)$, $\Yspec(x)$ or $Y_1 + \Yspec(x)$], perturbed thermalisation [creating $\DiffO\Mspec(x)$ and $Y_1(x)-\Yspec(x)$] and boosting [producing $\boostO\Mspec(x)$ or $\boostO\Yspec(x)$ depending on the background spectrum]. Fully exploiting this time dependence will require more sophisticated analysis tools and a further optimised numerics (see Sect.~\ref{sec:MCMC_prospects}).
The variety of spectral modes has strong potential for discriminating and isolating the physical origin of any energy injection in the primordial plasma, even if it is clear that additional optimisation of the analysis may be needed (see Sect.~\ref{sec:forecasts_basis_discussion}).

In our analysis we clearly isolated the main effects: Early energy injection see source terms arise from Compton scattering, perturbed scattering, perturbed emission and direct anisotropic heating. At late times the main sources are Doppler and potential driving, which are well-known in connection with the acoustic peaks seen in the regular temperature power spectra. Each of these were individually illustrated in detail, showing that new insight can be gained deep into the pre-recombination era (e.g., see Figs.~\ref{fig:switches_zh_5e+4}, \ref{fig:perturbed_decay_cases_pert1} and \ref{fig:zh_power_spectra_switches}) by accessing the full spectro-spatial information.

The probably most significant conclusion from this work is that through measurements of SD anisotropies CMB imagers like \Planck \citep{Planck:2018nkj}, \Litebird \citep{LiteBIRD:2020khw} and \PICO \citep{NASAPICO:2019thw} can provide limits on the {\it average energy release} at various epochs. These limits are expected to be comparable with those from \COBEF in the case of \Planck, while \Litebird and \PICO could supersede \COBEF several times (see Sect.~\ref{sec:forecasts}). For \PICO, we see possible gains by more than one order of magnitude for energy injection at $z\lesssim \pot{5}{4}$ (see Fig.~\ref{fig:forecasts_zh}). Evidently, a \PIXIE-like CMB spectrometer, directly targeting the average distortion, could improve these limits many-fold \citep[e.g.,][]{Chluba2019BAAS, Chluba2021ExA}; however, the experimental methods and analysis techniques are quite different, and the complementarity of the constraints highlights the unique synergy between the approaches. 

Although our estimates are based on simple Fisher forecasts, similar methods have already been shown to be reliable \citep{Rotti2022}, building confidence in our results. Nevertheless, we plan to carry out a more rigorous analysis using detailed foreground simulations and including other experimental effects to further refine these results in future work. We do not expect the main conclusion to change: CMB imaging and distortion science have now been united to allow full spectro-spatial considerations of CMB physics.
Extended forecasts that consider the benefits of ground-based observations with The Simons Observatory \citep{SOWP2018}, CMB-S4 \citep{Abazajian2016S4SB} and other space-mission concepts \citep{PRISM2013WP,PRISM2013WPII,Delabrouille2018, Delabrouille2021} are also planned.

We close with a few words about the path forward. 
First and foremost, it would be important to improve the analytic understanding of the solutions. In addition, our solutions clearly show that in the tight-coupling regime the distortion dipole transfer functions all follow $y_{n, 1}\propto \mu_1\propto \Theta_1$ once the average distortion has been created (see Sect.~\ref{sec:transfer_functions}). With this tracking solution and approximation for the potential $\Phi$, the distortion monopoles can be modelled using WKB approximations. Some complications arise from the SD evolution by Compton scattering terms and the transformation to the optimal observation basis. However, we leave a more detailed study to future work.

A simple and immediate extension of this work is to investigate the effects of different initial conditions beyond adiabatic modes as well as the effects of different heating mechanisms. In this paper the case of perturbed decay as a trace of dark matter illustrated the importance of both these avenues of exploration. Firstly the adiabaticity of the initial perturbations are capable of cancelling or enhancing important thermalisation terms, providing an additional test to the usual CMB temperature power spectra. A comprehensive study in this direction would complement the work on initial conditions and SD physics for the average photon spectrum \citep{Chluba2013iso}. Secondly, modulating local heating by perturbed quantities with known effects on the temperature power spectra yield powerful and predictable correlations. If observable, these concrete enhancements to the CMB power spectra could provide \textit{smoking guns} of concrete heating mechanisms and therefore new Physics.

Furthermore, here we only illustrated the SD anisotropy physics for single energy injection and decaying particle scenarios, focusing on the distortion signals. However, other mechanisms can be considered. For instance, dark matter annihilation should similarly lead to anisotropic distortions. For $s$-wave (i.e., temperature-independent) annihilation, these signals ought to be small from the pre-recombination era, given existing constraints from \Planck \citep[e.g.,][]{Planck2015params}; however, due to perturbed decay, $\propto \delta_{\rm cdm}$, late time $y$-type anisotropies could be sourced by the non-linear growth of structure, enhancing the expected signal. Sommerfeld boosts of the annihilation rate \citep{Slatyer2010Sommer} or a varying temperature dependence of the annihilation cross section \citep[e.g., $p$-wave annihilation][]{Chluba2013fore} could further modify the signals. In addition, cosmic bubble collisions \citep{Aguirre2011, Kleban2013, Deng2020}, primordial black holes \citep{Carr2016, Abe2019, Deng2021, Ozsoy2021}, primordial magnetic fields \citep{Jedamzik1998,Miyamoto2014,Minoda2018, Saga2019} or dark matter scattering effects \citep{Dvorkin2014, Yacine2015DM, Munoz2015} could lead to anisotropic energy release, which can now be modelled more accurately using our novel approach. 
For this, the possible changes to the ionisation history \citep{Chen2004, Padmanabhan2005, Slatyer2009, Chluba2010a} should be taken into account using state of the art recombination codes like {\tt CosmoRec}, an extension that we plan for the near future.

The computations of the SD anisotropies from primordial non-Gaussianity \citep{Pajer2012, Ganc2012} could also be refined, accurately treating all the transfer effects \citep[see][for some previous analytic attempts]{Pajer2012b, Chluba2017muT}, which will be crucial for distinguishing these signals from contaminations and foregrounds.
Specifically, this latter problem deserves additional attention, as the SD anisotropies we considered here could act as a new foreground to extracting information about primordial non-Gaussianity as long as we have no significantly improved upper limit on average energy release from absolute spectrometers such as \PIXIE.
This can be appreciated from Fig.~\ref{fig:forecasts_muT_noise}, which demonstrates that at the level of $f_{\rm nl}=3000$, the possible anisotropic distortion signals due to average energy release are comparable. Without theoretical prior, it will be challenging to eliminate this uncertain contribution using CMB imaging alone. Even if the $\mu\times T$ signal from primordial non-Gaussianity differs from the signals discussed here, it will be hard to reach a cosmic-variance limited measurement suggested by theory \citep{Pajer2012, Cabass2018} unless we could limit the average energy release to $\Delta \rho/\rho\lesssim 10^{-11}$. Similar comments could impede polarisation-distortion correlation studies, which have the potential to shed new light on inflation physics \citep{Orlando2022}.
Attempts to extract information about the ISW effect from SZ cluster-induced $y\times T$ correlations \citep{Taburet2011,Creque2016} would also be affected.
We plan to investigate these aspect more carefully in future work, hoping that the new perspectives given here provide further motivation to think about an extended synergistic approach in the future of CMB exploration.

{\small
\section*{Acknowledgments}
We thank Eiichiro Komatsu, Aditya Rotti and Rashid Sunyaev for stimulating discussion.
We furthermore thank Colin Hill, Rishi Khatri, Anthony Lewis, Atsuhisa Ota, Enrico Pajer and Nils Sch\"oneberg for comments on the manuscript.
This work was supported by the ERC Consolidator Grant {\it CMBSPEC} (No.~725456).
TK was also supported by STFC grant ST/T506291/1.
JC was furthermore supported by the Royal Society as a Royal Society University Research Fellow at the University of Manchester, UK (No.~URF/R/191023).
AR acknowledges support by the project "Combining Cosmic Microwave Background and Large Scale Structure data: an Integrated Approach for Addressing Fundamental Questions in Cosmology", funded by the MIUR Progetti di Ricerca di Rilevante Interesse Nazionale (PRIN) Bando 2017 - grant 2017YJYZAH.
}

{\small
\bibliographystyle{JHEP}
\bibliography{bibliography,Lit}

\providecommand{\href}[2]{#2}\begingroup\raggedright\begin{thebibliography}{100}

\bibitem{WMAP_params}
C.~L. {Bennett}, M.~{Halpern}, G.~{Hinshaw}, et~al., {\it {First-Year Wilkinson
  Microwave Anisotropy Probe (WMAP) Observations: Preliminary Maps and Basic
  Results}},  {\em \apjs} {\bf 148} (Sept., 2003) 1--27.

\bibitem{Planck2015params}
{Planck Collaboration}, P.~A.~R. {Ade}, N.~{Aghanim}, et~al., {\it {Planck 2015
  results. XIII. Cosmological parameters}},  {\em \aap} {\bf 594} (Sept., 2016)
  A13, [\href{http://arxiv.org/abs/1502.01589}{{\tt arXiv:1502.01589}}].

\bibitem{Sunyaev1970}
R.~A. {Sunyaev} and Y.~B. {Zeldovich}, {\it {Small-Scale Fluctuations of Relic
  Radiation}},  {\em \apss} {\bf 7} (1970) 3--+.

\bibitem{Peebles1970}
P.~J.~E. {Peebles} and J.~T. {Yu}, {\it {Primeval Adiabatic Perturbation in an
  Expanding Universe}},  {\em \apj} {\bf 162} (Dec., 1970) 815--+.

\bibitem{Silk1968}
J.~{Silk}, {\it {Cosmic Black-Body Radiation and Galaxy Formation}},  {\em
  \apj} {\bf 151} (Feb., 1968) 459--+.

\bibitem{Hu1995CMBanalytic}
W.~{Hu} and N.~{Sugiyama}, {\it {Anisotropies in the cosmic microwave
  background: an analytic approach}},  {\em \apj} {\bf 444} (May, 1995)
  489--506, [\href{http://arxiv.org/abs/astro-ph/9407093}{{\tt
  astro-ph/9407093}}].

\bibitem{Ma1995}
C.-P. {Ma} and E.~{Bertschinger}, {\it {Cosmological Perturbation Theory in the
  Synchronous and Conformal Newtonian Gauges}},  {\em \apj} {\bf 455} (Dec.,
  1995) 7--+, [\href{http://arxiv.org/abs/astro-ph/9506072}{{\tt
  astro-ph/9506072}}].

\bibitem{Seljak1997}
U.~{Seljak} and M.~{Zaldarriaga}, {\it {Signature of Gravity Waves in the
  Polarization of the Microwave Background}},  {\em Physical Review Letters}
  {\bf 78} (Mar., 1997) 2054--2057,
  [\href{http://arxiv.org/abs/astro-ph/9609169}{{\tt astro-ph/9609169}}].

\bibitem{Hu1997}
W.~{Hu} and M.~{White}, {\it {CMB anisotropies: Total angular momentum
  method}},  {\em \prd} {\bf 56} (July, 1997) 596--615,
  [\href{http://arxiv.org/abs/astro-ph/9702170}{{\tt astro-ph/9702170}}].

\bibitem{Zeldovich1969}
Y.~B. {Zeldovich} and R.~A. {Sunyaev}, {\it {The Interaction of Matter and
  Radiation in a Hot-Model Universe}},  {\em \apss} {\bf 4} (July, 1969)
  301--316.

\bibitem{Sunyaev1970mu}
R.~A. {Sunyaev} and Y.~B. {Zeldovich}, {\it {The interaction of matter and
  radiation in the hot model of the Universe, II}},  {\em \apss} {\bf 7} (Apr.,
  1970) 20--30.

\bibitem{Illarionov1975}
A.~F. {Illarionov} and R.~A. {Sunyaev}, {\it {Comptonization, characteristic
  radiation spectra, and thermal balance of low-density plasma}},  {\em Soviet
  Astronomy} {\bf 18} (Feb., 1975) 413--419.

\bibitem{Illarionov1975b}
A.~F. {Illarionov} and R.~A. {Sunyaev}, {\it {Comptonization, the
  background-radiation spectrum, and the thermal history of the universe}},
  {\em Soviet Astronomy} {\bf 18} (June, 1975) 691--699.

\bibitem{Danese1982}
L.~{Danese} and G.~{de Zotti}, {\it {Double Compton process and the spectrum of
  the microwave background}},  {\em \aap} {\bf 107} (Mar., 1982) 39--42.

\bibitem{Burigana1993}
C.~{Burigana}, {\it {Distortions of the CMB Spectrum by Continuous Heating}},
  in {\em Observational Cosmology} ({G.~L.~Chincarini, A.~Iovino, T.~Maccacaro,
  \& D.~Maccagni}, ed.), vol.~51 of {\em Astronomical Society of the Pacific
  Conference Series}, pp.~554--+, Jan., 1993.

\bibitem{Hu1993}
W.~{Hu} and J.~{Silk}, {\it {Thermalization and spectral distortions of the
  cosmic background radiation}},  {\em \prd} {\bf 48} (July, 1993) 485--502.

\bibitem{Chluba:spectro_spatial_I}
J.~{Chluba}, T.~{Kite}, and A.~{Ravenni}, {\it {Spectro-spatial evolution of
  the CMB I: discretisation of the thermalisation Green's function}},  {\em
  arXiv e-prints} (Oct., 2022) arXiv:2210.09327,
  [\href{http://arxiv.org/abs/2210.09327}{{\tt arXiv:2210.09327}}].

\bibitem{Chluba:spectro_spatial_II}
J.~{Chluba}, A.~{Ravenni}, and T.~{Kite}, {\it {Spectro-spatial evolution of
  the CMB II: generalised Boltzmann hierarchy}},  {\em arXiv e-prints} (Oct.,
  2022) arXiv:2210.15308, [\href{http://arxiv.org/abs/2210.15308}{{\tt
  arXiv:2210.15308}}].

\bibitem{Chluba2013Green}
J.~{Chluba}, {\it {Green's function of the cosmological thermalization
  problem}},  {\em \mnras} {\bf 434} (Sept., 2013) 352--357,
  [\href{http://arxiv.org/abs/1304.6120}{{\tt arXiv:1304.6120}}].

\bibitem{Lucca2020}
M.~{Lucca}, N.~{Sch{\"o}neberg}, D.~C. {Hooper}, J.~{Lesgourgues}, and
  J.~{Chluba}, {\it {The synergy between CMB spectral distortions and
  anisotropies}},  {\em \jcap} {\bf 2020} (Feb., 2020) 026,
  [\href{http://arxiv.org/abs/1910.04619}{{\tt arXiv:1910.04619}}].

\bibitem{Chluba:2x2}
J.~{Chluba}, R.~{Khatri}, and R.~A. {Sunyaev}, {\it {CMB at 2 {x} 2 order: the
  dissipation of primordial acoustic waves and the observable part of the
  associated energy release}},  {\em \mnras} {\bf 425} (Sept., 2012)
  1129--1169, [\href{http://arxiv.org/abs/1202.0057}{{\tt arXiv:1202.0057}}].

\bibitem{Danese1977}
L.~{Danese} and G.~{de Zotti}, {\it {The relic radiation spectrum and the
  thermal history of the Universe}},  {\em Nuovo Cimento Rivista Serie} {\bf 7}
  (Sept., 1977) 277--362.

\bibitem{Balashev2015}
S.~A. {Balashev}, E.~E. {Kholupenko}, J.~{Chluba}, A.~V. {Ivanchik}, and D.~A.
  {Varshalovich}, {\it {Spectral Distortions of the CMB Dipole}},  {\em \apj}
  {\bf 810} (Sept., 2015) 131, [\href{http://arxiv.org/abs/1505.06028}{{\tt
  arXiv:1505.06028}}].

\bibitem{deZotti2015}
G.~{De Zotti}, M.~{Negrello}, G.~{Castex}, A.~{Lapi}, and M.~{Bonato}, {\it
  {Another look at distortions of the Cosmic Microwave Background spectrum}},
  {\em \jcap} {\bf 3} (Mar., 2016) 047,
  [\href{http://arxiv.org/abs/1512.04816}{{\tt arXiv:1512.04816}}].

\bibitem{Chluba2011therm}
J.~{Chluba} and R.~A. {Sunyaev}, {\it {The evolution of CMB spectral
  distortions in the early Universe}},  {\em \mnras} {\bf 419} (Jan., 2012)
  1294--1314, [\href{http://arxiv.org/abs/1109.6552}{{\tt arXiv:1109.6552}}].

\bibitem{Acharya2021large}
S.~K. {Acharya} and J.~{Chluba}, {\it {CMB spectral distortions from continuous
  large energy release}},  {\em arXiv e-prints} (Dec., 2021) arXiv:2112.06699,
  [\href{http://arxiv.org/abs/2112.06699}{{\tt arXiv:2112.06699}}].

\bibitem{CAMB}
A.~{Lewis}, A.~{Challinor}, and A.~{Lasenby}, {\it {Efficient Computation of
  Cosmic Microwave Background Anisotropies in Closed Friedmann-Robertson-Walker
  Models}},  {\em \apj} {\bf 538} (Aug., 2000) 473--476,
  [\href{http://arxiv.org/abs/astro-ph/9911177}{{\tt astro-ph/9911177}}].

\bibitem{CLASSCODE}
J.~{Lesgourgues}, {\it {The Cosmic Linear Anisotropy Solving System (CLASS) I:
  Overview}},  {\em ArXiv:1104.2932} (Apr., 2011)
  [\href{http://arxiv.org/abs/1104.2932}{{\tt arXiv:1104.2932}}].

\bibitem{Chluba2010}
J.~{Chluba}, G.~M. {Vasil}, and L.~J. {Dursi}, {\it {Recombinations to the
  Rydberg states of hydrogen and their effect during the cosmological
  recombination epoch}},  {\em \mnras} {\bf 407} (Sept., 2010) 599--612,
  [\href{http://arxiv.org/abs/1003.4928}{{\tt arXiv:1003.4928}}].

\bibitem{Chluba2014}
J.~{Chluba}, {\it {Refined approximations for the distortion visibility
  function and {$\mu$}-type spectral distortions}},  {\em \mnras} {\bf 440}
  (Apr., 2014) 2544--2563, [\href{http://arxiv.org/abs/1312.6030}{{\tt
  arXiv:1312.6030}}].

\bibitem{Chluba2015GreensII}
J.~{Chluba}, {\it {Green's function of the cosmological thermalization problem
  - II. Effect of photon injection and constraints}},  {\em \mnras} {\bf 454}
  (Dec., 2015) 4182--4196, [\href{http://arxiv.org/abs/1506.06582}{{\tt
  arXiv:1506.06582}}].

\bibitem{Chluba2013PCA}
J.~{Chluba} and D.~{Jeong}, {\it {Teasing bits of information out of the CMB
  energy spectrum}},  {\em \mnras} {\bf 438} (Mar., 2014) 2065--2082,
  [\href{http://arxiv.org/abs/1306.5751}{{\tt arXiv:1306.5751}}].

\bibitem{Chluba2010b}
J.~{Chluba} and R.~M. {Thomas}, {\it {Towards a complete treatment of the
  cosmological recombination problem}},  {\em \mnras} {\bf 412} (Apr., 2011)
  748--764, [\href{http://arxiv.org/abs/1010.3631}{{\tt arXiv:1010.3631}}].

\bibitem{Burigana1991}
C.~{Burigana}, L.~{Danese}, and G.~{de Zotti}, {\it {Formation and evolution of
  early distortions of the microwave background spectrum - A numerical study}},
   {\em \aap} {\bf 246} (June, 1991) 49--58.

\bibitem{Senatore2009}
L.~{Senatore}, S.~{Tassev}, and M.~{Zaldarriaga}, {\it {Non-gaussianities from
  perturbing recombination}},  {\em \jcap} {\bf 9} (Sept., 2009) 38,
  [\href{http://arxiv.org/abs/0812.3658}{{\tt arXiv:0812.3658}}].

\bibitem{Zegeye2022}
D.~{Zegeye}, K.~{Inomata}, and W.~{Hu}, {\it {Spectral distortion anisotropy
  from inflation for primordial black holes}},  {\em \prd} {\bf 105} (May,
  2022) 103535, [\href{http://arxiv.org/abs/2112.05190}{{\tt
  arXiv:2112.05190}}].

\bibitem{Chluba2014TRR}
J.~{Chluba}, {\it {Tests of the CMB temperature-redshift relation, CMB spectral
  distortions and why adiabatic photon production is hard}},  {\em \mnras} {\bf
  443} (Sept., 2014) 1881--1888, [\href{http://arxiv.org/abs/1405.1277}{{\tt
  arXiv:1405.1277}}].

\bibitem{Steigman2007}
G.~{Steigman}, {\it {Primordial Nucleosynthesis in the Precision Cosmology
  Era}},  {\em Annual Review of Nuclear and Particle Science} {\bf 57} (Nov.,
  2007) 463--491, [\href{http://arxiv.org/abs/0712.1100}{{\tt
  arXiv:0712.1100}}].

\bibitem{Steigman2009}
G.~Steigman, {\it Tracking the post-bbn evolution of deuterium},  {\em arXiv}
  {\bf astro-ph.GA} (Jan, 2009) [\href{http://arxiv.org/abs/0901.4333v}{{\tt
  arXiv:0901.4333v}}].

\bibitem{Chluba2013fore}
J.~{Chluba}, {\it {Distinguishing different scenarios of early energy release
  with spectral distortions of the cosmic microwave background}},  {\em \mnras}
  {\bf 436} (Dec., 2013) 2232--2243,
  [\href{http://arxiv.org/abs/1304.6121}{{\tt arXiv:1304.6121}}].

\bibitem{Sarkar1984}
S.~{Sarkar} and A.~M. {Cooper}, {\it {Cosmological and experimental constraints
  on the tau neutrino}},  {\em Physics Letters B} {\bf 148} (Nov., 1984)
  347--354.

\bibitem{Hu1993b}
W.~{Hu} and J.~{Silk}, {\it {Thermalization constraints and spectral
  distortions for massive unstable relic particles}},  {\em Physical Review
  Letters} {\bf 70} (May, 1993) 2661--2664.

\bibitem{Bolliet2020PI}
B.~{Bolliet}, J.~{Chluba}, and R.~{Battye}, {\it {Spectral distortion
  constraints on photon injection from low-mass decaying particles}},  {\em
  arXiv e-prints} (Dec., 2020) arXiv:2012.07292,
  [\href{http://arxiv.org/abs/2012.07292}{{\tt arXiv:2012.07292}}].

\bibitem{Chen2004}
X.~{Chen} and M.~{Kamionkowski}, {\it {Particle decays during the cosmic dark
  ages}},  {\em \prd} {\bf 70} (Aug., 2004) 043502--+,
  [\href{http://arxiv.org/abs/astro-ph/0310473}{{\tt astro-ph/0310473}}].

\bibitem{Padmanabhan2005}
N.~{Padmanabhan} and D.~P. {Finkbeiner}, {\it {Detecting dark matter
  annihilation with CMB polarization: Signatures and experimental prospects}},
  {\em \prd} {\bf 72} (July, 2005) 023508--+,
  [\href{http://arxiv.org/abs/astro-ph/0503486}{{\tt astro-ph/0503486}}].

\bibitem{callin:how_to_CMB}
P.~{Callin}, {\it {How to calculate the CMB spectrum}},  {\em arXiv e-prints}
  (June, 2006) astro--ph/0606683,
  [\href{http://arxiv.org/abs/astro-ph/0606683}{{\tt astro-ph/0606683}}].

\bibitem{Fixsen1996}
D.~J. {Fixsen}, E.~S. {Cheng}, J.~M. {Gales}, et~al., {\it {The Cosmic
  Microwave Background Spectrum from the Full COBE FIRAS Data Set}},  {\em
  \apj} {\bf 473} (Dec., 1996) 576--+,
  [\href{http://arxiv.org/abs/astro-ph/9605054}{{\tt astro-ph/9605054}}].

\bibitem{Fixsen2009}
D.~J. {Fixsen}, {\it {The Temperature of the Cosmic Microwave Background}},
  {\em \apj} {\bf 707} (Dec., 2009) 916--920,
  [\href{http://arxiv.org/abs/0911.1955}{{\tt arXiv:0911.1955}}].

\bibitem{Remazeilles:2010hq}
M.~Remazeilles, J.~Delabrouille, and J.-F. Cardoso, {\it {CMB and SZ effect
  separation with Constrained Internal Linear Combinations}},  {\em Mon. Not.
  Roy. Astron. Soc.} {\bf 410} (2011) 2481,
  [\href{http://arxiv.org/abs/1006.5599}{{\tt arXiv:1006.5599}}].

\bibitem{Cooray:2000xh}
A.~Cooray, W.~Hu, and M.~Tegmark, {\it {Large scale Sunyaev-Zel'dovich effect:
  Measuring statistical properties with multifrequency maps}},  {\em Astrophys.
  J.} {\bf 540} (2000) 1--13,
  [\href{http://arxiv.org/abs/astro-ph/0002238}{{\tt astro-ph/0002238}}].

\bibitem{Hill:2013baa}
J.~C. Hill and E.~Pajer, {\it {Cosmology from the thermal Sunyaev-Zel’dovich
  power spectrum: Primordial non-Gaussianity and massive neutrinos}},  {\em
  Phys. Rev.} {\bf D88} (2013), no.~6 063526,
  [\href{http://arxiv.org/abs/1303.4726}{{\tt arXiv:1303.4726}}].

\bibitem{Ravenni:2020rzd}
A.~Ravenni, M.~Rizzato, S.~Radinovi\'c, et~al., {\it {Breaking degeneracies
  with the Sunyaev-Zeldovich full bispectrum}},  {\em JCAP} {\bf 06} (2021)
  026, [\href{http://arxiv.org/abs/2008.12947}{{\tt arXiv:2008.12947}}].

\bibitem{Abitbol:2017vwa}
M.~H. Abitbol, J.~Chluba, J.~C. Hill, and B.~R. Johnson, {\it {Prospects for
  Measuring Cosmic Microwave Background Spectral Distortions in the Presence of
  Foregrounds}},  {\em Mon. Not. Roy. Astron. Soc.} {\bf 471} (2017), no.~1
  1126--1140, [\href{http://arxiv.org/abs/1705.01534}{{\tt arXiv:1705.01534}}].

\bibitem{Tegmark:1999ke}
M.~Tegmark, D.~J. Eisenstein, W.~Hu, and A.~de~Oliveira-Costa, {\it
  {Foregrounds and forecasts for the cosmic microwave background}},  {\em
  Astrophys. J.} {\bf 530} (2000) 133--165,
  [\href{http://arxiv.org/abs/astro-ph/9905257}{{\tt astro-ph/9905257}}].

\bibitem{Dunkley:2013vu}
J.~Dunkley et~al., {\it {The Atacama Cosmology Telescope: likelihood for
  small-scale CMB data}},  {\em JCAP} {\bf 07} (2013) 025,
  [\href{http://arxiv.org/abs/1301.0776}{{\tt arXiv:1301.0776}}].

\bibitem{Remazeilles2022}
M.~{Remazeilles}, A.~{Ravenni}, and J.~{Chluba}, {\it {Leverage on small-scale
  primordial non-Gaussianity through cross-correlations between CMB E-mode and
  {\ensuremath{\mu}}-distortion anisotropies}},  {\em \mnras} {\bf 512} (May,
  2022) 455--470, [\href{http://arxiv.org/abs/2110.14664}{{\tt
  arXiv:2110.14664}}].

\bibitem{Remazeilles:2018kqd}
M.~Remazeilles and J.~Chluba, {\it {Extracting foreground-obscured
  $\mu$-distortion anisotropies to constrain primordial non-Gaussianity}},
  {\em Mon. Not. Roy. Astron. Soc.} {\bf 478} (2018), no.~1 807--824,
  [\href{http://arxiv.org/abs/1802.10101}{{\tt arXiv:1802.10101}}].

\bibitem{Rotti2022}
A.~{Rotti}, A.~{Ravenni}, and J.~{Chluba}, {\it {Non-Gaussianity constraints
  from Planck spectral distortion cross-correlations}},  {\em arXiv e-prints}
  (May, 2022) arXiv:2205.15971, [\href{http://arxiv.org/abs/2205.15971}{{\tt
  arXiv:2205.15971}}].

\bibitem{Planck:2018nkj}
{\bf Planck} Collaboration, N.~Aghanim et~al., {\it {Planck 2018 results. I.
  Overview and the cosmological legacy of Planck}},  {\em Astron. Astrophys.}
  {\bf 641} (2020) A1, [\href{http://arxiv.org/abs/1807.06205}{{\tt
  arXiv:1807.06205}}].

\bibitem{LiteBIRD:2020khw}
{\bf LiteBIRD} Collaboration, M.~Hazumi et~al., {\it {LiteBIRD: JAXA's new
  strategic L-class mission for all-sky surveys of cosmic microwave background
  polarization}},  {\em Proc. SPIE Int. Soc. Opt. Eng.} {\bf 11443} (2020)
  114432F, [\href{http://arxiv.org/abs/2101.12449}{{\tt arXiv:2101.12449}}].

\bibitem{NASAPICO:2019thw}
{\bf NASA PICO} Collaboration, S.~Hanany et~al., {\it {PICO: Probe of Inflation
  and Cosmic Origins}},  \href{http://arxiv.org/abs/1902.10541}{{\tt
  arXiv:1902.10541}}.

\bibitem{Ganc2012}
J.~{Ganc} and E.~{Komatsu}, {\it {Scale-dependent bias of galaxies and
  {$\mu$}-type distortion of the cosmic microwave background spectrum from
  single-field inflation with a modified initial state}},  {\em \prd} {\bf 86}
  (July, 2012) 023518, [\href{http://arxiv.org/abs/1204.4241}{{\tt
  arXiv:1204.4241}}].

\bibitem{Ravenni2017}
A.~{Ravenni}, M.~{Liguori}, N.~{Bartolo}, and M.~{Shiraishi}, {\it {Primordial
  non-Gaussianity with {\ensuremath{\mu}}-type and y-type spectral distortions:
  exploiting Cosmic Microwave Background polarization and dealing with
  secondary sources}},  {\em \jcap} {\bf 2017} (Sept., 2017) 042,
  [\href{http://arxiv.org/abs/1707.04759}{{\tt arXiv:1707.04759}}].

\bibitem{Taburet2011}
N.~{Taburet}, C.~{Hern{\'a}ndez-Monteagudo}, N.~{Aghanim}, M.~{Douspis}, and
  R.~A. {Sunyaev}, {\it {The ISW-tSZ cross-correlation: integrated Sachs-Wolfe
  extraction out of pure cosmic microwave background data}},  {\em \mnras} {\bf
  418} (Dec., 2011) 2207--2218, [\href{http://arxiv.org/abs/1012.5036}{{\tt
  arXiv:1012.5036}}].

\bibitem{Creque2016}
C.~{Creque-Sarbinowski}, S.~{Bird}, and M.~{Kamionkowski}, {\it
  {Cross-correlation between thermal Sunyaev-Zeldovich effect and the
  integrated Sachs-Wolfe effect}},  {\em \prd} {\bf 94} (Sept., 2016) 063519,
  [\href{http://arxiv.org/abs/1606.00839}{{\tt arXiv:1606.00839}}].

\bibitem{Kogut2016SPIE}
A.~{Kogut}, J.~{Chluba}, D.~J. {Fixsen}, S.~{Meyer}, and D.~{Spergel}, {\it
  {The Primordial Inflation Explorer (PIXIE)}},  in {\em SPIE Conference
  Series}, vol.~9904 of {\em Proc.SPIE}, p.~99040W, July, 2016.

\bibitem{Kogut2019BAAS}
A.~{Kogut}, M.~H. {Abitbol}, J.~{Chluba}, et~al., {\it {CMB Spectral
  Distortions: Status and Prospects}},  in {\em Bulletin of the American
  Astronomical Society}, vol.~51, p.~113, Sept., 2019.
\newblock \href{http://arxiv.org/abs/1907.13195}{{\tt arXiv:1907.13195}}.

\bibitem{abitbol_pixie}
M.~H. {Abitbol}, J.~{Chluba}, J.~C. {Hill}, and B.~R. {Johnson}, {\it
  {Prospects for Measuring Cosmic Microwave Background Spectral Distortions in
  the Presence of Foregrounds}},  {\em \mnras} (May, 2017)
  [\href{http://arxiv.org/abs/1705.01534}{{\tt arXiv:1705.01534}}].

\bibitem{Chluba2019BAAS}
J.~{Chluba}, A.~{Kogut}, S.~P. {Patil}, et~al., {\it {Spectral Distortions of
  the CMB as a Probe of Inflation, Recombination, Structure Formation and
  Particle Physics}},  {\em \baas} {\bf 51} (May, 2019) 184,
  [\href{http://arxiv.org/abs/1903.04218}{{\tt arXiv:1903.04218}}].

\bibitem{Chluba2021ExA}
J.~{Chluba}, M.~H. {Abitbol}, N.~{Aghanim}, et~al., {\it {New horizons in
  cosmology with spectral distortions of the cosmic microwave background}},
  {\em Experimental Astronomy} {\bf 51} (June, 2021) 1515--1554,
  [\href{http://arxiv.org/abs/1909.01593}{{\tt arXiv:1909.01593}}].

\bibitem{SOWP2018}
{The Simons Observatory Collaboration}, P.~{Ade}, J.~{Aguirre}, et~al., {\it
  {The Simons Observatory: Science goals and forecasts}},  {\em
  ArXiv:1808.07445} (Aug., 2018) [\href{http://arxiv.org/abs/1808.07445}{{\tt
  arXiv:1808.07445}}].

\bibitem{Abazajian2016S4SB}
K.~N. {Abazajian}, P.~{Adshead}, Z.~{Ahmed}, et~al., {\it {CMB-S4 Science Book,
  First Edition}},  {\em ArXiv:1610.02743} (Oct., 2016)
  [\href{http://arxiv.org/abs/1610.02743}{{\tt arXiv:1610.02743}}].

\bibitem{PRISM2013WP}
{PRISM Collaboration}, P.~{Andre}, C.~{Baccigalupi}, et~al., {\it {PRISM
  (Polarized Radiation Imaging and Spectroscopy Mission): A White Paper on the
  Ultimate Polarimetric Spectro-Imaging of the Microwave and Far-Infrared
  Sky}},  {\em ArXiv:1306.2259} (June, 2013)
  [\href{http://arxiv.org/abs/1306.2259}{{\tt arXiv:1306.2259}}].

\bibitem{PRISM2013WPII}
P.~{Andr{\'e}}, C.~{Baccigalupi}, A.~{Banday}, et~al., {\it {PRISM (Polarized
  Radiation Imaging and Spectroscopy Mission): an extended white paper}},  {\em
  \jcap} {\bf 2} (Feb., 2014) 6, [\href{http://arxiv.org/abs/1310.1554}{{\tt
  arXiv:1310.1554}}].

\bibitem{Delabrouille2018}
J.~{Delabrouille}, P.~{de Bernardis}, F.~R. {Bouchet}, et~al., {\it {Exploring
  cosmic origins with CORE: Survey requirements and mission design}},  {\em
  \jcap} {\bf 4} (Apr., 2018) 014, [\href{http://arxiv.org/abs/1706.04516}{{\tt
  arXiv:1706.04516}}].

\bibitem{Delabrouille2021}
J.~{Delabrouille}, M.~H. {Abitbol}, N.~{Aghanim}, et~al., {\it {Microwave
  spectro-polarimetry of matter and radiation across space and time}},  {\em
  Experimental Astronomy} {\bf 51} (June, 2021) 1471--1514.

\bibitem{Chluba2013iso}
J.~{Chluba} and D.~{Grin}, {\it {CMB spectral distortions from small-scale
  isocurvature fluctuations}},  {\em \mnras} {\bf 434} (Sept., 2013)
  1619--1635, [\href{http://arxiv.org/abs/1304.4596}{{\tt arXiv:1304.4596}}].

\bibitem{Slatyer2010Sommer}
T.~R. {Slatyer}, {\it {The Sommerfeld enhancement for dark matter with an
  excited state}},  {\em \jcap} {\bf 2010} (Feb., 2010) 028,
  [\href{http://arxiv.org/abs/0910.5713}{{\tt arXiv:0910.5713}}].

\bibitem{Aguirre2011}
A.~{Aguirre} and M.~C. {Johnson}, {\it {A status report on the observability of
  cosmic bubble collisions}},  {\em Reports on Progress in Physics} {\bf 74}
  (July, 2011) 074901, [\href{http://arxiv.org/abs/0908.4105}{{\tt
  arXiv:0908.4105}}].

\bibitem{Kleban2013}
M.~{Kleban}, T.~S. {Levi}, and K.~{Sigurdson}, {\it {Observing the multiverse
  with cosmic wakes}},  {\em \prd} {\bf 87} (Feb., 2013) 041301,
  [\href{http://arxiv.org/abs/1109.3473}{{\tt arXiv:1109.3473}}].

\bibitem{Deng2020}
H.~{Deng}, {\it {Spiky CMB distortions from primordial bubbles}},  {\em \jcap}
  {\bf 2020} (May, 2020) 037, [\href{http://arxiv.org/abs/2003.02485}{{\tt
  arXiv:2003.02485}}].

\bibitem{Carr2016}
B.~{Carr}, F.~{K{\"u}hnel}, and M.~{Sandstad}, {\it {Primordial black holes as
  dark matter}},  {\em \prd} {\bf 94} (Oct., 2016) 083504,
  [\href{http://arxiv.org/abs/1607.06077}{{\tt arXiv:1607.06077}}].

\bibitem{Abe2019}
K.~T. {Abe}, H.~{Tashiro}, and T.~{Tanaka}, {\it {Thermal Sunyaev-Zel'dovich
  anisotropy due to primordial black holes}},  {\em \prd} {\bf 99} (May, 2019)
  103519, [\href{http://arxiv.org/abs/1901.06809}{{\tt arXiv:1901.06809}}].

\bibitem{Deng2021}
H.~{Deng}, {\it {{\ensuremath{\mu}}-distortion around stupendously large
  primordial black holes}},  {\em \jcap} {\bf 2021} (Nov., 2021) 054,
  [\href{http://arxiv.org/abs/2106.09817}{{\tt arXiv:2106.09817}}].

\bibitem{Ozsoy2021}
O.~{{\"O}zsoy} and G.~{Tasinato}, {\it {CMB {\ensuremath{\mu}} T cross
  correlations as a probe of primordial black hole scenarios}},  {\em \prd}
  {\bf 104} (Aug., 2021) 043526, [\href{http://arxiv.org/abs/2104.12792}{{\tt
  arXiv:2104.12792}}].

\bibitem{Jedamzik1998}
K.~{Jedamzik}, V.~{Katalini{\'c}}, and A.~V. {Olinto}, {\it {Damping of cosmic
  magnetic fields}},  {\em \prd} {\bf 57} (Mar., 1998) 3264--3284,
  [\href{http://arxiv.org/abs/astro-ph/9606080}{{\tt astro-ph/9606080}}].

\bibitem{Miyamoto2014}
K.~{Miyamoto}, T.~{Sekiguchi}, H.~{Tashiro}, and S.~{Yokoyama}, {\it {CMB
  distortion anisotropies due to the decay of primordial magnetic fields}},
  {\em \prd} {\bf 89} (Mar., 2014) 063508,
  [\href{http://arxiv.org/abs/1310.3886}{{\tt arXiv:1310.3886}}].

\bibitem{Minoda2018}
T.~{Minoda}, K.~{Hasegawa}, H.~{Tashiro}, K.~{Ichiki}, and N.~{Sugiyama}, {\it
  {Thermal Sunyaev-Zel'dovich Effect in the IGM due to Primordial Magnetic
  Fields}},  {\em Galaxies} {\bf 6} (Dec., 2018) 143,
  [\href{http://arxiv.org/abs/1812.09813}{{\tt arXiv:1812.09813}}].

\bibitem{Saga2019}
S.~{Saga}, A.~{Ota}, H.~{Tashiro}, and S.~{Yokoyama}, {\it {Secondary CMB
  temperature anisotropies from magnetic reheating}},  {\em \mnras} {\bf 490}
  (Dec., 2019) 4419--4427, [\href{http://arxiv.org/abs/1904.09121}{{\tt
  arXiv:1904.09121}}].

\bibitem{Dvorkin2014}
C.~{Dvorkin}, K.~{Blum}, and M.~{Kamionkowski}, {\it {Constraining dark
  matter-baryon scattering with linear cosmology}},  {\em \prd} {\bf 89} (Jan.,
  2014) 023519, [\href{http://arxiv.org/abs/1311.2937}{{\tt arXiv:1311.2937}}].

\bibitem{Yacine2015DM}
Y.~{Ali-Ha{\"i}moud}, J.~{Chluba}, and M.~{Kamionkowski}, {\it {Constraints on
  dark matter interactions with standard model particles from CMB spectral
  distortions}},  {\em ArXiv e-prints} (June, 2015)
  [\href{http://arxiv.org/abs/1506.04745}{{\tt arXiv:1506.04745}}].

\bibitem{Munoz2015}
J.~B. {Mu{\~n}oz}, E.~D. {Kovetz}, and Y.~{Ali-Ha{\"\i}moud}, {\it {Heating of
  baryons due to scattering with dark matter during the dark ages}},  {\em
  \prd} {\bf 92} (Oct., 2015) 083528,
  [\href{http://arxiv.org/abs/1509.00029}{{\tt arXiv:1509.00029}}].

\bibitem{Slatyer2009}
T.~R. Slatyer, N.~Padmanabhan, and D.~P. Finkbeiner, {\it Cmb constraints on
  wimp annihilation: Energy absorption during the recombination epoch},  {\em
  Physical Review D (Particles, Fields, Gravitation, and Cosmology)} {\bf 80}
  (2009), no.~4 043526.

\bibitem{Chluba2010a}
J.~{Chluba}, {\it {Could the cosmological recombination spectrum help us
  understand annihilating dark matter?}},  {\em \mnras} {\bf 402} (Feb., 2010)
  1195--1207, [\href{http://arxiv.org/abs/0910.3663}{{\tt arXiv:0910.3663}}].

\bibitem{Pajer2012}
E.~{Pajer} and M.~{Zaldarriaga}, {\it {New Window on Primordial
  Non-Gaussianity}},  {\em Physical Review Letters} {\bf 109} (July, 2012)
  021302, [\href{http://arxiv.org/abs/1201.5375}{{\tt arXiv:1201.5375}}].

\bibitem{Pajer2012b}
E.~{Pajer} and M.~{Zaldarriaga}, {\it {A hydrodynamical approach to CMB
  {$\mu$}-distortion from primordial perturbations}},  {\em \jcap} {\bf 2}
  (Feb., 2013) 36, [\href{http://arxiv.org/abs/1206.4479}{{\tt
  arXiv:1206.4479}}].

\bibitem{Chluba2017muT}
J.~{Chluba}, E.~{Dimastrogiovanni}, M.~A. {Amin}, and M.~{Kamionkowski}, {\it
  {Evolution of CMB spectral distortion anisotropies and tests of primordial
  non-Gaussianity}},  {\em \mnras} {\bf 466} (Apr., 2017) 2390--2401,
  [\href{http://arxiv.org/abs/1610.08711}{{\tt arXiv:1610.08711}}].

\bibitem{Cabass2018}
G.~{Cabass}, E.~{Pajer}, and D.~{van der Woude}, {\it {Spectral distortion
  anisotropies from single-field inflation}},  {\em \jcap} {\bf 2018} (Aug.,
  2018) 050, [\href{http://arxiv.org/abs/1805.08775}{{\tt arXiv:1805.08775}}].

\bibitem{Orlando2022}
G.~{Orlando}, P.~D. {Meerburg}, and S.~P. {Patil}, {\it {Primordial tensor
  bispectra in {\ensuremath{\mu}}-CMB cross-correlations}},  {\em \jcap} {\bf
  2022} (Feb., 2022) 004, [\href{http://arxiv.org/abs/2109.01095}{{\tt
  arXiv:2109.01095}}].

\bibitem{Audren2014}
B.~{Audren}, J.~{Lesgourgues}, G.~{Mangano}, P.~D. {Serpico}, and T.~{Tram},
  {\it {Strongest model-independent bound on the lifetime of Dark Matter}},
  {\em \jcap} {\bf 2014} (Dec., 2014) 028--028,
  [\href{http://arxiv.org/abs/1407.2418}{{\tt arXiv:1407.2418}}].

\end{thebibliography}\endgroup
}

\newpage

\appendix

\begin{figure}
\centering
\includegraphics[width=1.0\columnwidth]{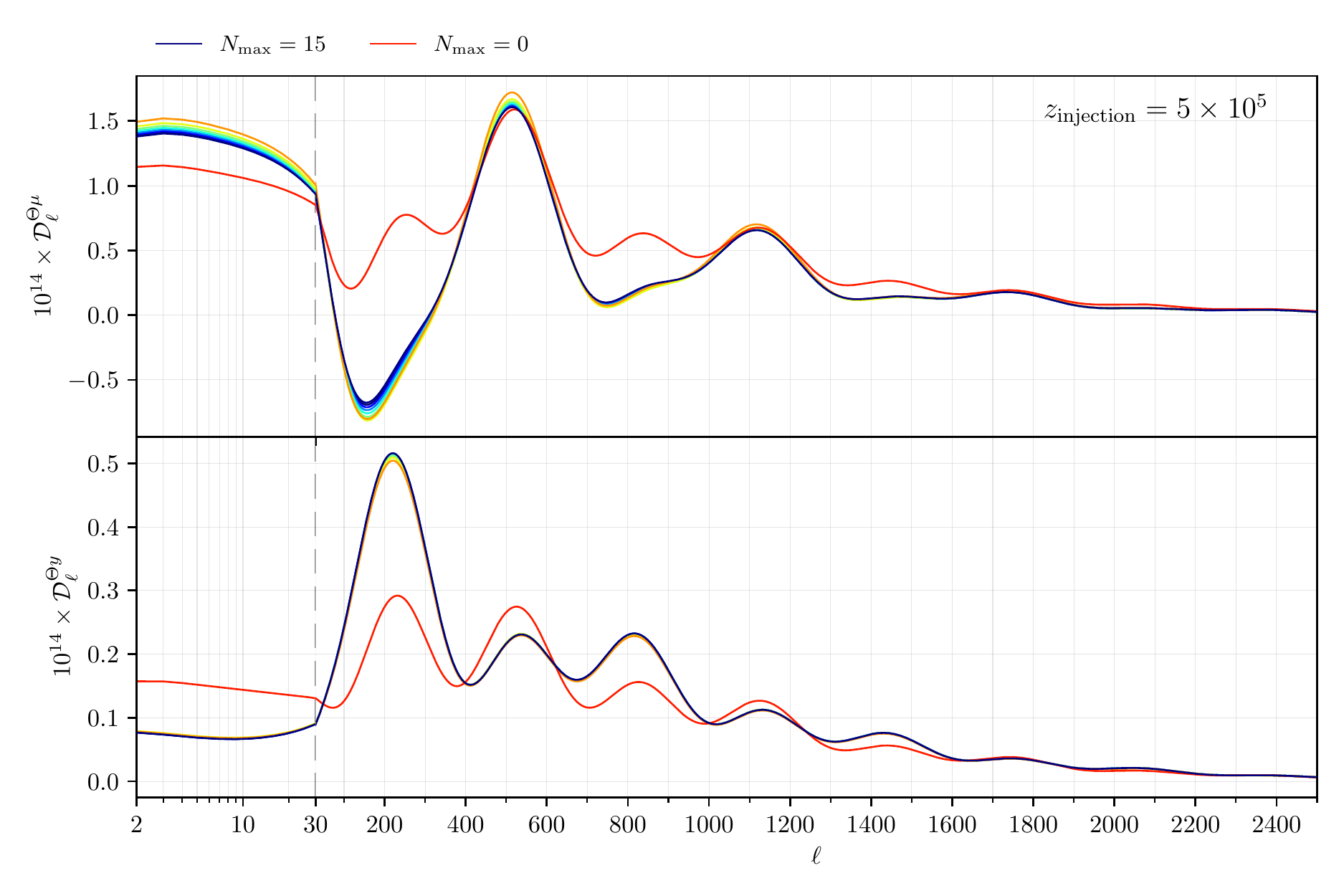}
\\[-1mm]
\includegraphics[width=1.0\columnwidth]{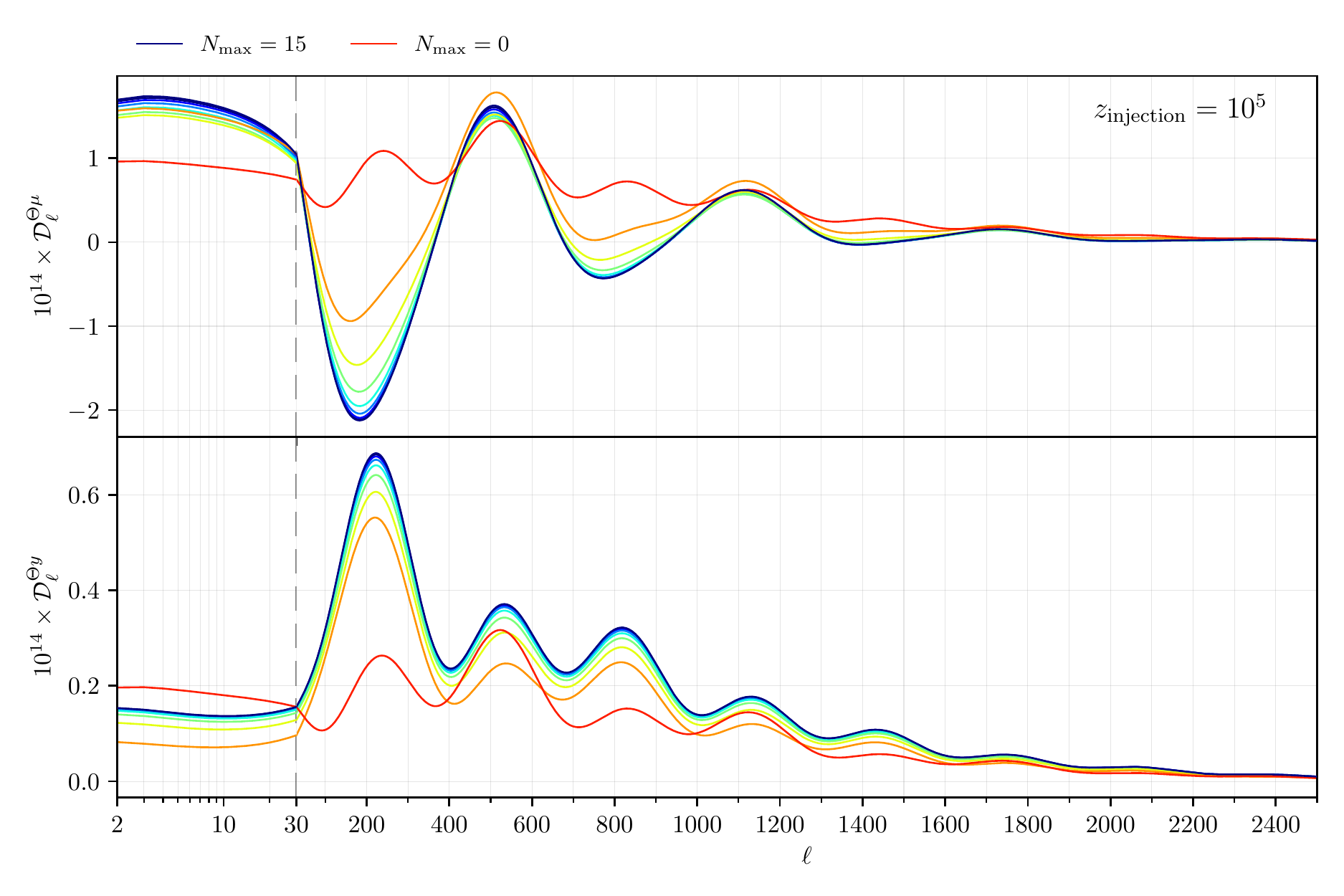}
\\[-1mm]
\caption{Two figures illustrating convergence of the power spectra for increasing $N_{\rm max}$. Shown are the $\Theta\times \mu$ and $\Theta\times y$ spectra at early injection redshifts. The vertical dashed line shows a division between linearly-spaced $\ell$ values (left) and log-spaced values (right).}
\label{fig:power_spectrum_convergence_1}
\end{figure}
\begin{figure}
\centering
\includegraphics[width=1.0\columnwidth]{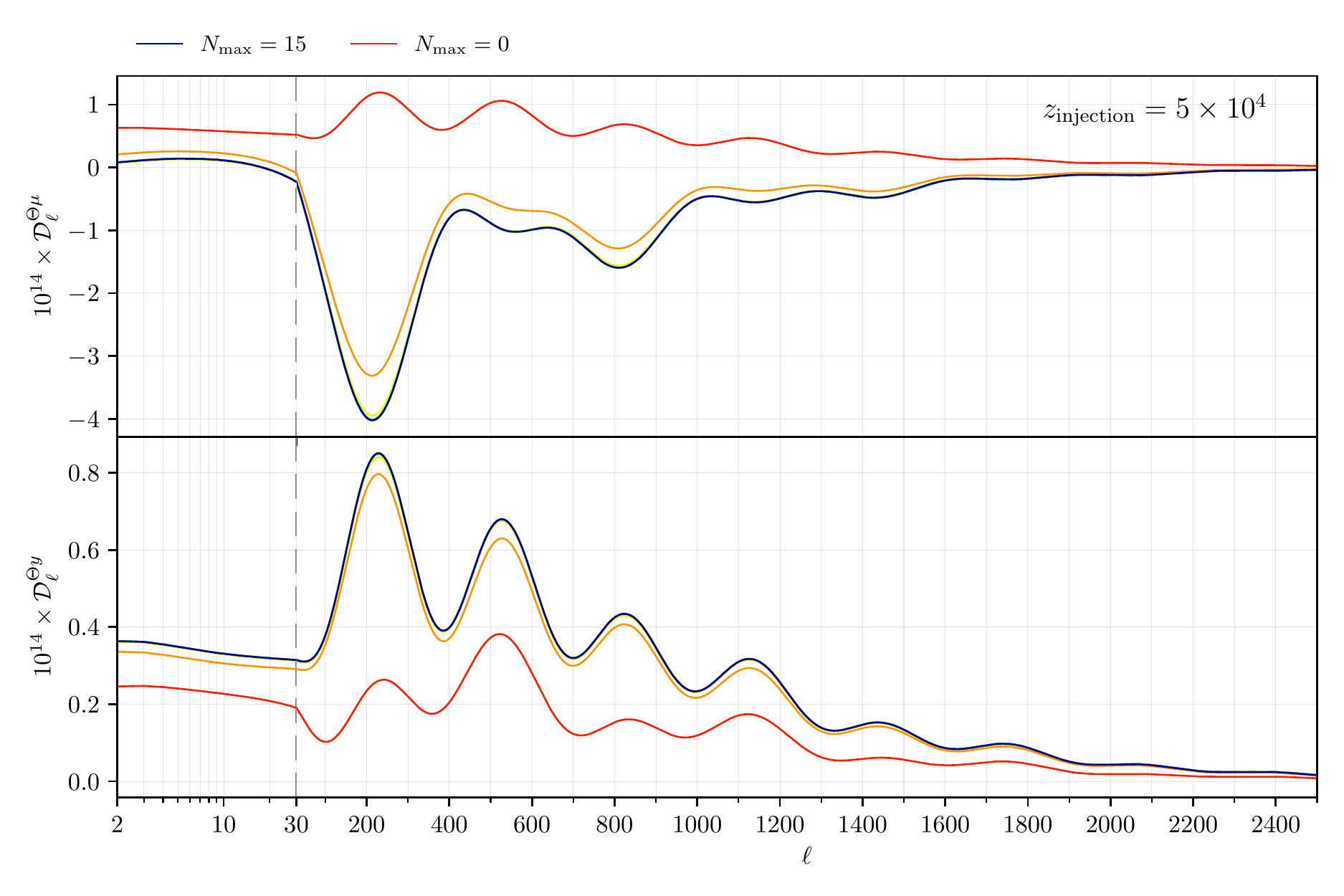}
\\[-1mm]
\includegraphics[width=1.0\columnwidth]{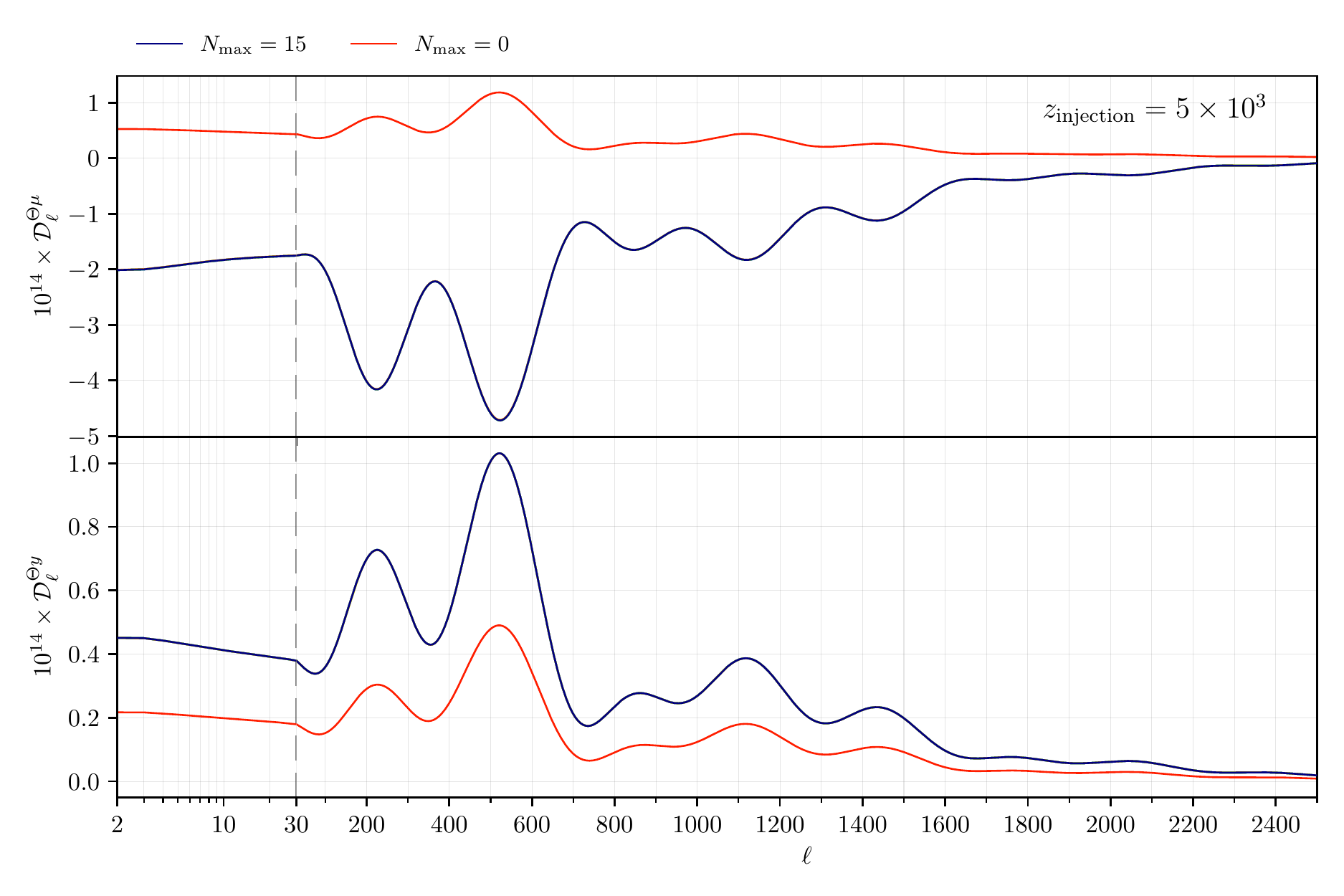}
\\[-1mm]
\caption{As for Fig.~\ref{fig:power_spectrum_convergence_1} but for late injection redshifts.}
\label{fig:power_spectrum_convergence_2}
\end{figure}

\section{Perturbed decay term for decaying particle scenarios}
\label{app:pert_decay}
For the average heating rate by particle decay we simply used
\bealf{
\frac{\id a^4\rho^{(0)}_\gamma }{a^4 \id t}=-\frac{\id a^3\rho^{(0)}_X }{a^3 \id t}=m_X c^2\,\Gamma_X N_X^{(0)},
}
which follows from the Collision term for the decaying particle. Here, we assumed rapid transfer of energy to the photon field though Compton scattering. This then give the relevant heating term 
\bealf{
\frac{\id \mathcal{Q}^{(0)}}{\id t}&=\frac{\id a^4\rho^{(0)}_\gamma}{a^4 \rho^{(0)}_\gamma \id t}=\frac{m_X c^2\,\Gamma_X N_X^{(0)}}{\rho^{(0)}_\gamma}
}
for the background evolution with $N_X^{(0)}=N_X^{(0)}(t=0)\,\expf{-\Gamma_X t}$.

To obtain the perturbed decay term for decaying particle scenarios, we can similarly use the evolution equation given by \cite{Audren2014} 
\bealf{
\frac{\partial}{\partial \eta}\delta^{(1)}_X&=-k \beta_X - 3 \frac{\partial}{\partial \eta}\Phi^{(1)} - a \Gamma_X \,\Psi^{(1)}
}
for the perturbations in a decaying particle component, $\delta^{(1)}_X=N^{(1)}_X/N^{(0)}_X$. We switched to our convention for the sign of $\Phi$, which is $-\Phi^{\rm A}$. Aside from the last term, this is simply like the standard dark matter equation \citep{Ma1995}. The last term gives rise to extra energy release from changes in the local time. Together with contributions from perturbations in the density of the decaying particle, the release of energy to the photon field then is given by
\bealf{
\frac{\id a^4\rho^{(1)}_\gamma }{a^4 \id t}=
\delta_X^{(1)}\,\frac{\id a^4\rho^{(0)}_\gamma }{a^4 \id t}
+m_Xc^2\,N^{(0)}_X\,\Gamma_X \,\Psi^{(1)}
\equiv \left[\delta_X^{(1)}+\Psi^{(1)}\right]\frac{\id a^4\rho^{(0)}_\gamma }{a^4 \id t}.
}
This implies
\bealf{
\frac{\id \mathcal{Q}^{(1)}}{\id t}&=\left[\delta_X^{(1)}+\Psi^{(1)}\right]\frac{\id \mathcal{Q}^{(0)}}{\id t},
}
which includes the particle density and local potential modulation effects.
If we furthermore assume that $\delta_X\simeq \delta_{\rm cdm}$, we obtain the expression in Eq.~\eqref{eq:perturbed_decay_heating}. This expression can also be directly obtained when thinking of the corrections from $\Psi^{(1)}$ to the background collision term \citep{Senatore2009}. A more in depth derivation is given in the Appendix of paper II, which also includes effects from the local heat capacity modulation.

\section{Power spectrum convergence}
\label{sec:power_spectrum_convergence}
Analogously to Fig.~\ref{fig:aniso_spec_convergence}, we can study the convergence of the power spectra, and build confidence in the extended basis. Again we show the three redshifts corresponding to the three eras ($z=\pot{5}{5}$, $z=\pot{5}{4}$ and $z=\pot{5}{3}$), but with one additional residual-era injection which is typically the poorest converged in this formalism (i.e., $z=10^5$, see paper I). We see in Fig.~\ref{fig:power_spectrum_convergence_1} and Fig.~\ref{fig:power_spectrum_convergence_2} that similar results hold as for studying individual spectra: the $y$-era injection requires only a single additional mode $y_1$ by construction, while the residual- and $\mu$-era injections requires around 3-5 modes. Only the second panel shows an appreciable difference between $N_{\rm max}=13$ and $N_{\rm max}=15$, however this only amounts to a $\lesssim 2\%$ difference. These levels of departures are at the limit of our computations, on par with other neglected effects.

\section{Physical effects in SD power spectra}
\label{app:auto_spectra_physics}
\begin{figure}
\centering
\includegraphics[width=0.85\columnwidth]{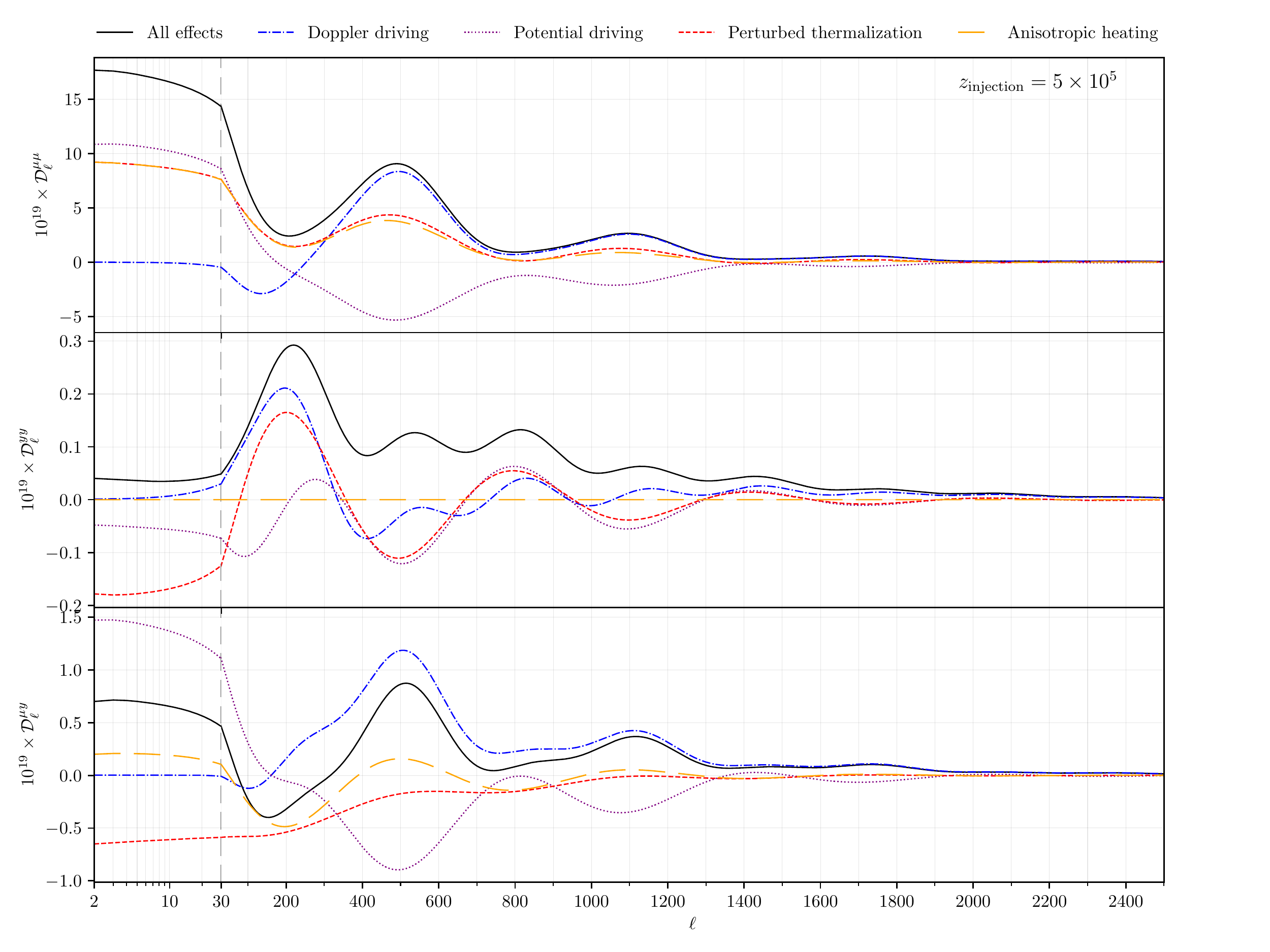}
\caption{A figure showing the distortion power spectra from injection at $z_{\rm injection}=\pot{5}{5}$ with different physical switches.}
\label{fig:SD_outer_power_spectra_switches_zh5e5}
\end{figure}
\begin{figure}
\centering
\includegraphics[width=0.85\columnwidth]{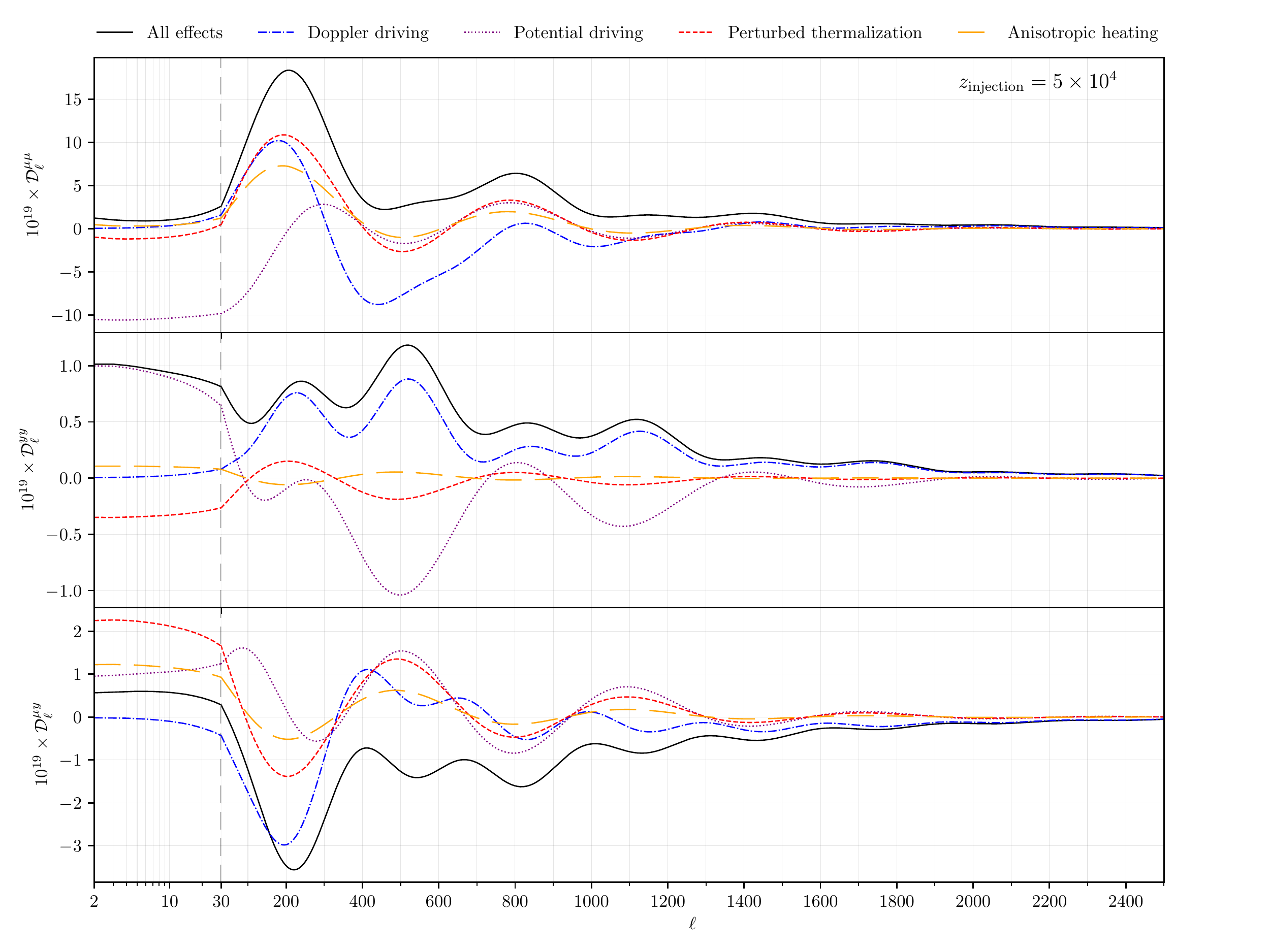}
\caption{As for Fig.~\ref{fig:SD_outer_power_spectra_switches_zh5e5} but for $z_{\rm injection}=\pot{5}{4}.$}
\label{fig:SD_outer_power_spectra_switches_zh5e4}
\end{figure}
\begin{figure}
\centering
\includegraphics[width=0.85\columnwidth]{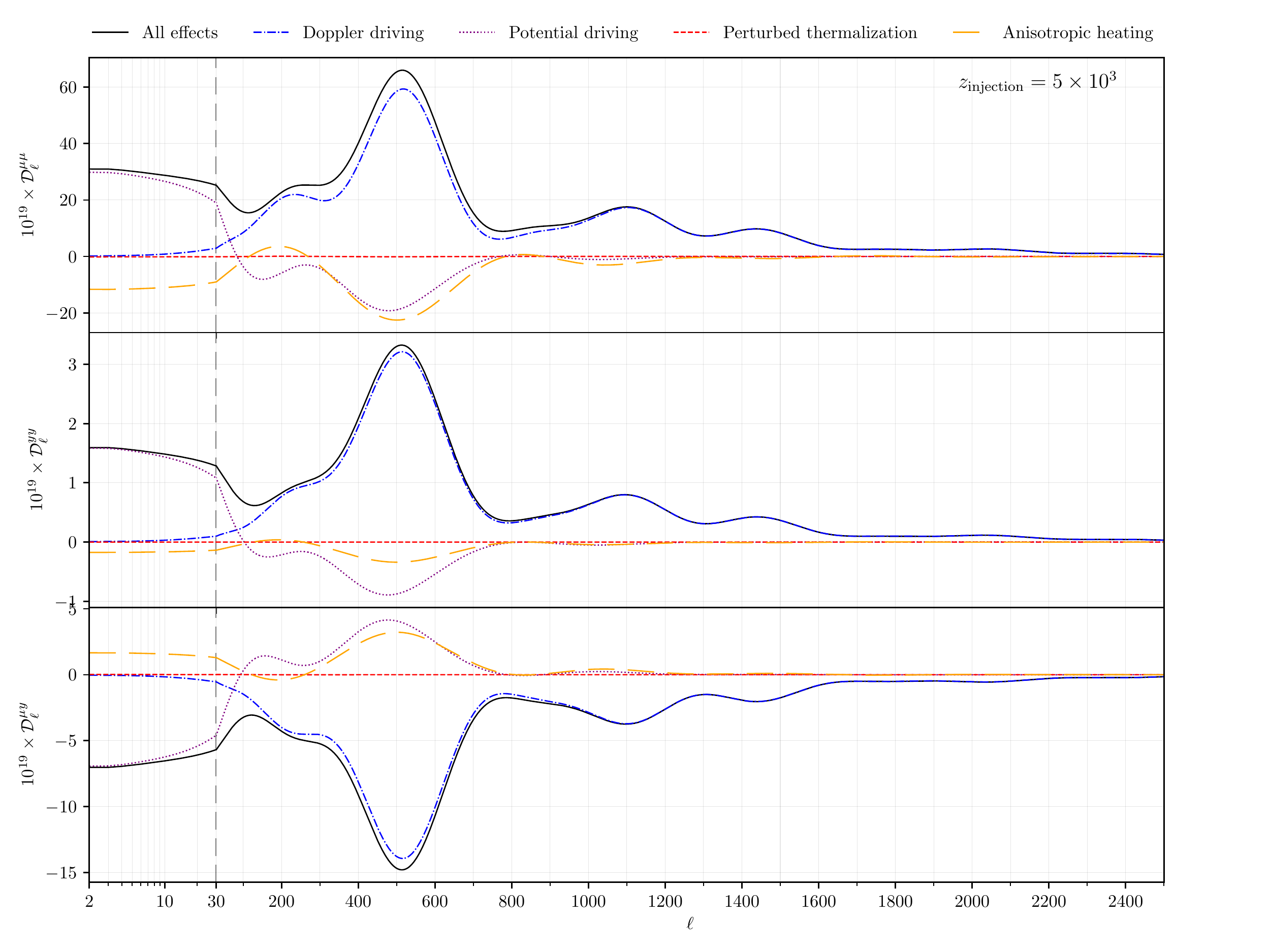}
\caption{As for Fig.~\ref{fig:SD_outer_power_spectra_switches_zh5e5} but for $z_{\rm injection}=\pot{5}{3}.$}
\label{fig:SD_outer_power_spectra_switches_zh5e3}
\end{figure}
For completeness, we include the pure distortion power spectra with various physical \textit{switches}, which helps illustrate the origin of concrete features. Many of these will be very analogous to those discussed in Sect.~\ref{sec:power_spectrum_switches}, but are useful to see in the absence of structure familiar from the temperature power spectra.

Referring to Figs.~\ref{fig:SD_outer_power_spectra_switches_zh5e5}, \ref{fig:SD_outer_power_spectra_switches_zh5e4} and \ref{fig:SD_outer_power_spectra_switches_zh5e3}, the first feature we note is the lack of thermalisation contributions for late injection. At early injection times thermalisation sources both $\mu^{(1)}$ and $y^{(1)}$ through the $M_{\rm D}\vek{y}^{(0)}_0$ and $(Y_1-\Yspec)$ terms. We independently verify perturbed photon emission only influences $\mu^{(1)}$, and is a small exclusively early time effect. Anisotropic heating exclusively sources $\mu^{(1)}$ at early times as expected, with other times sourcing a mix of both primary SEDs.
We can also again see a characteristic ceasing of potential boosts at the late time, since much of the energy is injected either sub-horizon or very close to horizon crossing.
Interestingly we see that the residual-era injection sees a remnant of the first power spectrum peak in both $\mu\times\mu$ \textit{and} $y\times y$. For the later injection times that Doppler peak is lost in both cases, and at early times it is only present in the $y\times y$ spectrum given that boosting is a subdominant contribution to the $\mu^{(1)}$ SED in the presence of anisotropic heating and perturbed thermalisation.

\begin{figure}
\centering
\includegraphics[width=0.85\columnwidth]{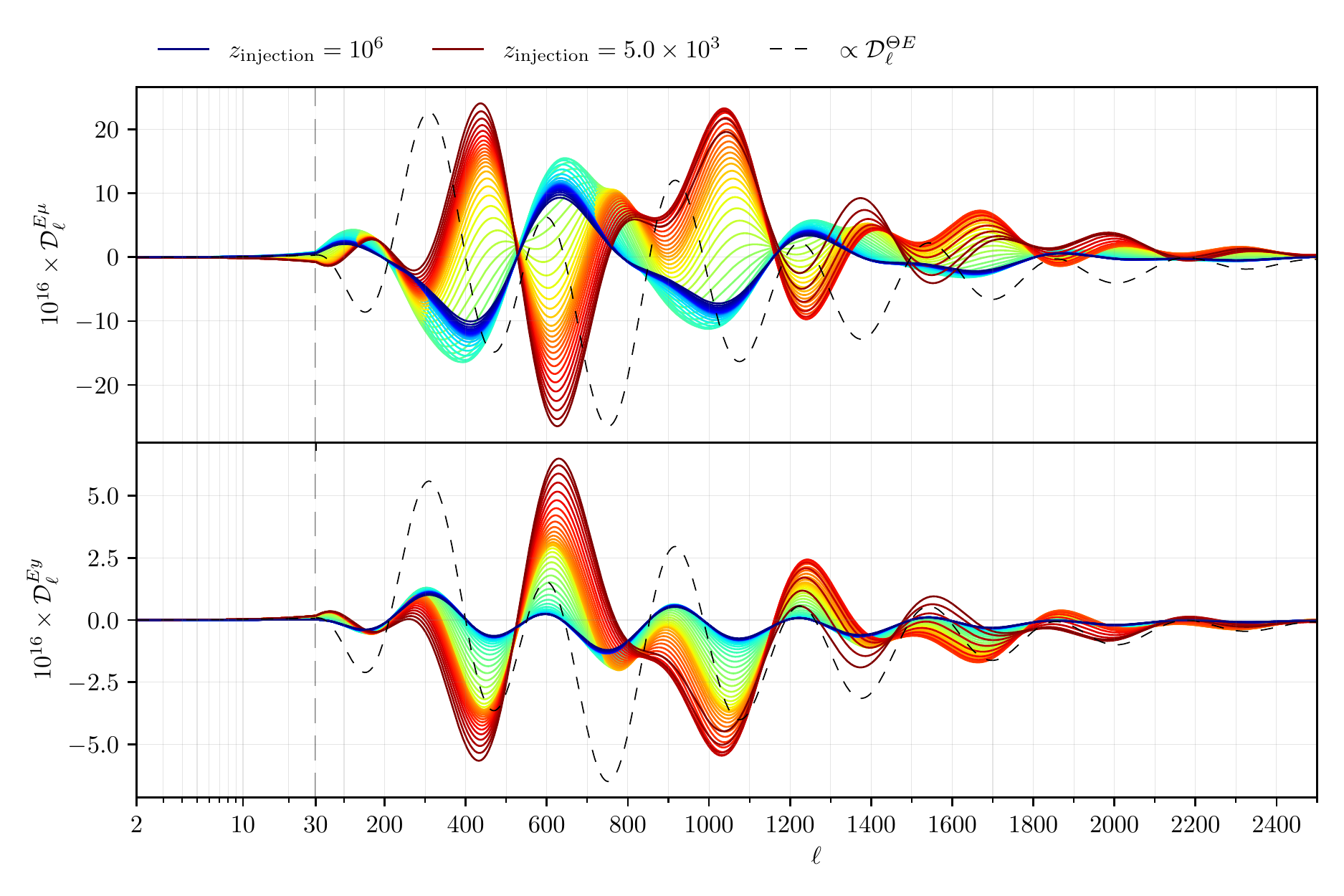}
\caption{The power spectra for $E \times \mu$ and $E \times y$ over a range of $50$ single-injection redshifts. Blue lines show early injection into the $\mu$-era and red lines show late injection in the $y$-era. The vertical dashed line shows a division between log-spaced $\ell$ values (left) and linear-spaced values (right). For reference, we show the familiar $\Theta\times E$ power spectrum (rescaled within each panel).}
\label{fig:SD_E_zh_correlation}
\end{figure}
\begin{figure}
\centering
\includegraphics[width=0.90\columnwidth]{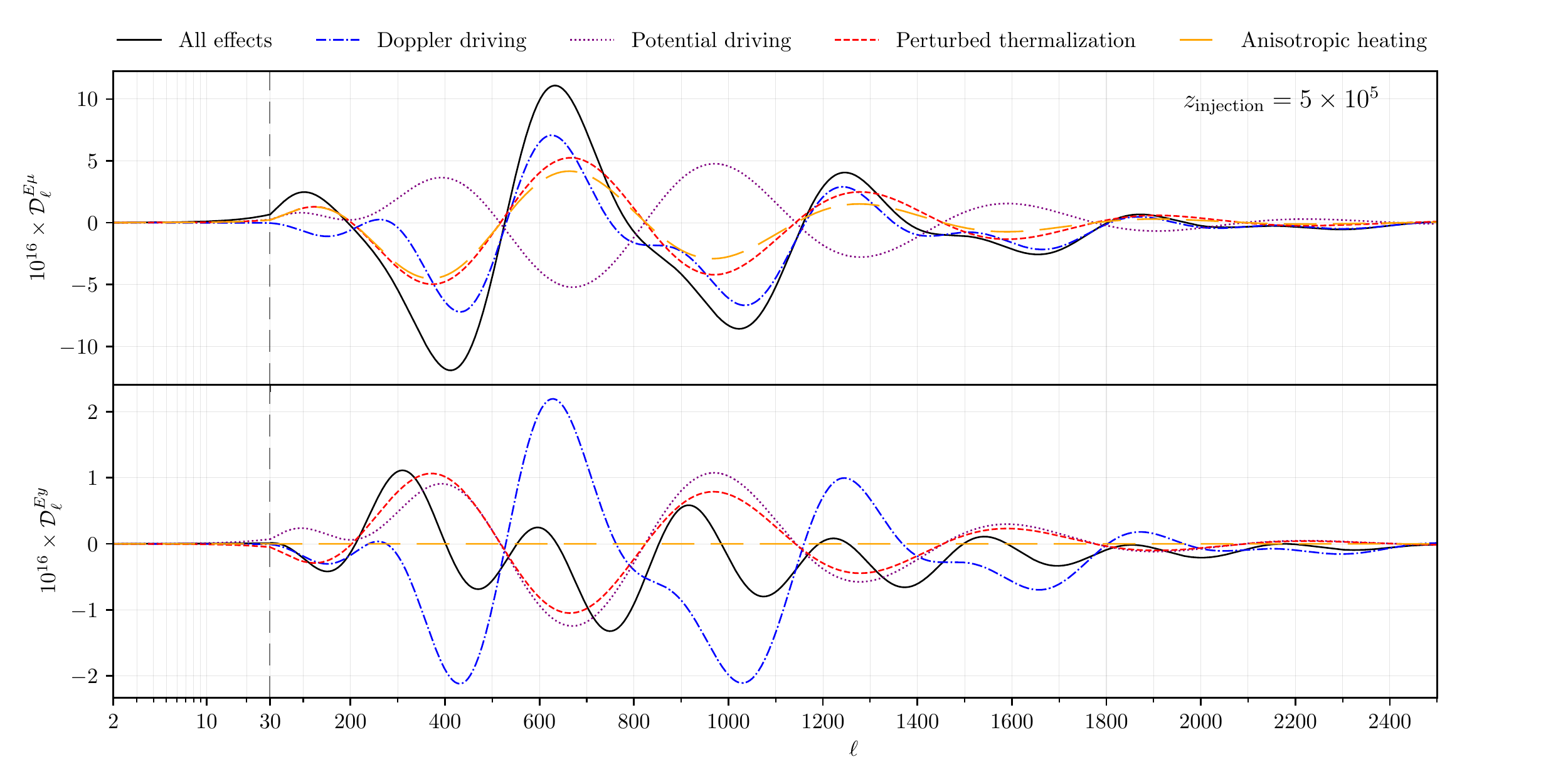}
\\[-1mm]
\includegraphics[width=0.90\columnwidth]{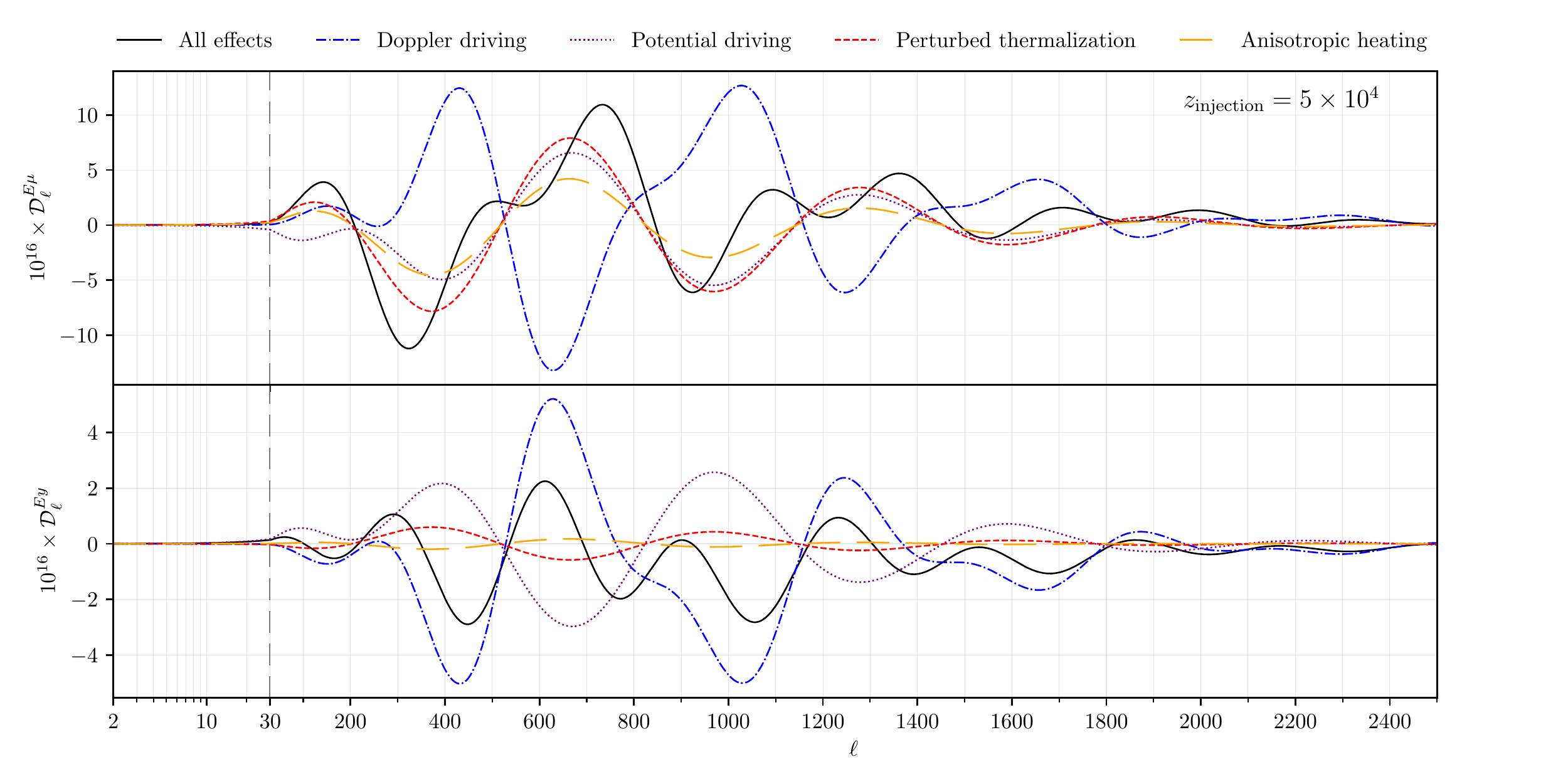}
\\[-1mm]
\includegraphics[width=0.90\columnwidth]{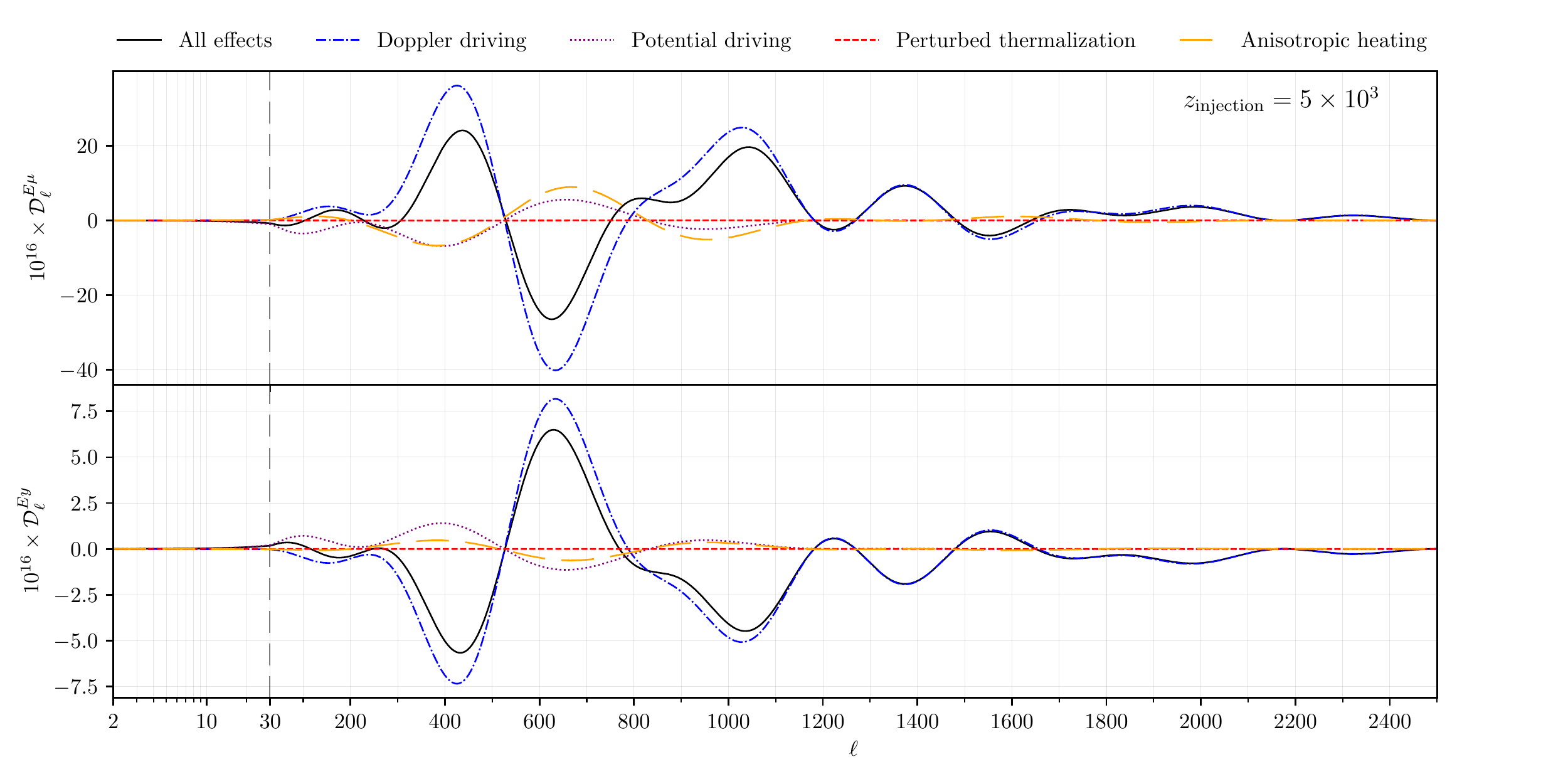}
\\[-1mm]
\caption{Three figures illustrating the $E \times \mu$ (top panel) and $E \times y$ (bottom panel) power spectra with various terms switched off. The figures from top to bottom show injection redshifts $\pot{5}{5}$, $\pot{5}{4}$ and $\pot{5}{3}$. The vertical dashed line shows a division between linearly-spaced $\ell$ values (left) and log-spaced values (right).}
\label{fig:SD_E_switches}
\end{figure}
\section{Correlations with $E$-modes}
\label{app:E_corr}
For completeness we show the correlations between distortion SEDs and temperature polarisation $E$-modes. While these are more difficult to interpret than the correlations with local temperature, they are a crucial contribution to the sensitivity within the forecasts. As seen in Fig.~\ref{fig:SD_E_zh_correlation} these have complex and strongly time dependent patterns. By inspecting Fig.~\ref{fig:SD_E_switches} this complexity can be attributed to a strong correlation driven through perturbed thermalisation and anisotropic heating which, unlike the cases seen earlier, are almost comparable in amplitude to the Doppler boosting term rather than being a small correction. Recalling that perturbed thermalisation and anisotropic heating largely cancel in the case of perturbed decay, this can explain the significant loss of power in the $\mu\times E$ (dashed blue lines) within Fig.~\ref{fig:forecasts_gamma_pert1} compared to Fig.~\ref{fig:forecasts_zh}.

As is very familiar by now, for late time injection these terms become less important and the signal relies more on the boosting mechanism. This together with the characteristic sign flip for $\mu^{(1)}$ for boosting early and late time injections yields a strong overall flip of the $\mu\times E$ power spectrum.

\end{document}